\documentclass[11pt,a4paper]{article}
\usepackage[utf8]{inputenc}
\usepackage{hyperref}
\usepackage{xcolor}
\hypersetup{colorlinks,linkcolor={blue},citecolor={red}}
\usepackage{subcaption}
\usepackage{graphicx}
\usepackage{amsmath}
\usepackage{amssymb}
\usepackage{relsize,exscale}
\usepackage{multicol}
\usepackage{booktabs}
\usepackage{makecell}
\usepackage{float}
\usepackage[a4paper, total={17cm, 25cm}]{geometry} 
\usepackage{caption}

\begin{document}

\title{Coherent Structure Dynamics of Heat Transfer in Wakes of an Inclined Elliptical Cylinder : A Novel Lagrangian Framework}

\author{
  Pratham Singh$^{1}$, Raghav Singhal$^2$, C. Jiten C. Kalita$^{1,*}$ \\
  \small $^1$Department of Mathematics, Indian Institute of Technology Guwahati, Assam 781039, India \\
  \small $^2$Department of Computer Science, University of California, Davis, CA 95616 USA \\
  \small $^*$Corresponding author: jiten@iitg.ac.in
}

\date{} 
\maketitle

\begin{abstract}
This work introduces a novel Lagrangian-based framework to analyze forced convective heat transfer in the unsteady wake of a heated elliptical cylinder inclined at angles ranging from $\theta = 0^\circ$ to $90^\circ $, in $15^\circ$ increments with $Pr = 0.71$ at a fixed Reynolds number of $Re = 100$. The framework correlates the temporal evolution of the surface-averaged Nusselt number with dynamic behavior of Lagrangian saddle points—formed at the intersection of repelling and attracting Lagrangian Coherent Structures (LCSs), extracted via Finite-Time Lyapunov Exponent (FTLE) fields.

The study is carried out within a precisely constructed observational domain, a previously unreported influential region in the near-wake where the trajectory analysis of the newly defined key saddle points (active saddle points) consistently align with the trends in surface heat transfer. This domain enables predictive identification of key transitional events in the Nusselt number profile, including local extrema and slope inflections, across all inclination angles.

The analysis reveals that oblique displacement of active saddle points enhances heat transfer by promoting the shedding of repelling LCSs, while parallel displacement leads to weakened heat transfer due to the delayed detachment of repelling coherent structures. The proposed framework enables the construction of a temporal function that closely replicates the monotonicity and transitional features of the Nusselt number evolution. Furthermore, threshold displacement metrics are defined for dominant repelling LCSs to correspond with peak heat transfer efficiency.

To the best of our knowledge, this is the first study to formulate a Lagrangian framework that directly links saddle point dynamics with the thermal performance of an inclined heated elliptical cylinder. The proposed methodology not only generalizes across a wide range of inclination angles but also provides a physically interpretable framework for predicting heat transfer enhancement based on coherent structure evolution in unsteady flows. 
\end{abstract}

\section{Introduction}

Forced convection heat transfer over bluff bodies plays a vital role in the thermal performance of many engineering systems. Bluff-body flows are encountered in a wide range of applications including heat exchangers, electronic cooling, nuclear reactor cores, building ventilation systems, and process engineering components. Cylindrical and elliptical geometries are particularly common due to their relevance in compact tube-bundle arrangements and cross-flow configurations \cite{Zukauskas1972, aark2015heat}. The unsteady, laminar wakes formed behind these bodies have a significant influence on heat transfer by periodically entraining and transporting heated or cooled fluid away from the body surface.

Historically, extensive research has focused on unsteady flow past bluff bodies and associated convective heat transfer across a range of cross-sectional geometries.  These studies relied heavily on Eulerian flow descriptions based on fixed spatial observations of flow variables such as streamlines, vorticity and pressure fields, and surface forces. These fixed spatial observations are made to extract time-averaged quantities such as drag coefficients, Strouhal numbers, and mean Nusselt numbers \cite{zdravkovich1997flow}.

Among the major contributions in this direction, Patel \textit{et al.}~\cite{patel1981flow} investigated the flow past an impulsively started elliptic cylinder at various angles of incidence ($\theta = 0^\circ$, $30^\circ$, $45^\circ$, and $90^\circ$). Ota \textit{et al.}~\cite{ota1983forced} conducted an experimental study on forced convection from an elliptical cylinder with an axis ratio of 1:2 over a Reynolds number range of 5000 to 90000. Shigeo kimura \cite{shigeo1988forced} addressed the transition from conduction to convection in forced heat transfer around elliptic cylinders—a geometry encompassing flat plates and circular cylinders as special cases. H. M. Badr \cite{badr1998forced} numerically studied laminar forced convection around an isothermal elliptical cylinder in uniform flow by solving the full Navier–Stokes and energy equations. Other contributions worth mentioning are those of Yue \textit{et al.}~\cite{investigation_yue_2010}, Paul \textit{et al.}~\cite{numerical_paul_2014}and Pawar \textit{et al.}~\cite{pawar2020forced}. Singhal \textit{et al.}~\cite{singhal2023comprehensive} performed a comprehensive numerical investigation of forced convective heat transfer from an inclined, heated elliptical cylinder subjected to uniform flow. 

Several studies have investigated the formation of coherent structures  \cite{shadden2005definition} in flow past the bluff bodies using Eulerian Diagnostics. G.Biswas \textit{et al.}~\cite{biswas2009effect} studied vortex shedding and heat transfer behind a heated circular cylinder at low Reynolds numbers ($Re = 10$ – $45$) using an SUPG-based finite element approach.  T.A smith and Y.Ventikos \cite{smith2021wing} conducted direct numerical simulations of flow over a flat-tipped wing at $Re = 10^4$, $4 \times 10^4$, and $10^5$ to analyze tip vortex formation and its interaction with boundary layer dynamics.  Coherent vortical structures were identified using the $Q$-criterion and the normalized $\hat{Q}$-criterion, revealing detailed vortex topology and transition mechanisms. While the approaches discussed above have yielded foundational insights and are robust in capturing the instantaneous structure of the flow, eulerian tools such as streamlines, vorticity are limited in their ability to track fluid particle trajectories, identify material transport barriers, and analyze long-term heat transport and mixing. Particularly in laminar wakes at low Reynolds numbers, where vortex shedding dominates the downstream region, time-averaged data can fail to capture the underlying mechanisms of heat and momentum exchange \cite{he2016evolution}. Huang \textit{et al.} \cite{huang2022lagrangian} demonstrated that regions with dense closed streamlines in unsteady flows do not necessarily indicate coherent vortices. While vorticity contours offer a more precise approximation of vortex boundaries than streamlines, their use remains subjective due to the absence of a universal criterion for contour selection. Additionally, confirming the coherence of a selected vorticity contour requires tracking its advective evolution over time. In other cases, eulerian Diagnostics often obscure the dynamic, time-dependent processes that govern scalar transport in unsteady flows. They often rely on user-specified thresholds, making the detected vortex boundaries subjective, and they lack information about the flow’s temporal evolution or history \cite{haller2005objective} and also lack objectivity \cite{haller2015lagrangian}. These limitations make it challenging to fully understand the Lagrangian dynamics and coherent structures that govern unsteady flow behavior.

To overcome these limitations, the Lagrangian approach has emerged as a robust alternative, offering a frame-invariant, temporally resolved perspective on fluid transport. Central to this approach is the concept of the flow map, which tracks the evolution of fluid particles over finite time intervals. The deformation gradient of this map, often quantified through the Finite-Time Lyapunov Exponent (FTLE), enables the identification of Lagrangian Coherent Structures (LCSs)---material lines that act as transport barriers or conduits \cite{haller2015lagrangian, haller2000lagrangian}. These LCSs delineate regions of high stretching and folding in the flow, thus playing a pivotal role in mixing, stirring, and scalar dispersion.
Notable contributions to the development of LCS theory in the context of flow past bluff bodies stem from extensive research efforts aimed at uncovering the underlying transport structures and organizing mechanisms in unsteady wakes.
Peng and Dabiri \cite{peng2009transport} introduced a dynamical systems approach to identify particle Lagrangian Coherent Structures (pLCS) as transport boundaries in inertial particle flows. Applying this to jellyfish–plankton interactions, they used experimental flow data and a modified Maxey–Riley equation to reveal prey capture zones. Their framework links particle dynamics to biological and mechanical factors and is extendable to a wide range of multiphase and reacting flows. ecent

Recent works in this direction worth mentioning are  \cite{huang2015detection, chen2016using, rockwood2017detecting}. Amongst these, Cao \textit{et al.} \cite{cao2021forced} applied the LCS framework to investigate forced convection heat transfer around a circular cylinder in the laminar flow regime. They simulated coupled flow and heat transfer fields and computed LCSs via FTLE analysis. By correlating thermal and vorticity patterns, lift and drag coefficients, and Nusselt number with the evolving LCSs at $Pr = 0.7$ and $Re = 20$–$180$, they provided new insights into mass and energy transport mechanisms by providing an insightful connection of attracting/ repelling LCSs with convective heat transfer and vortex dynamics. Their study demonstrated the potential of LCS-based analysis in advancing the understanding of convection heat transfer phenomena.

The elliptical cylinder offers a versatile geometry for studying forced convection heat transfer, bridging the gap between circular cylinders and flat plates. Its tunable aspect ratio and inclination enable systematic exploration of flow separation, vortex shedding, and thermal transport under varying conditions \cite{singhal2023comprehensive}. Unlike circular cylinders, elliptical geometries allow enhanced heat transfer performance when aligned with the flow and exhibit richer unsteady dynamics at oblique incidences \cite{ota1983forced}. 

In this study, we leverage LCS-based tools to analyze the transient heat transfer in forced convection over a heated elliptical cylinder at $Re = 100$ for various angles of attack ranging from $\theta=0^\circ$ to $\theta=90^\circ$ with an increment of $\theta=15^\circ$. We, for the first time, propose a novel Lagrangian Coherent Structure framework that systematically correlates the time evolution of the surface-averaged Nusselt number with the trajectory of Lagrangian saddle points in the unsteady wake. These dynamically significant saddle points \cite{huang2015detection, rockwood2017detecting} are formed at the intersections of attracting and repelling LCSs, akin to the intersection of stable and unstable manifolds in dynamical systems theory \cite{rockwood2017detecting, green2010using}. This is probably the first demonstration of a direct correlation between Lgrangian saddle point trajectories and the temporal evolution of the surface-averaged Nusselt number in bluff body wakes. To this end, we compute both forward-time and backward-time Finite-Time Lyapunov Exponent (FTLE) fields, filtering them with adaptive thresholds to isolate the most dynamically relevant structures. These structures are then tracked across a vortex shedding cycle to understand their role in mediating convective heat transfer. A key feature of our framework lies in defining a novel concept of active saddle points and an observational domain. 

Furthermore, we perform a sensitivity analysis to explore how variations in the FTLE extraction threshold affects the clarity and accuracy of LCS extraction, the positional sensitivity of identified saddle points, and the geometry of the observational domain. This ensures that the proposed framework remains robust across a range of visualization parameters. Our research contributes to the growing body of work seeking to bridge the gap between Lagrangian kinematics and heat transfer physics. By integrating rigorous saddle point dynamics with thermal field characterization, this study not only enhances the fundamental understanding of heat transport in bluff body wakes but also offers practical insights into flow control and thermal optimization strategies for real-world engineering systems.

The structure of the paper is as follows: Section~\ref{sec:numerical} outlines the numerical methodology used to extract velocity fields and isotherms, which serve as the input for LCS computation. Section~\ref{sec:ftle} provides a brief overview of the Finite-Time Lyapunov Exponent (FTLE) as a Lagrangian diagnostic, along with the ridge extraction procedure used to identify coherent structures. Section~\ref{sec:framework} presents the proposed framework by introducing key definitions—such as active saddle points and the observational domain—and applies it across a range of inclination angles to analyze the temporal evolution of the surface-averaged Nusselt number in relation to saddle point dynamics. Finally, Section~\ref{sec:conclusion} summarizes the key findings and offers concluding remarks.

\section{The Problem, the Governing Equations and Numerical Methodology}\label{sec:numerical}

\begin{figure}[H] 
    \centering
    \includegraphics[width=1.1\textwidth]{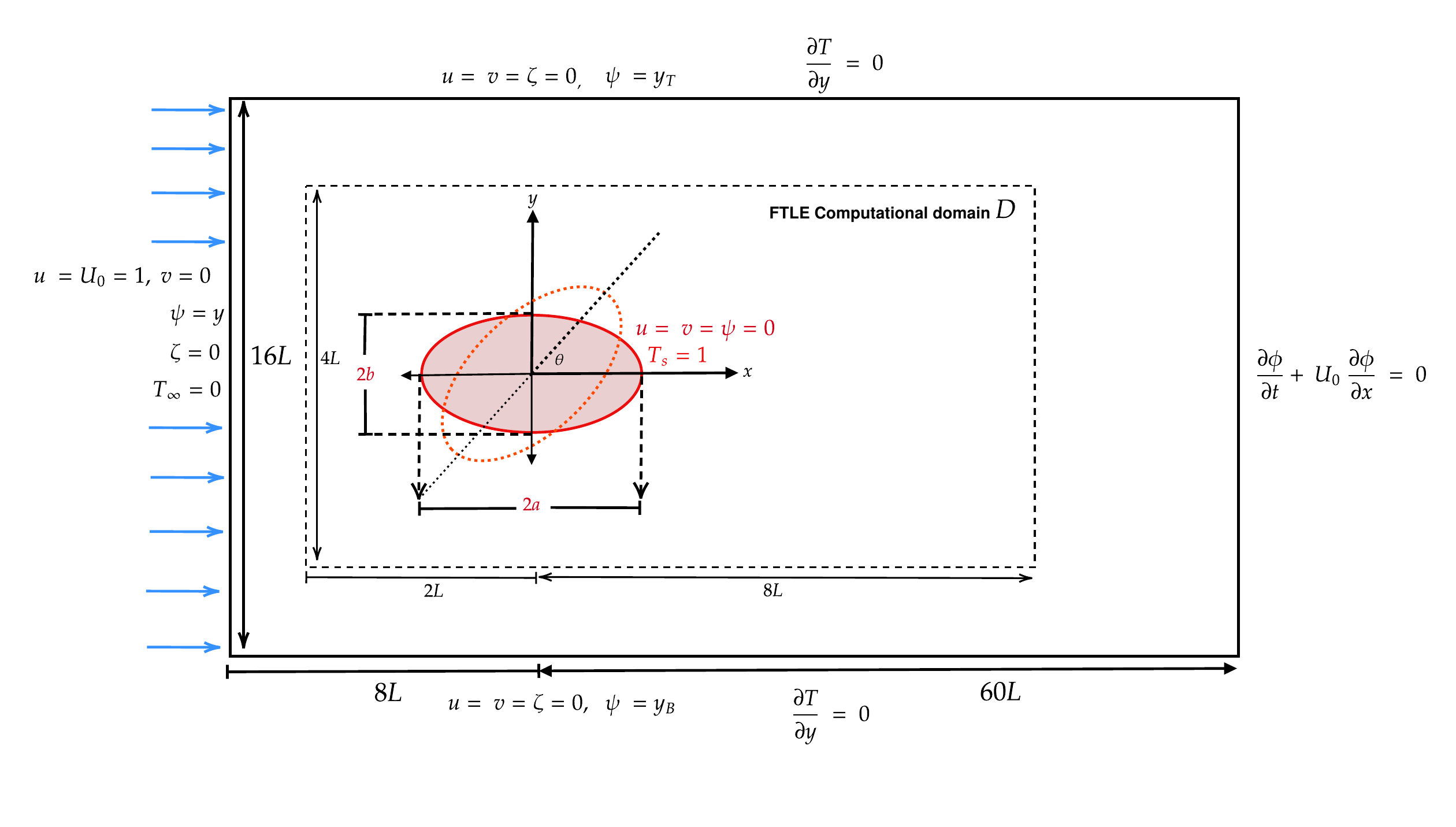}
    \caption{Schematic Representation for Flow Description}
    \label{fig:flowmapx}
\end{figure}

The problem considered here is that of the two-dimensional, laminar flow past a heated elliptical cylinder immersed in an incompressible fluid inclined to the uniform free stream. The fluid has constant thermophysical properties, and buoyancy effects due to gravity are neglected. The dimensionless equations governing the flow are the streamfunction-vorticity ($\psi$-$\zeta$) form of the Navier-Stokes equations and the energy equation, given by:

\begin{align}
    \nabla^2 \psi &= -\zeta, \\
  \label{sv}  \frac{\partial \zeta}{\partial t} + u \frac{\partial \zeta}{\partial x} + v \frac{\partial \zeta}{\partial y} &= \frac{1}{Re} \nabla^2 \zeta, \\ 
    \frac{\partial T}{\partial t} + u \frac{\partial T}{\partial x} + v \frac{\partial T}{\partial y} &= \frac{1}{Re \cdot Pr} \nabla^2 T,
\end{align}
where $u$, $v$ are the velocities along $x$ and $y$ directions respectively, $t$ is the time, $T$ the temperature, $Re = U_0 L/\nu$ is the Reynolds number with $U_0$ being the free-stream (or characteristic) velocity, $L$ the characteristic length assumed to be the length of the minor axis and $\nu$ the kinematic viscosity. Also  $Pr = \nu/\alpha$ is the Prandtl number with $\alpha$ being the thermal diffusivity. Throughout all the simulations, $Re$ and $Pr$ are fixed as $100$ and $0.71$ respectively, and an aspect ratio $AR=2/3$, the ratio of the semi-minor and the semi-major axes $b$ and $a$ of the elliptic cylinder has been maintained.

The streamfunction $\psi$ and the vorticity $\zeta$ are related to the velocities $u$ and $v$ by
\begin{equation}
u= \frac{\partial \psi}{\partial y}, \quad v = -\frac{\partial \psi}{\partial x}, \quad \zeta = \frac{\partial v}{\partial x} - \frac{\partial u}{\partial y}.
\end{equation}

The surface temperature of the cylinder is kept constant at $T_s = 1$, while the free-stream temperature is $T_\infty = 0$. The temperature difference between the cylinder surface and the ambient fluid, $\Delta T = T_s - T_\infty$, is assumed small enough not to affect fluid properties significantly.

The initial and boundary conditions are as follows:
\begin{itemize}
    \item At $t = 0$, the flow field is initialized with $u = 1$, $v = 0$, and $T = 0$.
    \item At the inlet, the velocity and temperature remain uniform: $u = U_0=1$, $v = 0$, and $T = 0$.
    \item At the outlet, convective boundary conditions are applied:
    \[
        \frac{\partial \Phi}{\partial t} + U_0 \frac{\partial \Phi}{\partial x} = 0, \quad \text{for } \Phi = u, v, \psi, \zeta, T.
    \]
    \item The top and bottom domain boundaries are treated with free-slip and adiabatic conditions:
\begin{align*}
u = v = \zeta = 0, \quad \frac{\partial T}{\partial y} = 0, \quad \psi = y_T &\quad \text{on the top (T) boundary,} \\
\psi = y_B &\quad \text{on the bottom (B) boundary.}
\end{align*}

    \item On the elliptical cylinder surface, no-slip and constant temperature conditions are enforced:
    \[
        u = v = \psi = 0, \quad T = 1.\] The vortcity $\zeta$ is computed using equation \eqref{sv} by adopting the same strategy employed in \cite{singhal2022efficient}.
    
\end{itemize}

The computational domain in space is chosen as the rectangular region $[-8,60] \times [-8,8]$ on the $xy$-plane such that the location of the center of the ellipse is  $(x,y)=(0,0)$ with the length of its minor axis being $2b=1.0$. The schematic of the flow is shown in Fig. \ref{fig:flowmapx}. Following \cite{singhal2022efficient}, we have chosen a non-uniform grid of size $551 \times 441$ with varying time step $\Delta t=0.0025$ to $0.01$ depending upon the angle of inclination in our computation. Once the velocities $u$ and $v$ are known, the data is post-processed further in the FTLE (see Fig. \ref{fig:flowmapx}) domain $D$ on a uniform grid of size $951 \times 401$ set over the region $[-2, 8] \times [-2,2]$ in order to detect the LCSs. For the Reynolds number chosen in the study, the flow is unsteady and is characterised by the formation of the well-known von Karman vortex street. Here, two rows are vortices are being shed alternately, once from the upper and next from the lower surface of the cylinder. The onset of vortex shedding depends upon the angle of inclination of the cylinder to the free-stream direction. In the analysis that follows, it is assumed that the non-dimensional time to complete one cycle of vortex shedding is $T$.  

Building upon the methodology developed by Kalita \textit{et al.} \cite{kalita2002class}, Singhal and Kalita proposed a Higher Order Compact (HOC) explicit jump Immersed Interface Method (IIM) for 2D transient problems involving bluff bodies immersed in incompressible viscous flows on Cartesian mesh in  \cite{singhal2022efficient}. The above problem utilized the unconditionally stable IIM scheme \cite{kalita2002class} developed in  \cite{singhal2022efficient} to simulate and analyze forced convection heat transfer over an elliptic cylinder at varying angle of incidences  \cite{singhal2023comprehensive}. The Detailed solution methodology is provided in the work of Singhal and Kalita \cite{singhal2022efficient}.

\subsection{Nusselt number}
The non-dimesnional parameter Nusselt number is the dimensionless temperature gradient at the surface providing a measure of the convective heat transfer occurring at the surface (Incopera and Dewitt in \cite{incropera1996fundamentals}) of heated bodies. In this study, the Nusselt number characterizes the rate of heat transfer across the fluid around the heated elliptic cylinder. The quantitative parameter indicating heat transfer, i.e. the local Nusselt number $Nu_l$, is defined as

\begin{equation}
Nu_{l}= - \frac{\partial T}{\partial n}
\end{equation}
where \( n \) is the direction normal to the boundary of the bluff body. A novel approach in calculating $Nu_l$, using the flow information in the direction normal to the bluff body’s surface on a cartesian grid is given in \cite{singhal2023comprehensive}.

The surface averaged Nusselt number is given by
\begin{equation}
Nu_{av} = \frac{1}{W} \int_W Nu \, dS
\end{equation}
where \( W \) is the surface area of the cylinder. For simplicity, we will omit the subscript in $Nu_{av}$ and just use $Nu$ for surface avergaed Nusselt number.

In this paper, we utilize the velocity and temperature fields ($u$, $v$, $T$) obtained from the properly validated numerical simulations in \cite{singhal2022efficient} to compute the Finite-Time Lyapunov Exponents and to visualize Isotherms (Fig. \ref{fig:ftle-bftle-compare}) and surface avergaed Nusselt number evolution (Fig. \ref{fig:SurfaceNu}) in association with the FTLE plots.

\section{LCS and FTLE}\label{sec:ftle}

Lagrangian Coherent Structures (LCSs) are time-evolving material surfaces in unsteady flows that act as transport barriers, organizing the motion of fluid elements. They can be identified using the Finite-Time Lyapunov Exponent (FTLE) field, which quantifies the exponential separation of neighbouring fluid particles over a finite time interval. When computed in forward time, FTLE highlights repelling LCSs, which act like finite-time unstable manifolds; in backward time, it reveals attracting LCSs, analogous to finite-time stable manifolds. The ridges of the FTLE field—regions of locally maximal FTLE values—correspond to these LCSs and delineate zones of maximal material deformation. As such, LCSs provide a powerful framework for identifying coherent structures that govern mixing, separation, and transport in complex unsteady flows.

In this work, we consider a time-dependent dynamical system on a bounded domain $D \subseteq \mathbb{R}^2$ in a two-dimensional fluid flow, though similar considerations can be made in higher dimensional space. The position of a fluid particle $\mathbf{x}(t) = (x(t), y(t))\in D$ with velocity field
$\mathbf{v}(\mathbf{x}(t), t) : D \times I \rightarrow {\mathbb{R}^2}$, $I \subset \mathbb{R} $ is governed by the differential equation
\begin{equation}
\frac{d}{dt} \mathbf{x}(t) = \mathbf{v}(\mathbf{x}(t), t), \qquad \mathbf{x}(t_0; t_0, \mathbf{x}_0) = \mathbf{x}_0\label{eq:1}
\end{equation}
where $\mathbf{x}_0$ denotes the initial position of the particle $\mathbf{x}$ at initial time $t_0$. As $ \mathbf{x}(t)$ is dependent on the parameters $\mathbf{x}_0$ and $t_0$, we refer $ \mathbf{x}(t)$ as $\mathbf{x}(t; t_0, \mathbf{x}_0)$ whenever required. To obtain particle trajectory $\mathbf{x}(t; t_0, \mathbf{x}_0)$ from $t_0$ to time $t$, we integrate eq.\eqref{eq:1} as

\begin{equation}
\mathbf{x}(t; t_0, \mathbf{x}_0) = \mathbf{x}_0 + \int_{t_0}^{t} \mathbf{v}(\mathbf{x}(s; t_0, \mathbf{x}_0), s) \, ds, \label{eq:2}
\end{equation}

 and define a mapping in domain $D$ called  \textit{flow map} as

\begin{equation}
\mathbf{\Phi}_{t_0}^{t} : D \rightarrow D : \mathbf{x}_0 \mapsto \mathbf{\Phi}_{t_0}^{t}(\mathbf{x}_0) = \mathbf{x}(t; t_0, \mathbf{x}_0) \label{eq:3}
\end{equation}

The mapping $\mathbf{\Phi}_{t_0}^{t}(\mathbf{x}_0)$ maps a fluid particle from its initial position $ \mathbf{x}_0\in D$ at time $t_0$ to its position at a later time $t$ in $D$ under the influence of $\mathbf{v}(\mathbf{x}(t), t)$.

\begin{figure}[H] 
    \centering
    \includegraphics[width=0.7\textwidth]{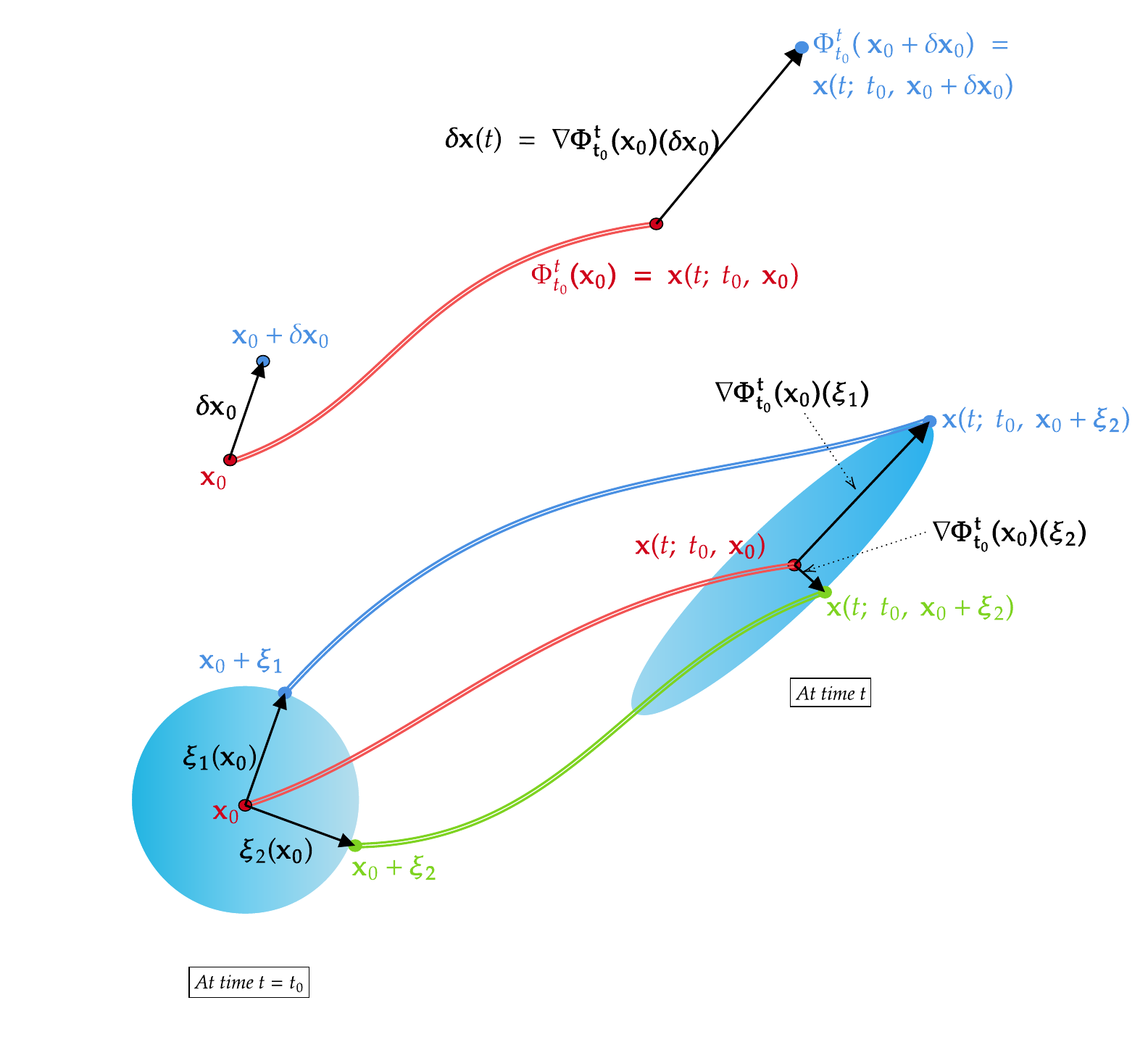}
    \caption{Schematic Represntation of Flow map and Perturbation Transport.}
    \label{fig:flowmap}
\end{figure}


The flow map Jacobian, denoted as $\nabla \mathbf{\Phi}_{t_0}^{t}(\mathbf{x}_0)$ is the gradient of the flow map which  quantifies how infinitesimal fluid elements deform over time. It encapsulates the linearized behavior of fluid particle trajectories originating from a small neighborhood around the initial position $\mathbf{x}_0$. At time $t_0$, the evolution of an infinitesimal perturbation $\delta \mathbf{x}(t_0) = \delta \mathbf{x}_0$, introduced in the vicinity ($\epsilon$-neighbourhood) of a reference particle $\mathbf{x}_0$ is governed by the deformation gradient of the flow map as illustrated in Fig.\ref{fig:flowmap}. Consider a second fluid particle $\mathbf{y}(t_0) = \mathbf{x}_0 + \delta \mathbf{x}_0 = \mathbf{y}_0$ at time $t_0$ with a slightly perturbed initial position than that of $ \mathbf{x}_0$, where $\delta \mathbf{x}_0$ denotes an infinitesimal displacement vector oriented in an arbitrary direction within the flow field as shown in Fig.\ref{fig:flowmap}. With the help of Taylor series approximation, we obtain the position of the advected perturbation at time t as:

\begin{equation}
\begin{aligned}
\delta \mathbf{x}(t) &= \mathbf{y}(t; t_0, \mathbf{y}_0) - \mathbf{x}(t; t_0, \mathbf{x}_0) 
= \boldsymbol{\Phi}_{t_0}^{t}(\mathbf{x}_0 + \delta \mathbf{x}_0) - \boldsymbol{\Phi}_{t_0}^{t}(\mathbf{x}_0) \\
&= \nabla \boldsymbol{\Phi}_{t_0}^{t}(\mathbf{x}_0) \, \delta \mathbf{x}_0 + o(|\delta \mathbf{x}_0|^2) \\
&\approx \nabla \boldsymbol{\Phi}_{t_0}^{t}(\mathbf{x}_0) \, \delta \mathbf{x}_0
\end{aligned}
\end{equation}

We see that $\delta \mathbf{x}(t)  \approx \nabla \mathbf{\Phi}_{t_0}^{t}(\mathbf{x}_0)  \delta \mathbf{x}(t_0)$ as $|\delta \mathbf{x}_0| \to  0$, and the magnitude of the advected perturbation is given by:
\begin{equation}
\begin{aligned}
|\delta \mathbf{x}(t)| 
&= \sqrt{ \left\langle  
\nabla \boldsymbol{\Phi}_{t_0}^{t}(\mathbf{x}_0) \, \delta \mathbf{x}_0,\ 
\nabla \boldsymbol{\Phi}_{t_0}^{t}(\mathbf{x}_0) \, \delta \mathbf{x}_0 
\right\rangle } \\
&= \sqrt{ 
\left\langle \delta \mathbf{x}_0,\ 
\left[ \nabla \boldsymbol{\Phi}_{t_0}^{t}(\mathbf{x}_0) \right]^T 
\nabla \boldsymbol{\Phi}_{t_0}^{t}(\mathbf{x}_0) \, \delta \mathbf{x}_0 
\right\rangle } \\
&= \sqrt{ 
\left\langle \delta \mathbf{x}_0,\ 
 \mathbf{C}_{t_0}^t(\mathbf{x}_0) \, \delta \mathbf{x}_0 
\right\rangle }
\end{aligned}
\end{equation}

where  $ \left[ .  \right]^T$ represents the matrix transpose, and $ \mathbf{C}_{t_0}^t(\mathbf{x}_0) = \left[ \nabla \mathbf{\Phi}_{t_0}^{t}(\mathbf{x}_0)  \right]^T\nabla \mathbf{\Phi}_{t_0}^{t}(\mathbf{x}_0) $ is a $d \times d$ symmetric positive definite matrix called the Cauchy-Green deformation tensor. Here $d$ is the dimension of the space under consideration.

This tensor describes how the position vector $\mathbf{x}(t; t_0, \mathbf{x}_0)$ is stretched in various directions as it evolves with the flow. Let the eigenvalues of the matrix $\mathbf{C}_{t_0}^t(\mathbf{x}_0)$ be $\lambda_i (\mathbf{x}_0)$ and the corresponding eigenvectors be $\xi_i (\mathbf{x}_0)$ for $i= 1,2, ..., d$. Unless otherwise stated, the notations $\lambda_i (\mathbf{x}_0)$ and  $\xi_i (\mathbf{x}_0)$ will omit explicit dependence on  $ \mathbf{x}_0$. Due to the incompressibility constraint, the product of the eigenvalues of the Cauchy--Green deformation tensor satisfies 
$\prod_{i=1}^{d} \lambda_i = 1$. In two dimensions $(d = 2)$, this condition leads to $\lambda_2 < 1 < \lambda_1$, indicating area-preserving but anisotropic stretching.
When the initial displacement vector  $\delta \mathbf{x}_0$ is oriented along the dominant eigenvector $\xi_1$ associated with the dominant eigenvalue $\lambda_1$ of the matrix $\mathbf{C}_{t_0}^t(\mathbf{x}_0)$, maximum stretching evidently occurs by the largest possible factor $\sqrt{\lambda_1}$ . Similarly initial perturbations oriented with the weakest eigenvector $\xi_1$ will be compressed by the factor  
$\sqrt{\lambda_1}$.  In Fig.\ref{fig:flowmap}, the initial perturbation is aligned in the direction of eigenvector $\xi_1$ corresponding to the dominant eigenvalue $\lambda_1$ and the expression for the maximum magnitude of the evolved pertubation is given by:

\begin{equation}
\max \|\delta \mathbf{x}(t)\| 
= \sqrt{\lambda_{\max}(\mathbf{C}_{t_0}^{t})( \mathbf{x_0})} \, \|\delta \mathbf{x}(t_0)\|, \label{eq:6}
\end{equation}

To study the exponential growth in the separation of nearby trajectories over the time interval $\Delta t = t - t_0$,  the maximum stretching factor $\sqrt{\lambda_{\max}(\mathbf{C}_{t_0}^{t})} $ is taken as:

\begin{equation}
\sqrt{\lambda_{\max}(\mathbf{C}_{t_0}^{t})} = e^{\sigma_{t_0}^{t}(\mathbf{x}_0) |\Delta t|} \label{eq:7}
\end{equation}
and the Finite-Time Lyapunov Exponent (FTLE) is defined as
\begin{equation}
\sigma_{t_0}^{t} = \frac{1}{|\Delta t|} \ln \sqrt{\lambda_{\max}(\mathbf{C}_{t_0}^{t})},
\label{eq:ftle}
\end{equation}
where $\lambda_{\max}(\mathbf{C}_{t_0}^{t})$ is the largest eigenvalue of the Cauchy--Green deformation tensor $\mathbf{C}_{t_0}^{t}$, and $\Delta t = t - t_0$ is the integration time.

\begin{figure*}[ht!]
\centering
     \includegraphics[width=13cm, height=9.5cm]{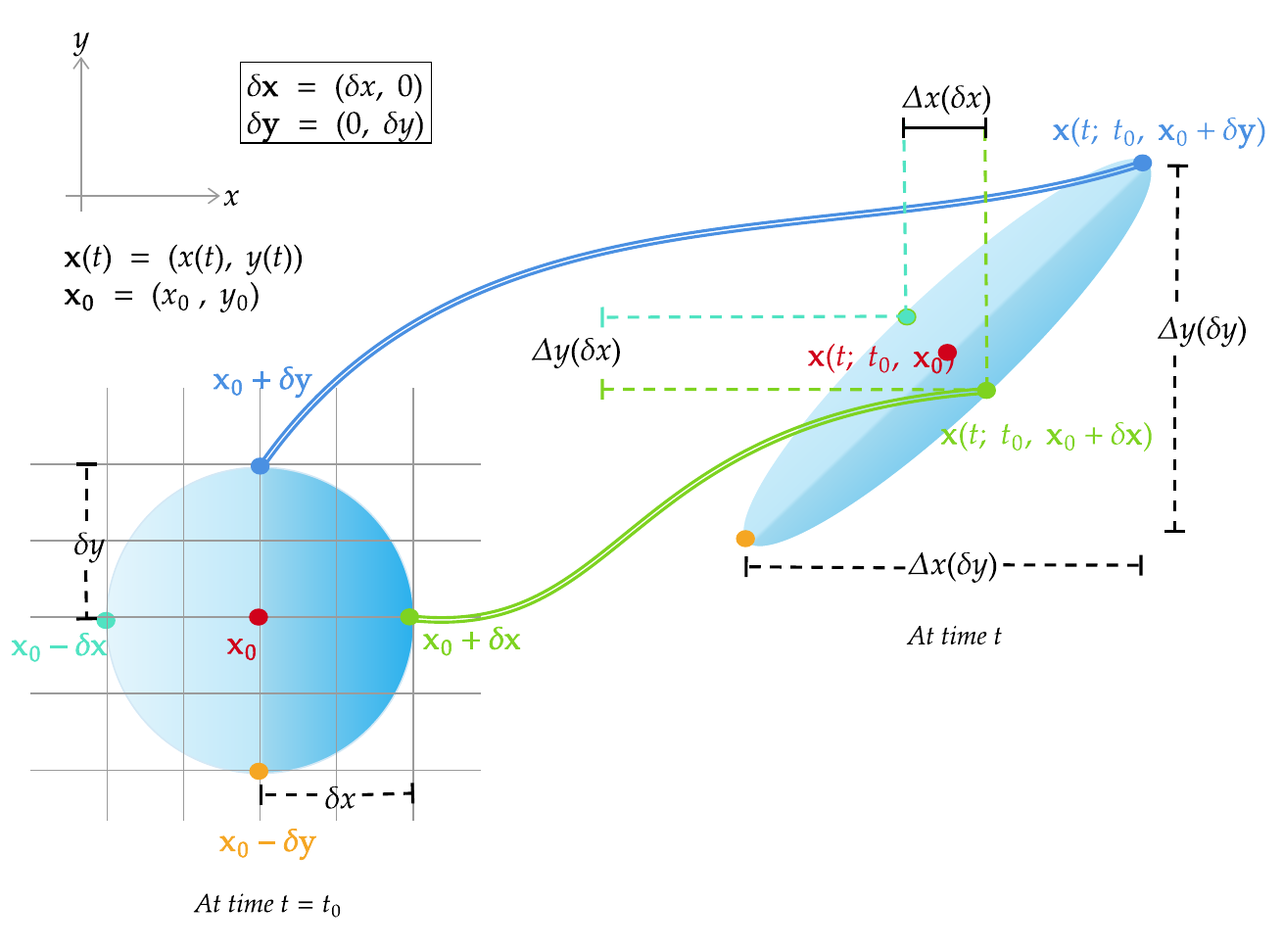}
     \caption{Schematic Representation for computing Flow Map Jacobian}
     \label{fig: jacobian}
\end{figure*}

\begin{figure}[htbp]
\centering

\begin{subfigure}[t]{0.495\textwidth}
    \centering
    \includegraphics[width=\linewidth]{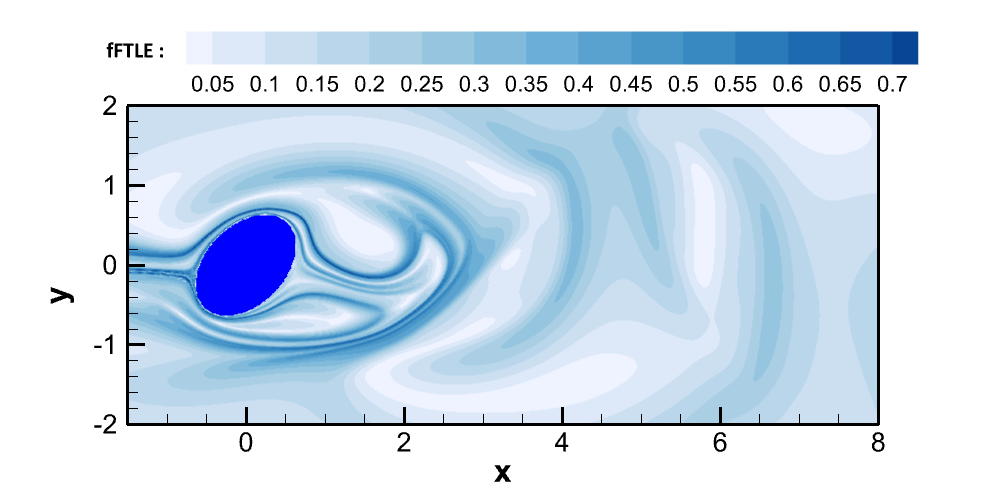}
    \caption*{(a) fFTLE scalar field}
    \includegraphics[width=\linewidth]{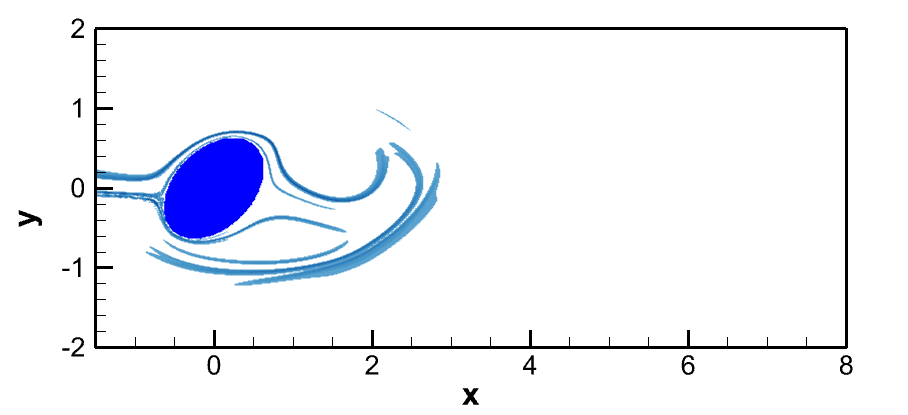}
    \caption*{(c) Ridges in fFTLE field with 50$\%$ threshold}
\end{subfigure}
\hfill
\begin{subfigure}[t]{0.495\textwidth}
    \centering
    \includegraphics[width=\linewidth]{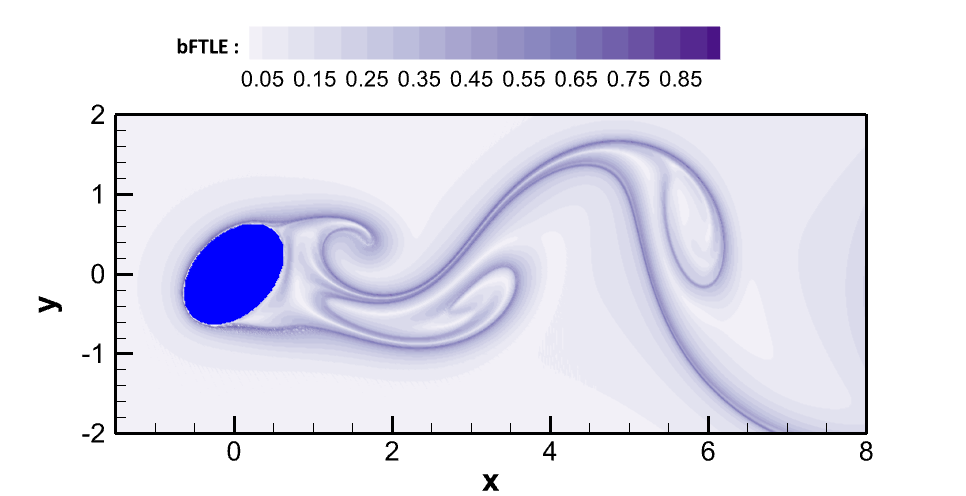}
  \caption*{(a) bFTLE scalar field}
    \includegraphics[width=\linewidth]{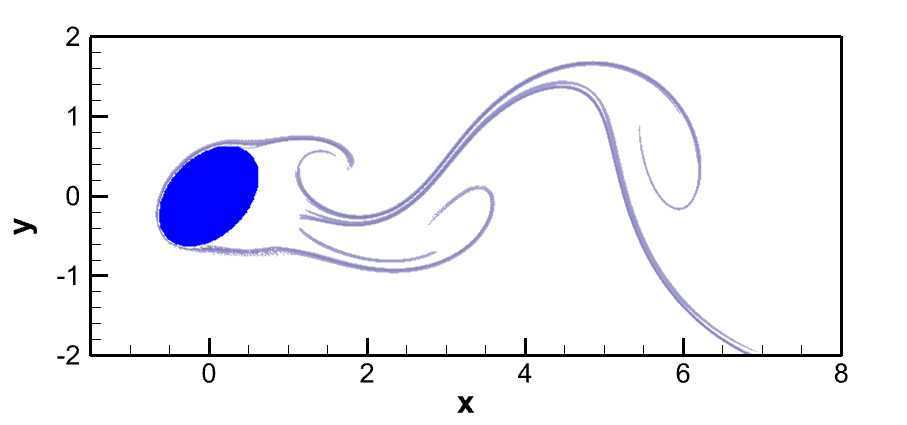}
\caption*{(d) Ridges in bFTLE field with 50$\%$ threshold}

\end{subfigure}

 \caption{(a)Forward-time FTLE (b) Backward-time FTLE (c) Filtered fFTLE field (d) Filtered bFTLE field}
    \label{fig: LCSimage}
\end{figure}

\begin{figure}[htbp]
\centering

\includegraphics[width=\textwidth]{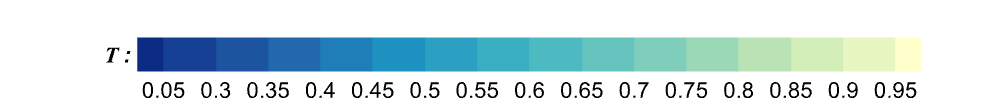}

\vspace{0.5em} 

\begin{subfigure}[t]{0.47\textwidth}
    \centering
    \includegraphics[width=\linewidth]{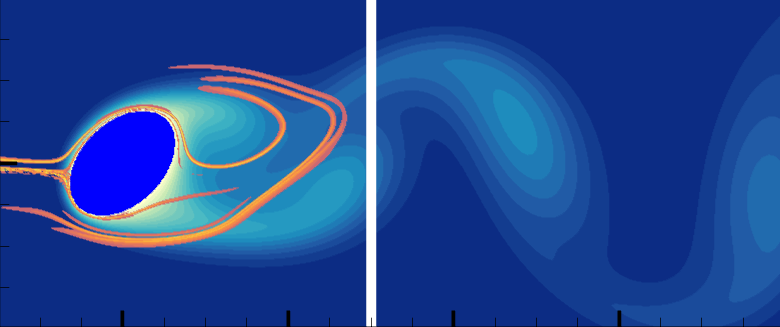}
    \caption*{$t = T$}
    \includegraphics[width=\linewidth]{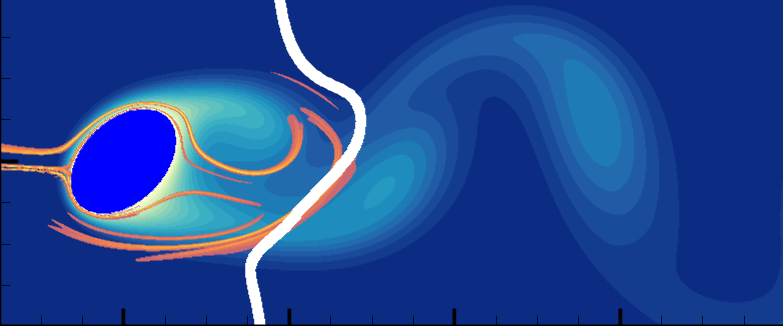}
    \caption*{$t = 0.75T$}
    \includegraphics[width=\linewidth]{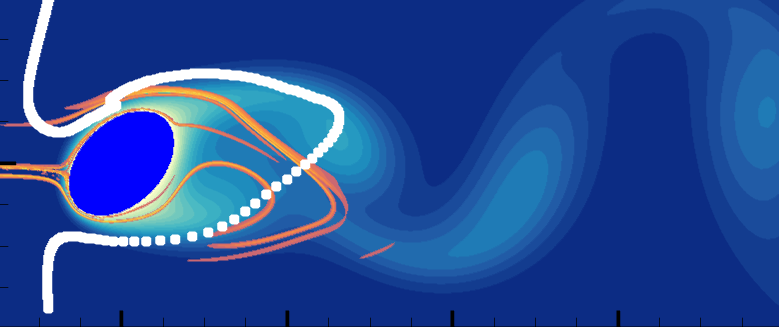}
    \caption*{$t = 0.5T$}
    \includegraphics[width=\linewidth]{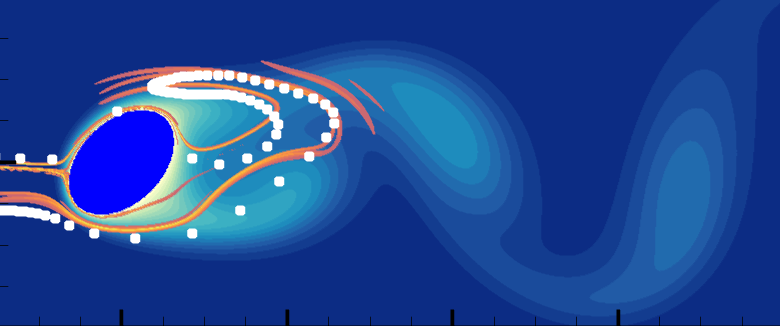}
    \caption*{$t = 0.25T$}
    \includegraphics[width=\linewidth]{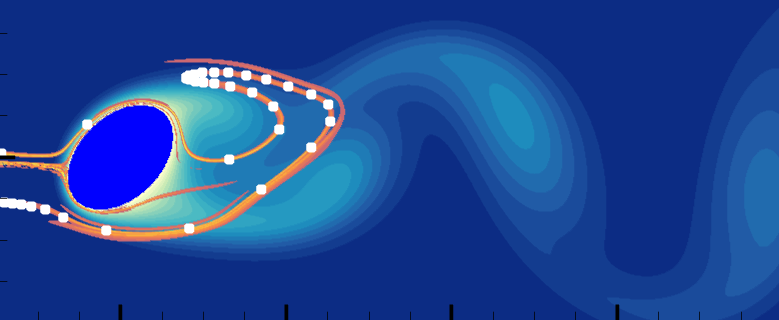}
    \caption*{$t = 0$}
    \caption{}
    \label{fig:fftletracers}
\end{subfigure}
\hfill
\begin{subfigure}[t]{0.47\textwidth}
    \centering
    \includegraphics[width=\linewidth]{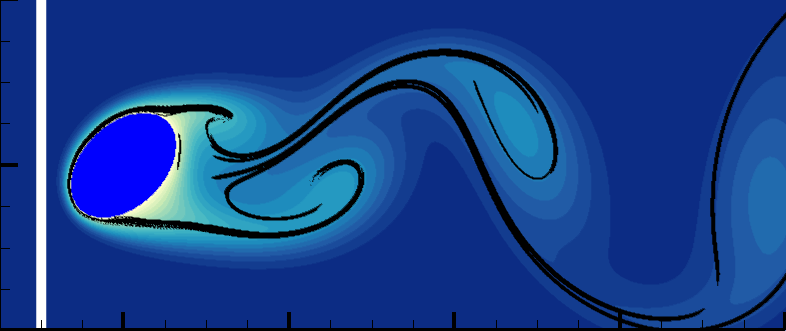}
    \caption*{$t = 0$}
    \includegraphics[width=\linewidth]{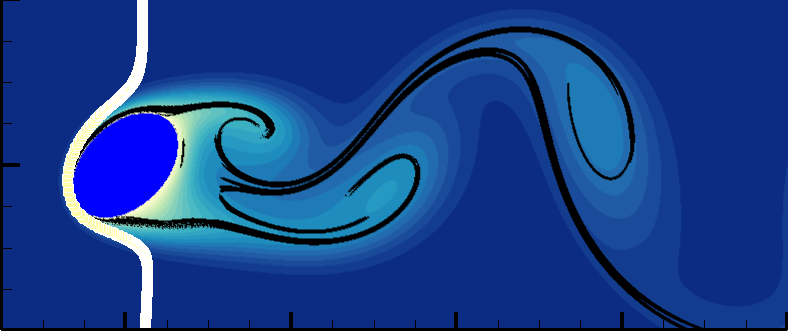}
    \caption*{$t = 0.25T$}
    \includegraphics[width=\linewidth]{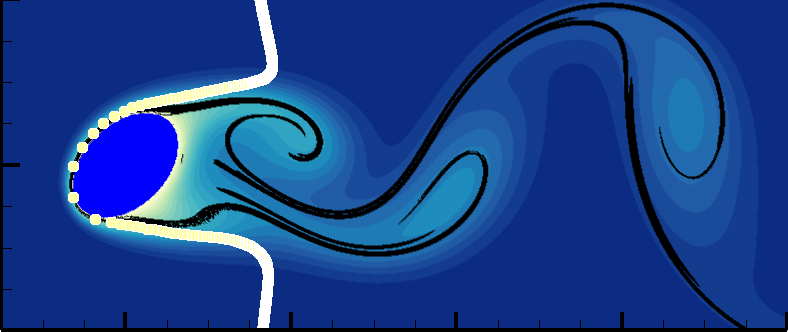}
    \caption*{$t = 0.5T$}
    \includegraphics[width=\linewidth]{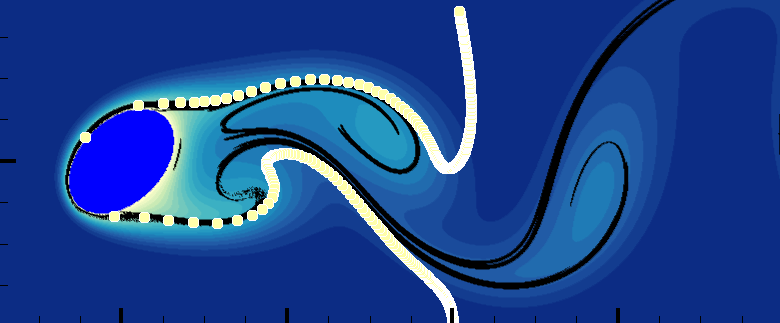}
    \caption*{$t = 0.75T$}
    \includegraphics[width=\linewidth]{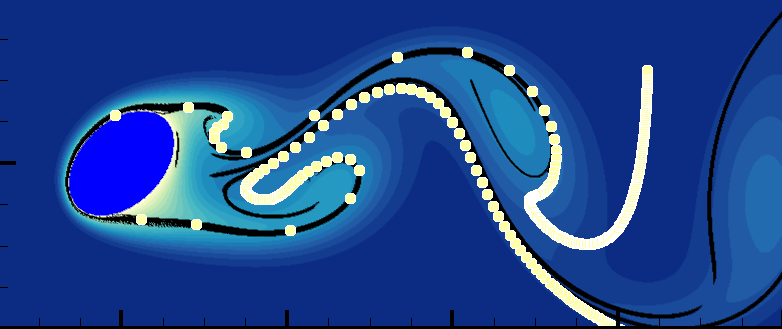}
    \caption*{$t = T$}
    \caption{}
    \label{fig:bftletracers}
\end{subfigure}

\caption{Evolution of passive tracer particles (white) and Isotherms (in background) at different phases within one shedding cycle of period $T$: (a) Repelling LCSs (fLCSs) in orange and (b) Attracting LCSs (bLCSs) in black.}
\label{fig:ftle-bftle-compare}
\end{figure}

The FTLE quantifies the rate at which nearby fluid particles diverge over a finite time interval. This scalar field is computed at each grid point within a discretized domain. For a given grid point selected as the initial particle position \( \mathbf{x}_0 \), a small perturbation is introduced to generate four neighboring positions at time \( t_0 \) as shown in in Fig.\ref{fig: jacobian}. These perturbed positions are constructed by applying positive and negative displacements along each principal coordinate direction. The flow map $\mathbf{\Phi}_{t_0}^{t}(\mathbf{x}_0)$ is then used to advect these particles from \( t_0 \) to a later time \( t \), yielding their new positions using  eq.~\eqref{eq:2}. 

To approximate the gradient of the flow map $\nabla \mathbf{\Phi}_{t_0}^{t}(\mathbf{x}_0)$, central finite difference approxmation using the advected positions of the perturbed particles is employed to get:
\[
\renewcommand{\arraystretch}{1.3}
\left(
\begin{array}{cc}
\dfrac{\Delta x(\delta x)}{2|\delta_x|} & \dfrac{\Delta x(\delta y)}{2|\delta_y|} \\[2pt]
\dfrac{\Delta y(\delta x)}{2|\delta_x|} & \dfrac{\Delta y(\delta y)}{2|\delta_y|}
\end{array}
\right)
\renewcommand{\arraystretch}{1.0}
\]

Here, \( \delta x \) and \( \delta y \) denote the initial displacements between the reference position \( \mathbf{x}_0 \) and its perturbed neighbors in the horizontal and vertical directions, respectively. These are typically chosen to match the grid spacing of the discretized domain. After advection of horizontally perturbed particles by the flow map, the horizontal and vertical displacement between the advected positions of the perturbed particles are denoted by $ \Delta x(\delta x) = x (t; t_0, \mathbf{x}_0 + \delta \mathbf{x}) - x (t; t_0, \mathbf{x}_0 - \delta \mathbf{x}) $ and $ \Delta y(\delta x) =  y (t; t_0, \mathbf{x}_0 + \delta \mathbf{x}) - y (t; t_0, \mathbf{x}_0 - \delta \mathbf{x} )$ respectively, representing the finite-time deformation experienced by the fluid element as shown in Fig. \ref{fig: jacobian} . Similarly, the horizontal and vertical displacement between the advected positions of the perturbed particles under vertical perturbation are denoted by  $ \Delta x(\delta y) = x (t; t_0, \mathbf{x}_0 + \delta \mathbf{y}) - x (t; t_0, \mathbf{x}_0 - \delta \mathbf{y}) $ and $ \Delta y(\delta y) =  y (t; t_0, \mathbf{x}_0 + \delta \mathbf{y}) - y (t; t_0, \mathbf{x}_0 - \delta \mathbf{y} )$ respectively.\\

This approach provides a finite-time approximation of the Cauchy–Green deformation tensor at each grid point, which quantifies local material stretching and rotation. Once the FTLE field is computed using eq.\eqref{eq:6} and eq.\eqref{eq:7}, LCSs can be extracted by identifying ridge-like features within the scalar FTLE contours. These ridges correspond to material lines in the flow that act as transport barriers. 
To extract well-defined FTLE ridges, numerical computation of FTLE is carried out with a sufficiently large integration time \cite{lekien2004dynamically} using a Runge-kutta integrator of fourth order. When integrated backward in time $(\Delta t < 0)$, the ridges in backward-time FTLE (bFTLE) scalar field reveals attracting LCSs denoted as bLCS—material lines that locally draw fluid elements together as time evolves. In contrast, forward-time integration $(\Delta t > 0)$ yields the forward-time FTLE (fFTLE) scalar field, whose ridges identify repelling LCSs denoted as fLCS—material lines that locally repels the fluid parcles as time evolves, leading to the separation of nearby trajectories. In this study, both bFTLE and fFTLE scalar fields are filtered by applying a threshold of $50\%$ to obtain bLCSs and fLCSs, respectively as illustrated in Fig. \ref{fig: LCSimage}, meaning only those FTLE values exceeding $50\%$ of the respective maximum FTLE are retained for visualization. It is important to note that the repelling structures identified by forward-time FTLE, when observed in backward time acts as attracting material lines due to the intrinsic time-reversal symmetry of the FTLE formulation. In Figs. \ref{fig:fftletracers} and \ref{fig:bftletracers}, we illustrate the evolution of fLCS and bLCS in one vortex shedding period $T$ for the problem considered here. As depicted in Fig. \ref{fig:fftletracers}, a stream of passive tracer particles is released at the downstream location $x = 3$, and their backward-time evolution illustrates how the particles are drawn toward the forward-time FTLE ridges (fLCS). This behavior underscores the repelling nature of fLCS structures in forward-time evolution. Similarly, in Fig.\ref{fig:bftletracers}, particles seeded upstream at $x = -1$ gradually envelops the backward-time FTLE ridges (bLCS), illustrating the attracting behavior of these LCSs as the flow evolves. These attracting and repelling LCSs collectively define the evolving kinematic skeleton of the flow and are especially useful in identifying transport dynamics, coherent vortex boundaries, and separation regions. Furthermore, combining forward and backward FTLE fields allows for the detection of dynamically significant features in the flow such as Lagrangian saddle point—intersection of fLCS and bLCS.

\begin{figure}[ht!]
\centering

\begin{subfigure}[t]{0.48\textwidth}
    \centering
    \includegraphics[width=\linewidth]{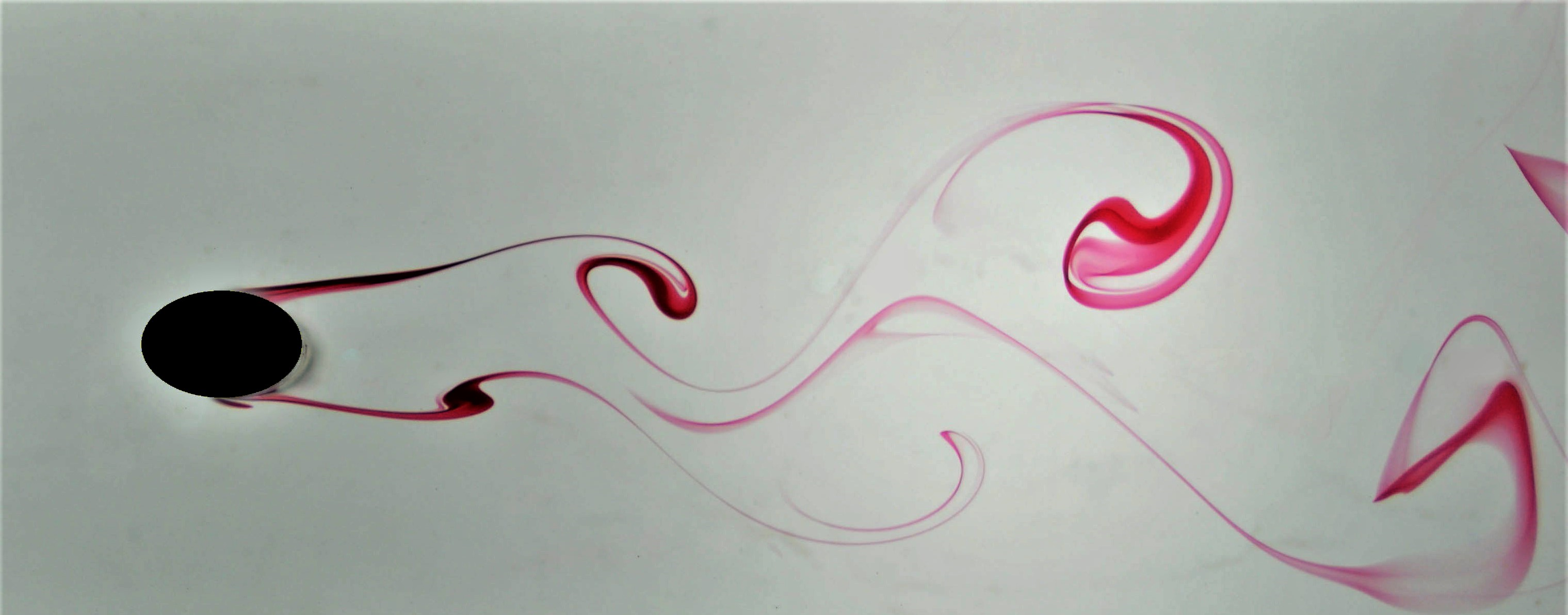}
    \caption*{(a) Experimental visualization }
    \includegraphics[width=\linewidth]{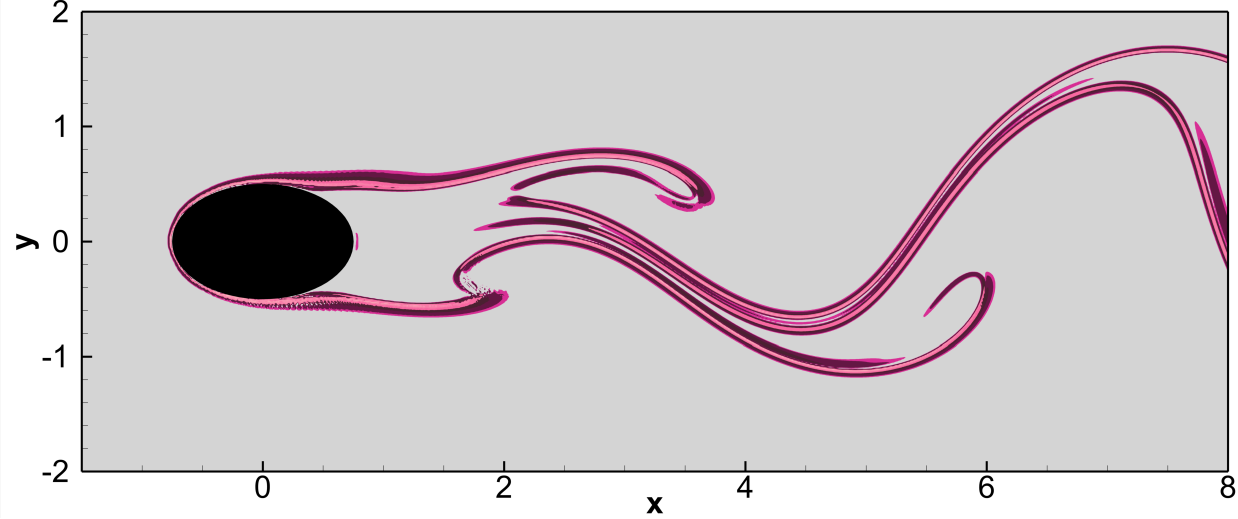}
    \caption*{(b) bLCS contours obtained from numerical simulation. }
\end{subfigure}

\caption{Comparison between streaklines for flow past an elliptic cylinder inclined at $\theta = 0^\circ$ with $AR = 0.67$ at $Re = 107$ between: (a) Experimental observations by Fonseca \textit{et al.} \cite{fonseca2013flow} and (b) bLCSs computed from numerical simulations.}
\label{fig:compare}
\end{figure}

In Fig.~\ref{fig:compare}, we present a comparative visualization between experimentally obtained streaklines from Fonseca \textit{et al.}~\cite{fonseca2013flow} and numerically computed attracting Lagrangian Coherent Structures (bLCSs) for flow past an elliptic cylinder inclined at $\theta = 0^\circ$ with an aspect ratio $AR = 0.67$ at $Re = 107$. The close structural resemblance between the experimentally generated streaklines and the computed bLCSs highlights the accuracy of the FTLE-based extraction of coherent structures. As shown in \cite{fonseca2013flow}, streaklines visualized using dye injection capture key flow features such as boundary layer separation, recirculation zones, and vortex shedding, all of which are closely associated with unsteady convective dynamics. This visual agreement in Fig.~\ref{fig:compare} not only reinforces the validity of the present numerical methodology for LCS extraction but also demonstrates the capability of LCS analysis to effectively identify and interpret the underlying transport mechanisms that govern convective heat transfer in bluff-body wakes.

\section{Results and Discussion}\label{sec:framework}

\subsection{LCSs and Saddle point framework to track temporal evolution of surface-averaged Nusselt number}

In this study, effective Lagrangian analysis tools such as Lagrangian Coherent structures and saddle point tracking analysis are used to investigate the forced convective heat transfer around a Heated Elliptical Cylinder in one vortex shedding time period $T$. Fig.~\ref{fig:SurfaceNu} presents the transient variation of the surface-averaged Nusselt number as a function of the re-scaled time $t^* = \frac{t - t_0}{T}$, where $t_0$ denotes the instant corresponding to the minimum Nusselt number within the shedding cycle. The subsequent discussions pertain to $t^* \in [0, 1]$. The results are shown for various angles of attack, ranging from $\theta = 0^\circ$ to $\theta = 90^\circ$ in increments of $15^\circ$, at a Reynolds number of $Re = 100$. Throughout this section, the term 'Nusselt number' is used to denote the surface-averaged Nusselt number unless stated otherwise. In Fig. \ref{fig:NU0}, the Nusselt number exhibits an oscillatory behavior, characterized by two local maxima and an intermediate minimum. The critical time instants corresponding to changes in the slope shifts in the $Nu$ plot have been marked in red on the plot. These points indicate transitions in the heat transfer dynamics and represent time instants of enhanced and weakened convective transport in the flow. In Figs. \ref{fig:NU15} and \ref{fig:NU30}, the Nusselt number trends differ from the case at $\theta=0^\circ$, as in these plots, there is only one local peak of  heat transfer rate and the local minimum is not distinctively observed. For the  $\theta=45^\circ, 60^\circ, 75^\circ, 90^\circ $ in Figs. \ref{fig:NU45} - \ref{fig:NU90}, the oscillatory trend of Nusselt number evolution plot is also characterized as that of $\theta=0^\circ$ with two local maxima and an intermediate minimum. Notably, the highest heat transfer rate occurs at the first local maximum for $\theta=45^\circ, 60^\circ$ whereas for $\theta=75^\circ , 90^\circ$, it is achieved at the second local maximum.\\

\begin{figure}[htbp]
    \centering
    \subfloat[$\theta = 0^\circ$\label{fig:NU0}]{%
        \includegraphics[width=0.44\linewidth]{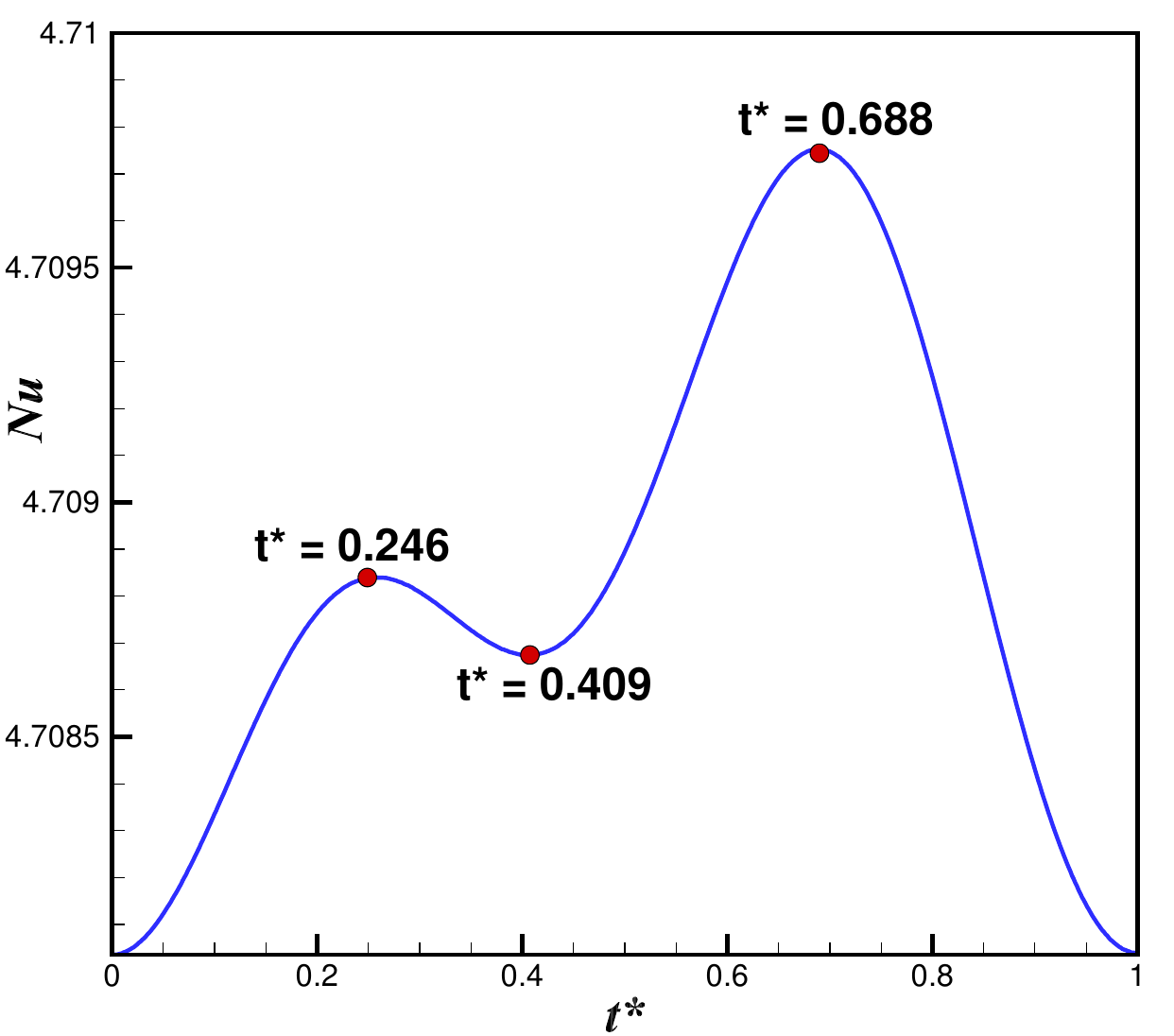}
    }\hfill
    \subfloat[$\theta = 15^\circ$\label{fig:NU15}]{%
        \includegraphics[width=0.44\linewidth]{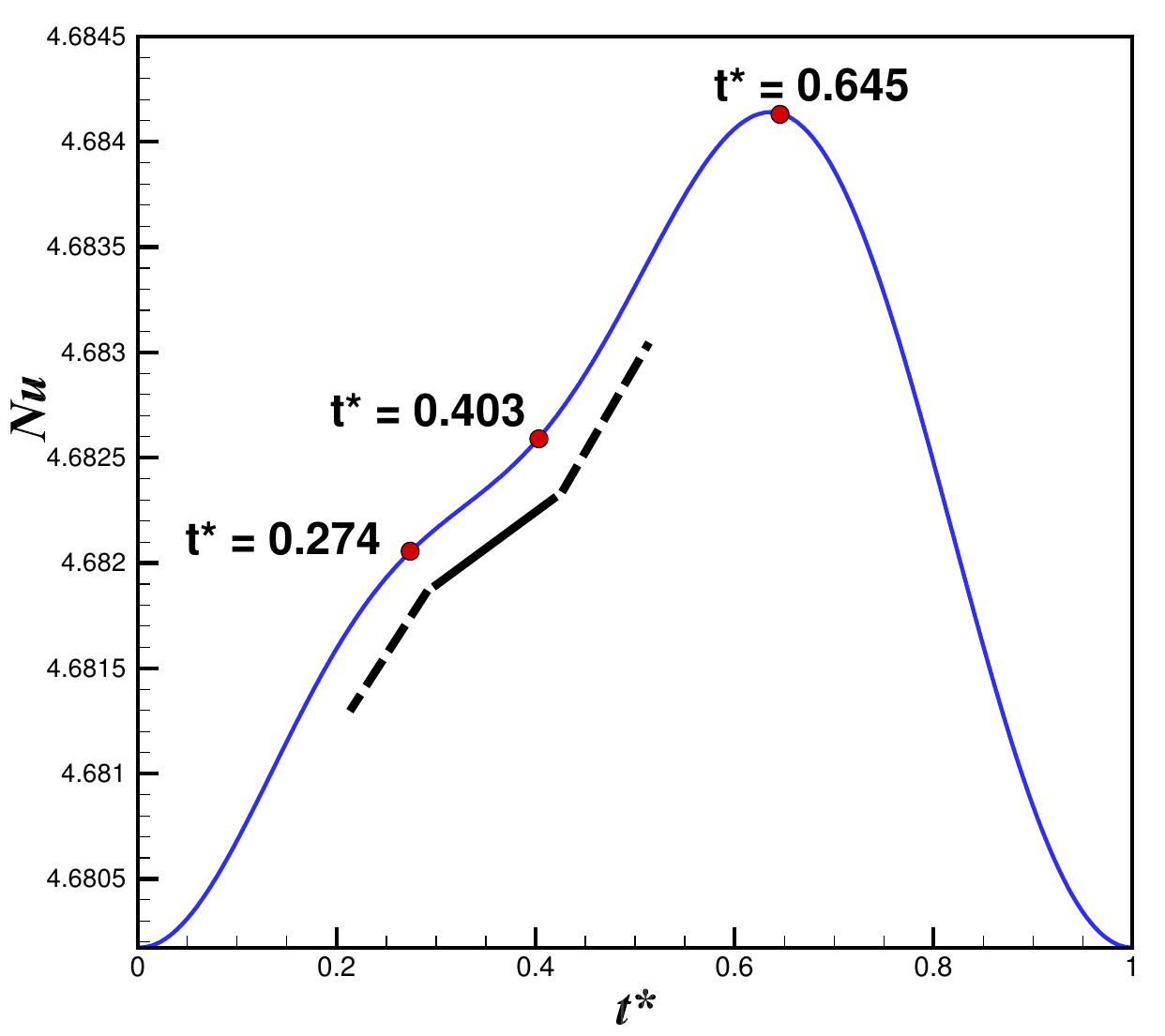}
    }\par\medskip

    \subfloat[$\theta = 30^\circ$\label{fig:NU30}]{%
        \includegraphics[width=0.44\linewidth]{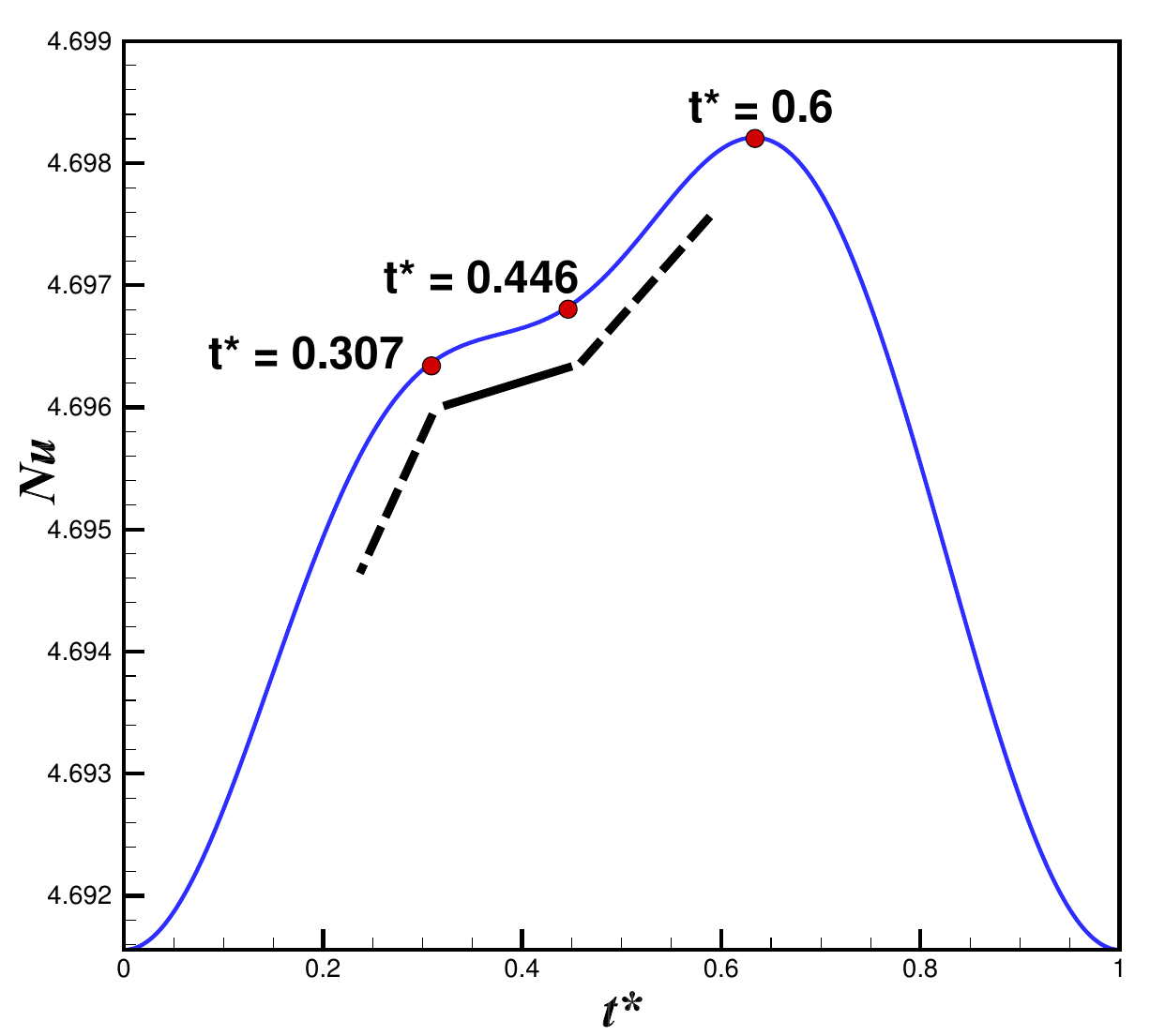}
    }\hfill
    \subfloat[$\theta = 45^\circ$\label{fig:NU45}]{%
        \includegraphics[width=0.44\linewidth]{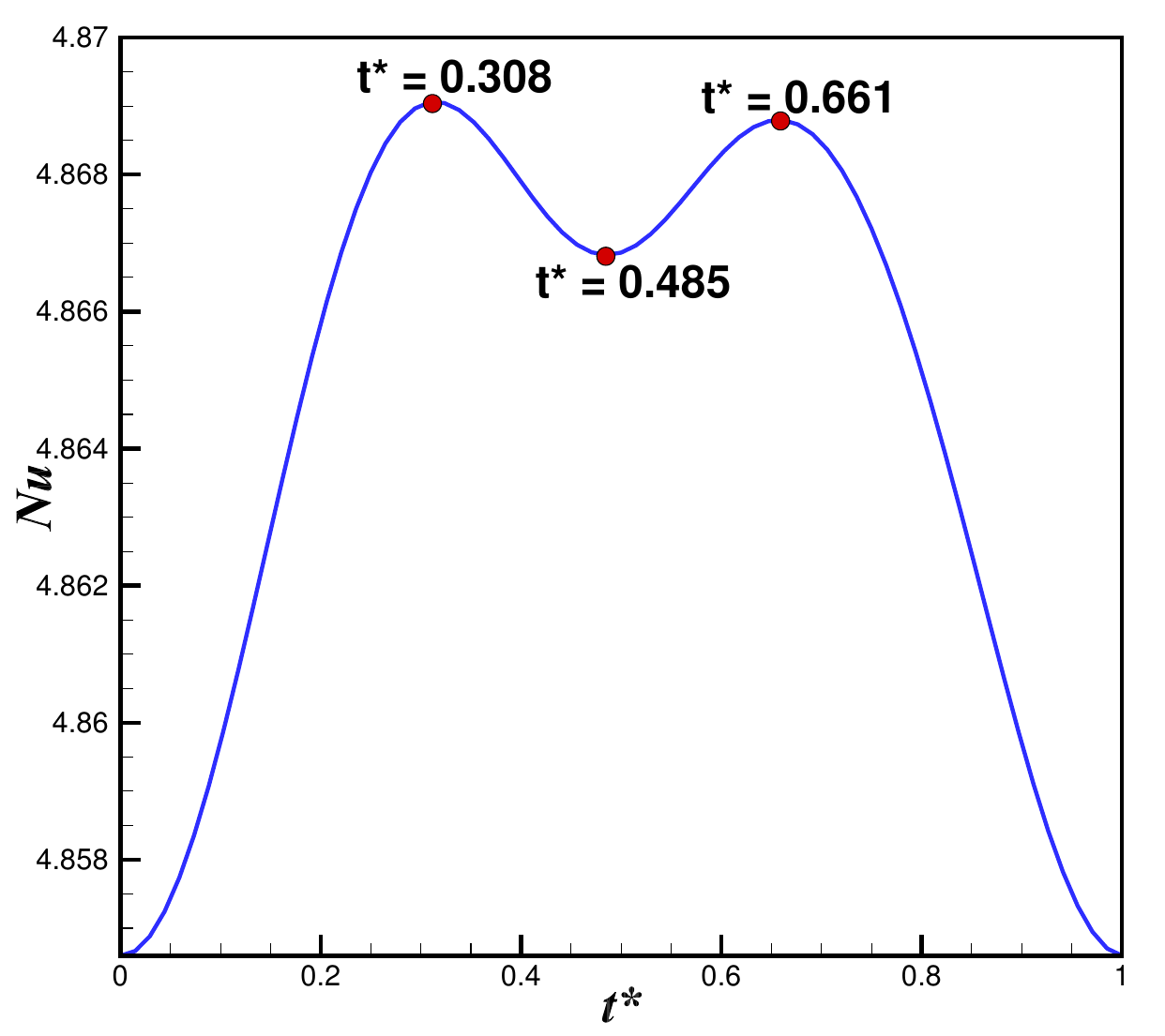}
    }

   
\end{figure}

\begin{figure}[htbp]
    \ContinuedFloat
    \centering
    \subfloat[$\theta = 60^\circ$\label{fig:NU60}]{%
        \includegraphics[width=0.45\linewidth]{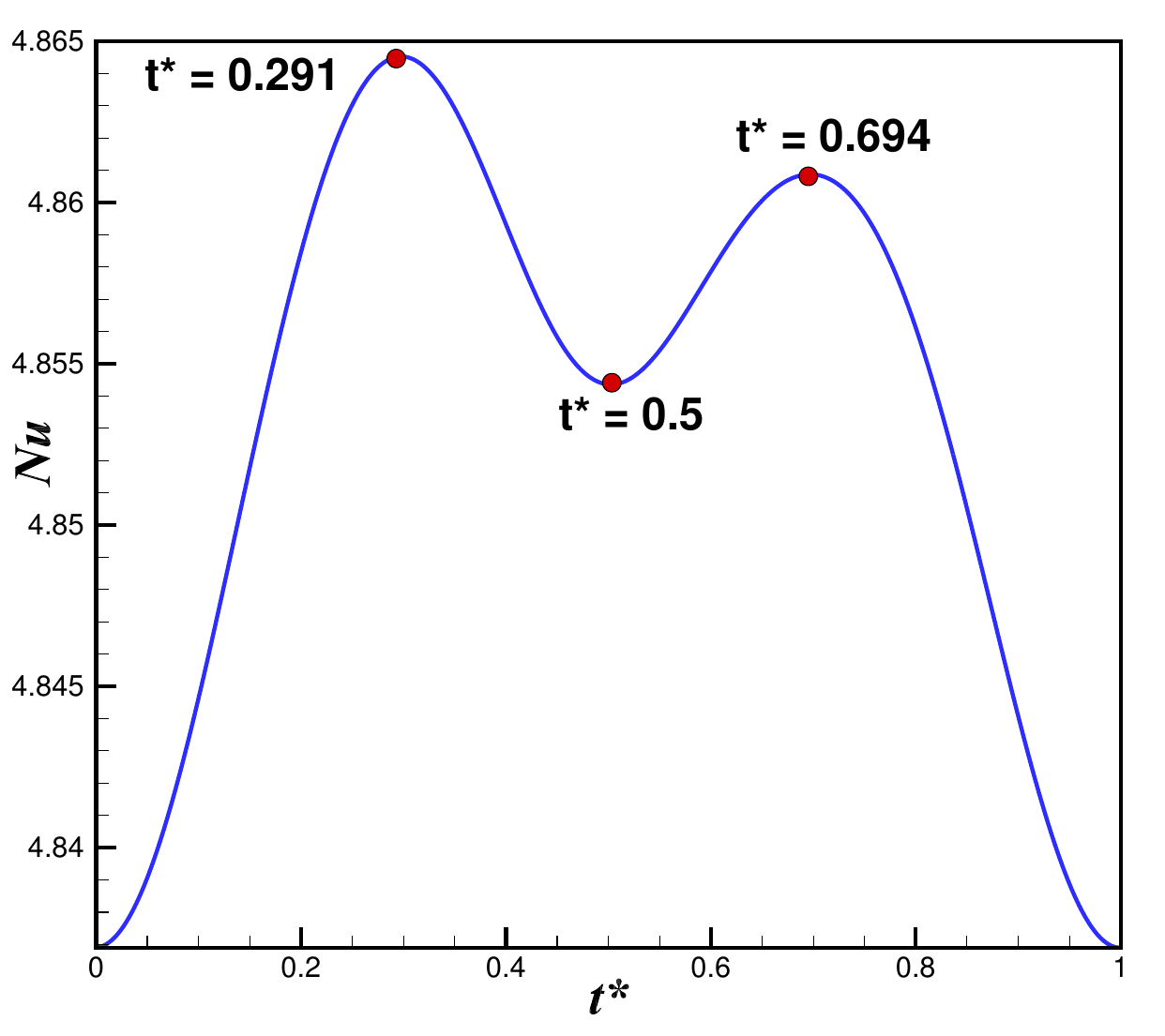}
    }\hfill
    \subfloat[$\theta = 75^\circ$\label{fig:NU75}]{%
        \includegraphics[width=0.45\linewidth]{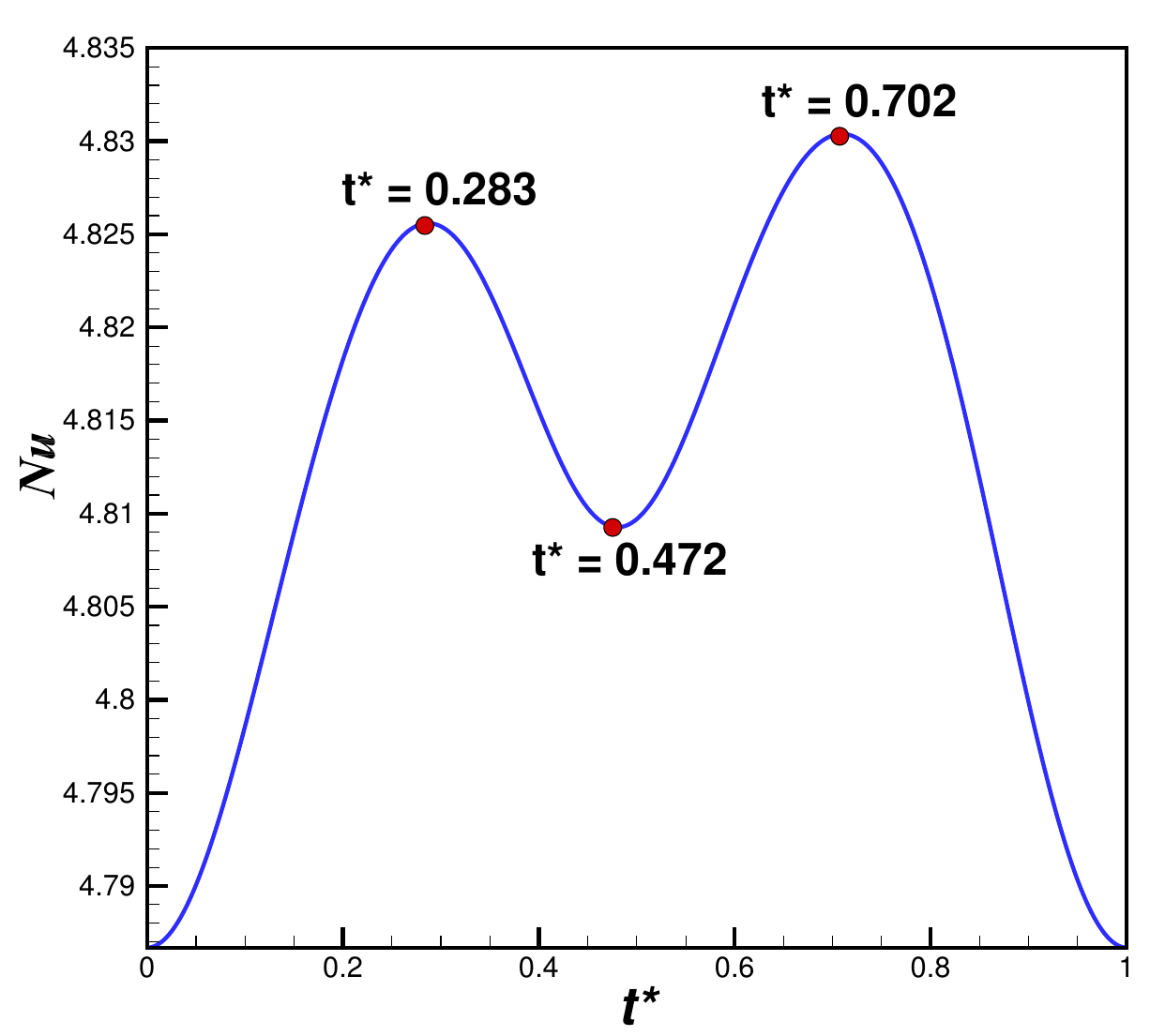}
    }\par\medskip

    \subfloat[$\theta = 90^\circ$\label{fig:NU90}]{%
        \includegraphics[width=0.45\linewidth]{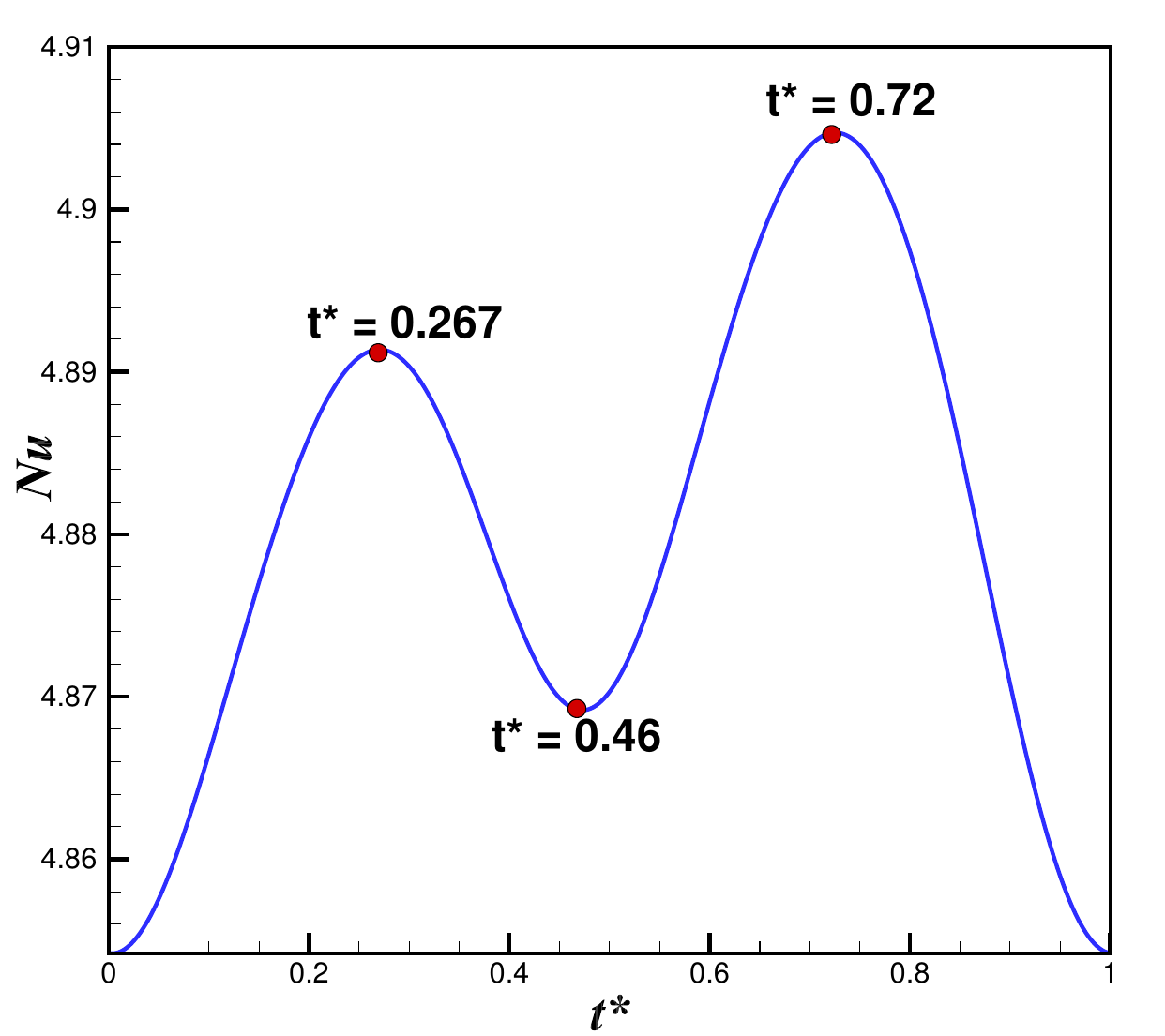}
    }

    \caption{Time variation of surface-averaged Nusselt number for $Re = 100$ at : (a) $\theta = 0^\circ$, (b) $\theta = 15^\circ$, (c) $\theta = 30^\circ$, (d) $\theta = 45^\circ$, (e) $\theta = 60^\circ$, (f) $\theta = 75^\circ$, (g) $\theta = 90^\circ$}
 \label{fig:SurfaceNu}
\end{figure}

We propose a novel theoretical framework to elucidate the correlation between the temporal evolution of Nusselt number and the dynamics of saddle point trajectory. The core of the framework lies in linking the slope transitions in the $Nu$ temporal profile with instantaneous transitions in saddle point trajectories by utilizing plots of Lagrangian Coherent Structures.  LCSs and isotherms are captured for $\theta=0^\circ, 15^\circ, 30^\circ, 45^\circ, 60^\circ, 75^\circ$ and $90^\circ$ over time period $T$ at various representative time instants $t^*$ at $Re = 100$. Selective illustrations are presented in Fig. \ref{fig: LCS0} to Fig. \ref{fig: LCS45} for $\theta=0^\circ, 15^\circ$ and $90^\circ$. For all angular variations involved in computation, the fLCSs and bLCSs are plotted for values above $50\%$ of maximum FTLE value. Fig. \ref{fig: LCS0IM1}  illustrates the plots of LCSs and isotherms at $t^*=0$ for the elliptical cylinder inclined at $\theta=0^\circ$. The temperature field and LCSs demonstrate a distinct asymmetry with respect to the centerline (red-dotted line in Fig. \ref{fig: LCS0IM1}), highlighting the unsteady nature of the wake, which varies periodically over time. For notational simplicity, we omit the term "LCS" in the labels fLCS/bLCS and instead use numerics to distinguish between different LCSs in the discussion that follows. At \( t^* = 0 \) in Fig.~\ref{fig: LCS0IM1}, a prominently thick repelling LCS extending into the near wake, positioned away from the cylinder surface, is labeled as f1. Another repelling LCS, emerging directly on the surface of the elliptical cylinder in the upper half region, is labeled as ff1. This addittional 'f' in the notation of LCS ff1 indicates that the location of LCS f1 at $t^* =0$ will be occupied by LCS ff1 after a time interval of $T$. A similar interpretation applies to the repelling LCS ff2 which emerges on the lower half of the cylinder below the centerline. The LCS f11 in Fig. \ref{fig: LCS0IM4} represent a newly formed branch of LCS f1. Same nomenclature is carried out with the attracting LCSs b1 and b2. In Fig. \ref{fig: LCS0IM6}, LCS b11 and LCS b21 represent the branch of LCS b1 and LCS b2 respectively. The intersection of attracting and repelling LCSs gives rise to structurally stable points known as \textit{Lagrangian saddle points}. This is analogous to intersection of unstable and stable manifolds creating hyperbolic fixed points in dynamical systems \cite{rockwood2017detecting, green2010using}. Specifically, intersection of LCS f1 and LCS b1 is labeled by the saddle point $s_1$, while saddle point $s_{\text{f1}}$ is formed by the intersection of LCS ff1 and b1.  Similarly, the intersection of LCS f2 and b2 represents saddle point $s_2$ and the intersection of LCS ff2 and b2 represents saddle point $s_{\text{f2}}$. These Lagrangian saddle points are highlighted with red dots in Fig. \ref{fig: LCS0IM1}. \\

To develop a framework for the systematic exploration of the influence of saddle point trajectories on heat transfer enhancement,  a well-defined domain of observation or observational domain is proposed. The region bounded by the horizontal lines $m$, $n$ and the $y$-axis is defined as the observational domain as shown in Fig.  \ref{fig: LCS0IM1}. LCS b1 is considered as a continuous smooth curve in a two-dimensional plane \cite{peacock2013lagrangian}. The location of the upper boundary $m$ of the domain is chosen such that it would correspond to the horizontal tangent at the first relative peak of LCS b1 at $t^* = 0$ as shown in the Fig. \ref{fig: LCS0IM1}. Similarly, the location of lower boundary $n$ is chosen analogous to the location of horizontal tangent at the first relative minimum of LCS b2 curve at $t^* = 0$ in two-dimensional plane as shown in Fig. \ref{fig: LCS0IM1}. An interesting observation highlighted in this study is that the region bounded by these horizontal tangents to attracting LCS b1 and LCS b2 constitutes the dominant zone in the flow field where the advection of saddle points significantly influences surface convective heat transfer. In contrast, saddle points that migrate beyond these boundaries exhibit negligible influence on the convective heat transfer, indicating that the primary contributions originate from within this confined region. As Surface averaged nusselt number measures the convective heat transfer occurring at the surface, advection of saddle points originated on the surface or lifted off from the surface of the eliptical cylinder contributes to the convective heat transfer rate significantly more than those farther away from the body's surface. This aspect will be further elaborated upon in the following discussion. Motivated by this observation and in order to examine the interplay between local heat transfer characteristic captured by the Nusselt number and the evolving topology of saddle points, we introduce the concept of "Active Saddle Point". It is defined as the saddle point in the observational domain that either originate on the surface of the elliptical cylinder within the current time period or have lifted off from the cylinder’s surface during the preceding period. This definition aims to extract key intersection points of attracting and repelling LCSs in the fluid flow that play a pivotal role in influencing the near-wall fluid dynamics and, consequently, the convective heat transfer behavior around the bluff body. Among many possible intersections of attracting and repelling LCSs, the active saddle points identified within the observational domain are labeled as $s_1, s_{\text{f1}}, s_2, s_{\text{f2}}$ as shown in Fig. \ref{fig: LCS0IM1}. Among these, $s_{\text{f1}}$ and $s_{\text{f2}}$ originate at the current time instant, while $s_1$ and  $s_2$ have lifted off from the surface of the elliptical cylinder during the preceding period. At the end of one vortex shedding time period T, i.e, at $t^*=1$, the saddle points $s_{\text{f1}},  s_{\text{f2}}$ take the initial position of saddle points $s_1$ and $s_2$.\\

In Fig. \ref{fig: LCS0IM1}, the repelling LCS ff1 encapsulates the front half of the elliptical cylinder inclined at $\theta=0^\circ$ around the centerline and extends along the body in the rear half and then separates from the surface at $t^*=0$. A less visually prominent repelling LCS ff2 also exhibits a similar behavior along the lower surface of the cylinder. The incoming flow is divided into two upper and lower surface of the elliptical cylinder due to the fomation of LCS ff1 and LCS ff2. The LCS f2 envelopes the cylinder and is formed below the LCS ff2. A thicker and more prominent repelling LCS f1 is formed slightly away from the upper surface of the cylinder and extends in the near wake region. This frontal encapsulation by the LCS f1 repels the mainstream flow to come in contact with the cylinder, which leads to the reduction in the convective heat transfer in the upper half. The black attracting LCS formed in the front of the cylinder has been divided into LCSs b1 and b2, encapsulating the surface of the cylinder above and below the centerline respectively. These attracting structures play a critical role in organizing the transport of hot fluid particles near the surface and contribute significantly to the development of thermal wake structures. The LCS b1 extends along the upper surface of the elliptical cylinder from the front region into the near wake, actively convecting the heat in the near wake, and LCS b2 extends into the far wake region. This asymmetrical behavior in the spatial distribution of LCSs leads to periodic fluctuations in the convective heat transfer rates. The configuration of LCSs at $t^* = 0$, as presented in Fig.~\ref{fig: LCS0} through~\ref{fig: LCS45}, corresponds to the time instant within the shedding cycle where the surface-averaged Nusselt number reaches its minimum for each respective inclination angle.\\

\begin{figure}[htbp]
    \centering
    \subfloat[$t^* = 0$\label{fig: LCS0IM1}]{%
        \includegraphics[width=0.49\linewidth]{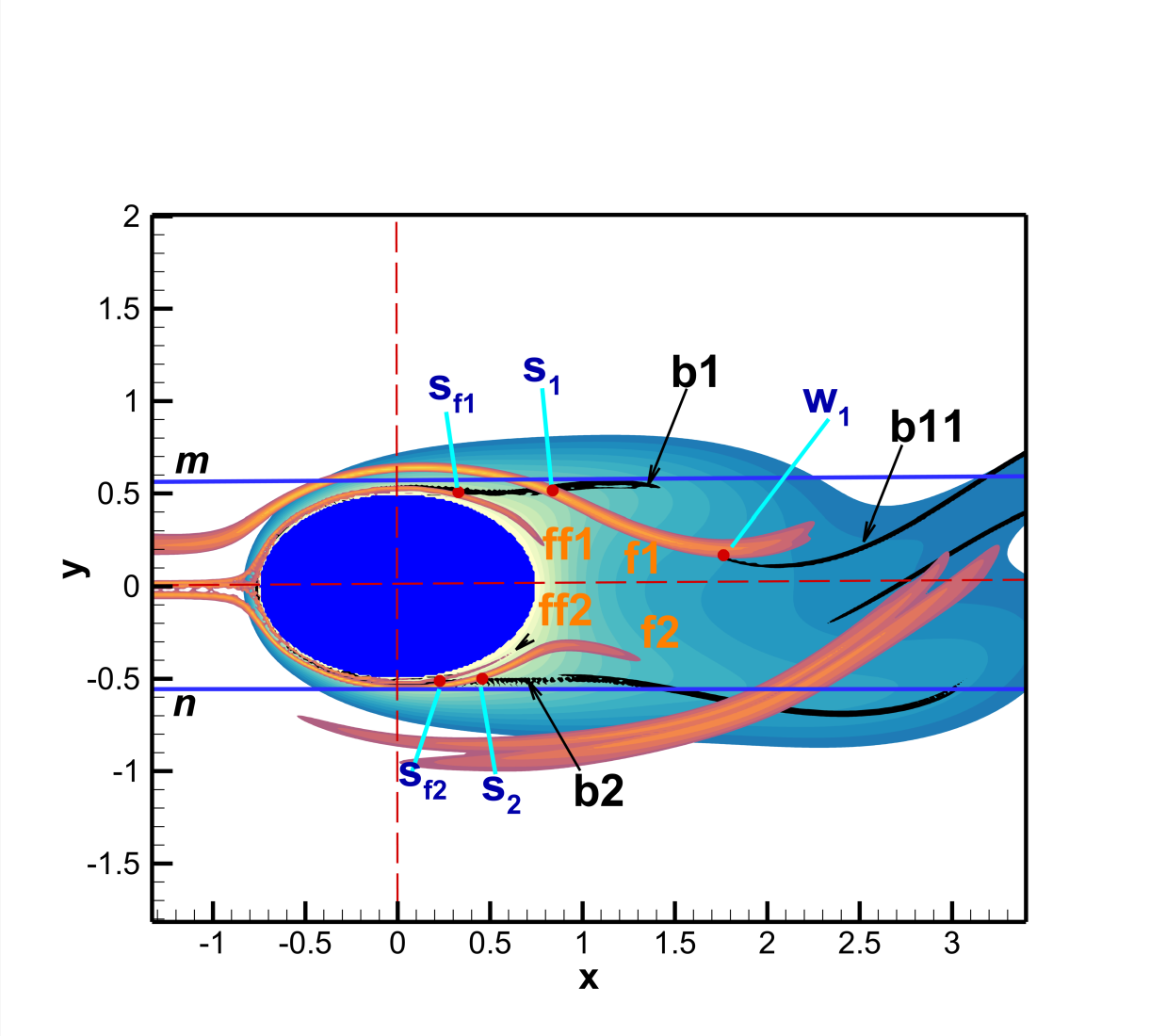}
    }\hfill
    \subfloat[$t^* = 0.114$\label{fig: LCS0IM2}]{%
        \includegraphics[width=0.49\linewidth]{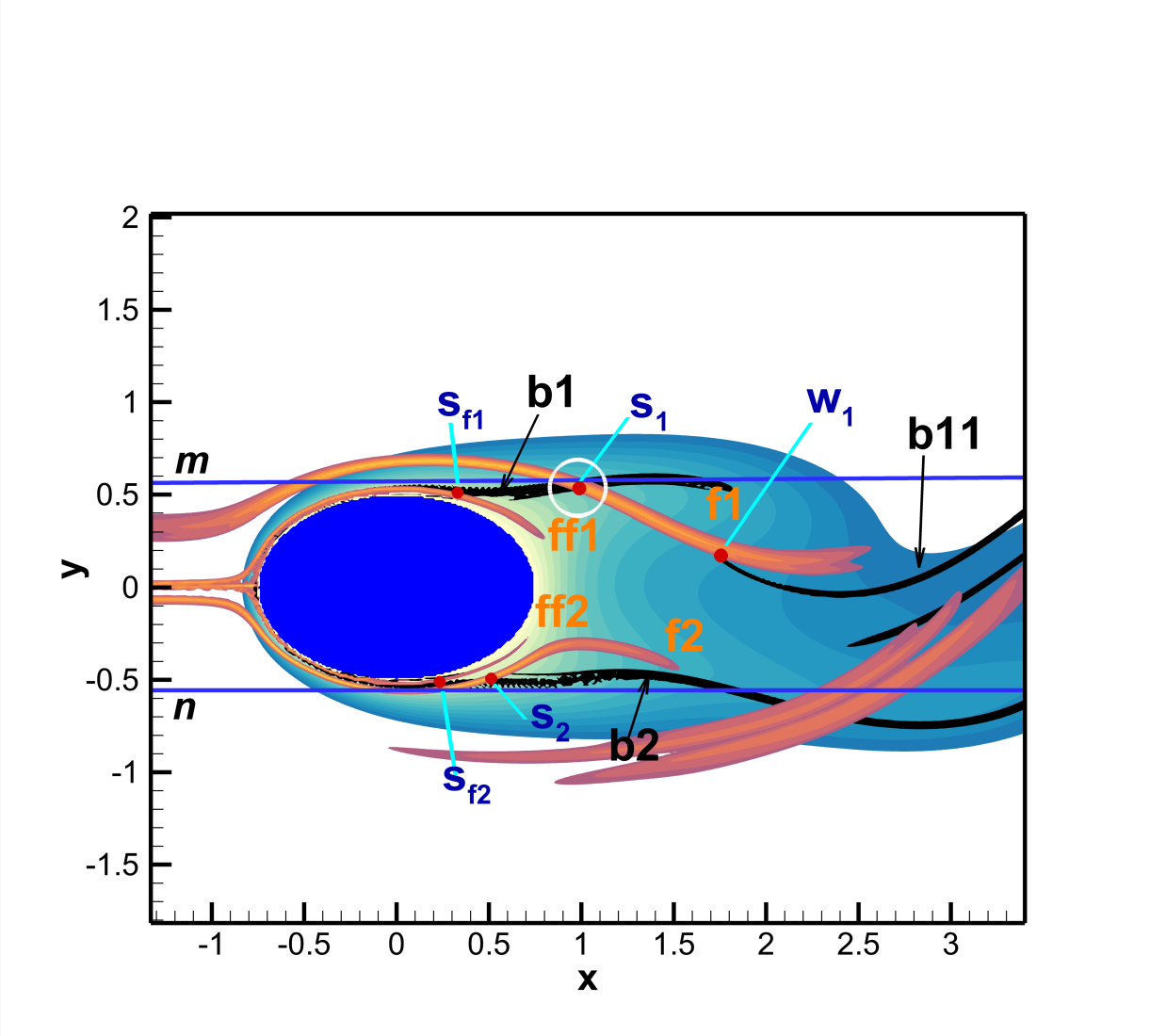}
    }\par\medskip

    \subfloat[$t^* = 0.246$\label{fig: LCS0IM3}]{%
        \includegraphics[width=0.49\linewidth]{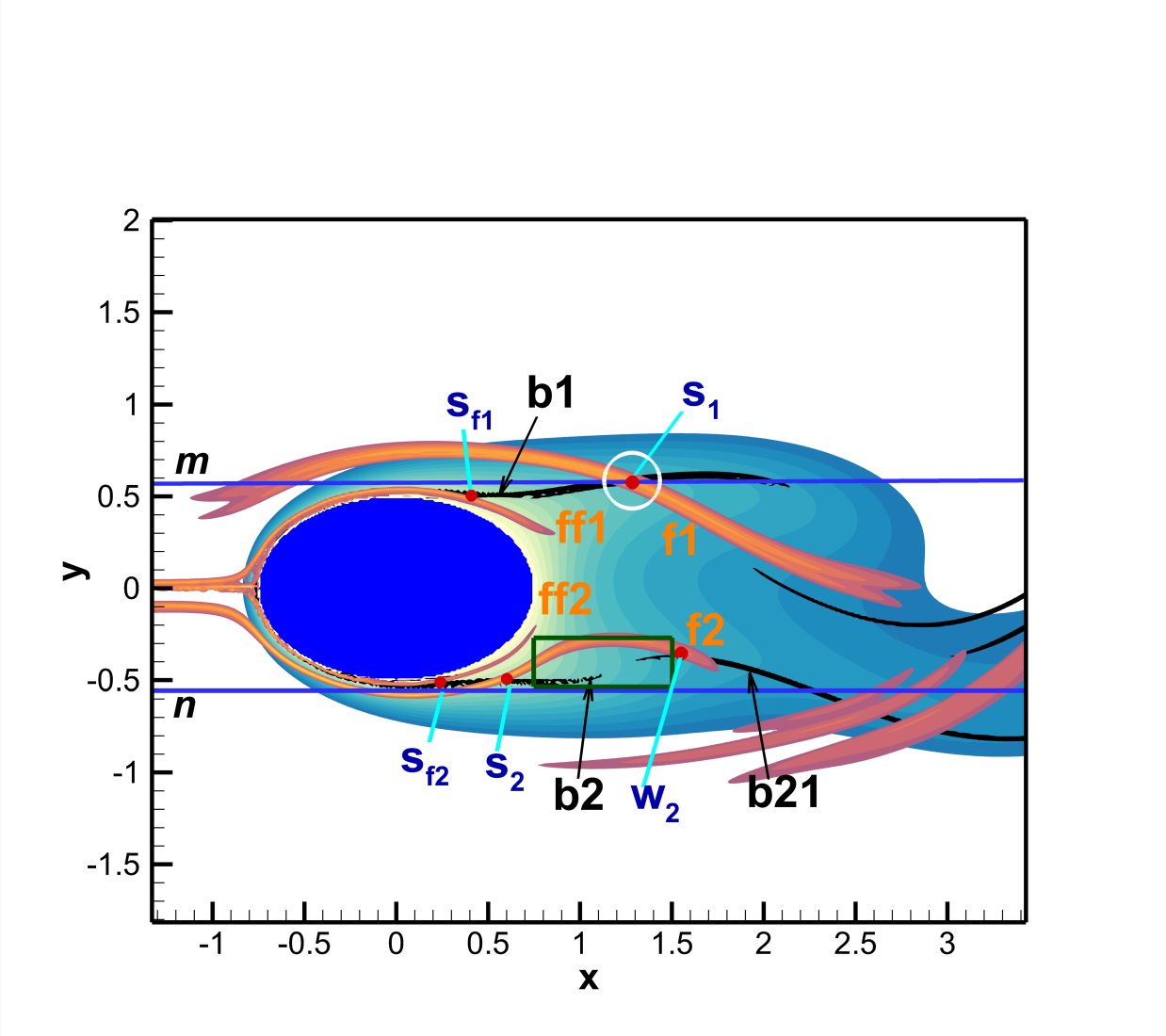}
    }\hfill
    \subfloat[$t^* = 0.409$\label{fig: LCS0IM4}]{%
        \includegraphics[width=0.49\linewidth]{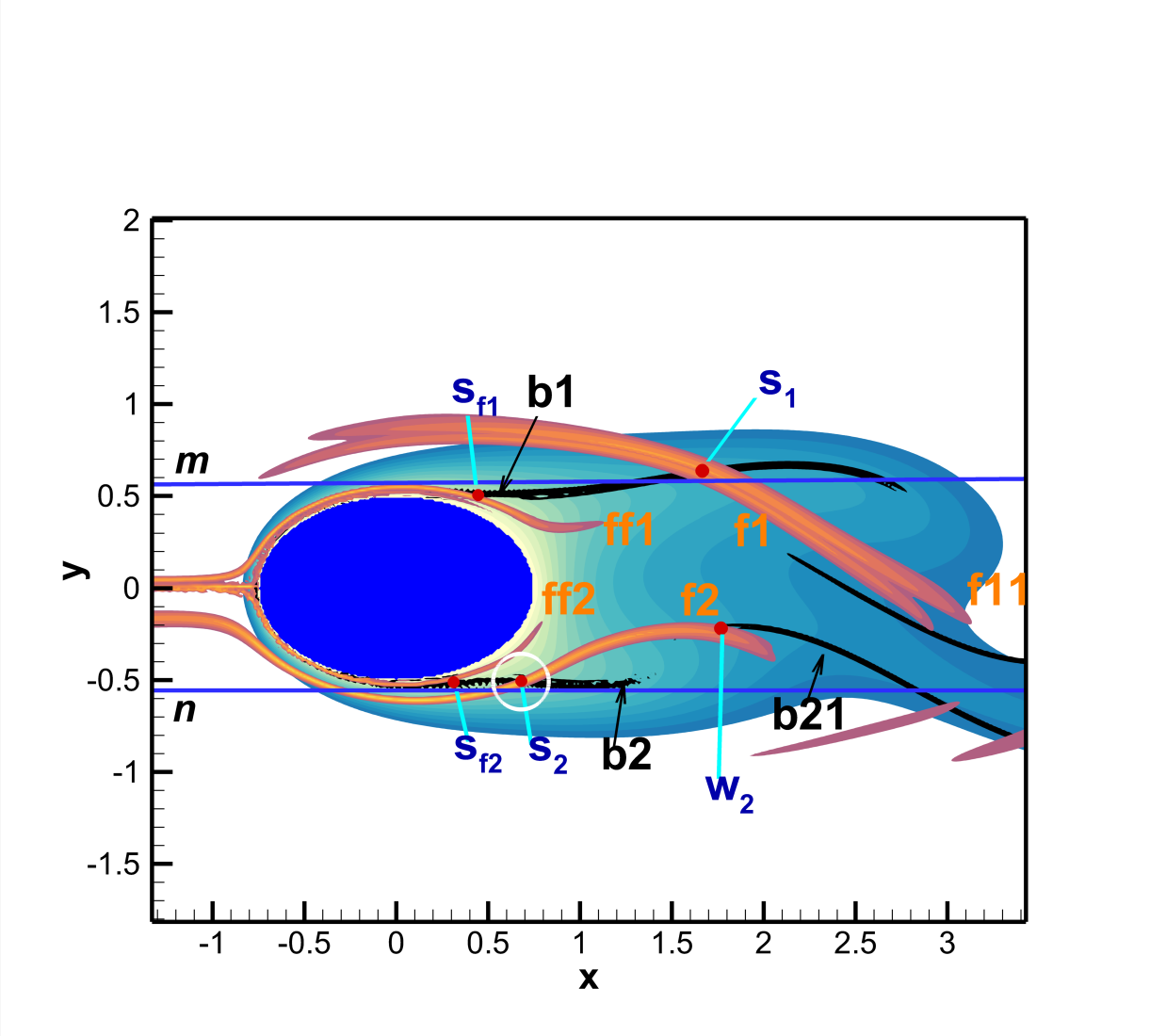}
    }\par\medskip

    \subfloat[$t^* = 0.54$\label{fig: LCS0IM5}]{%
        \includegraphics[width=0.49\linewidth]{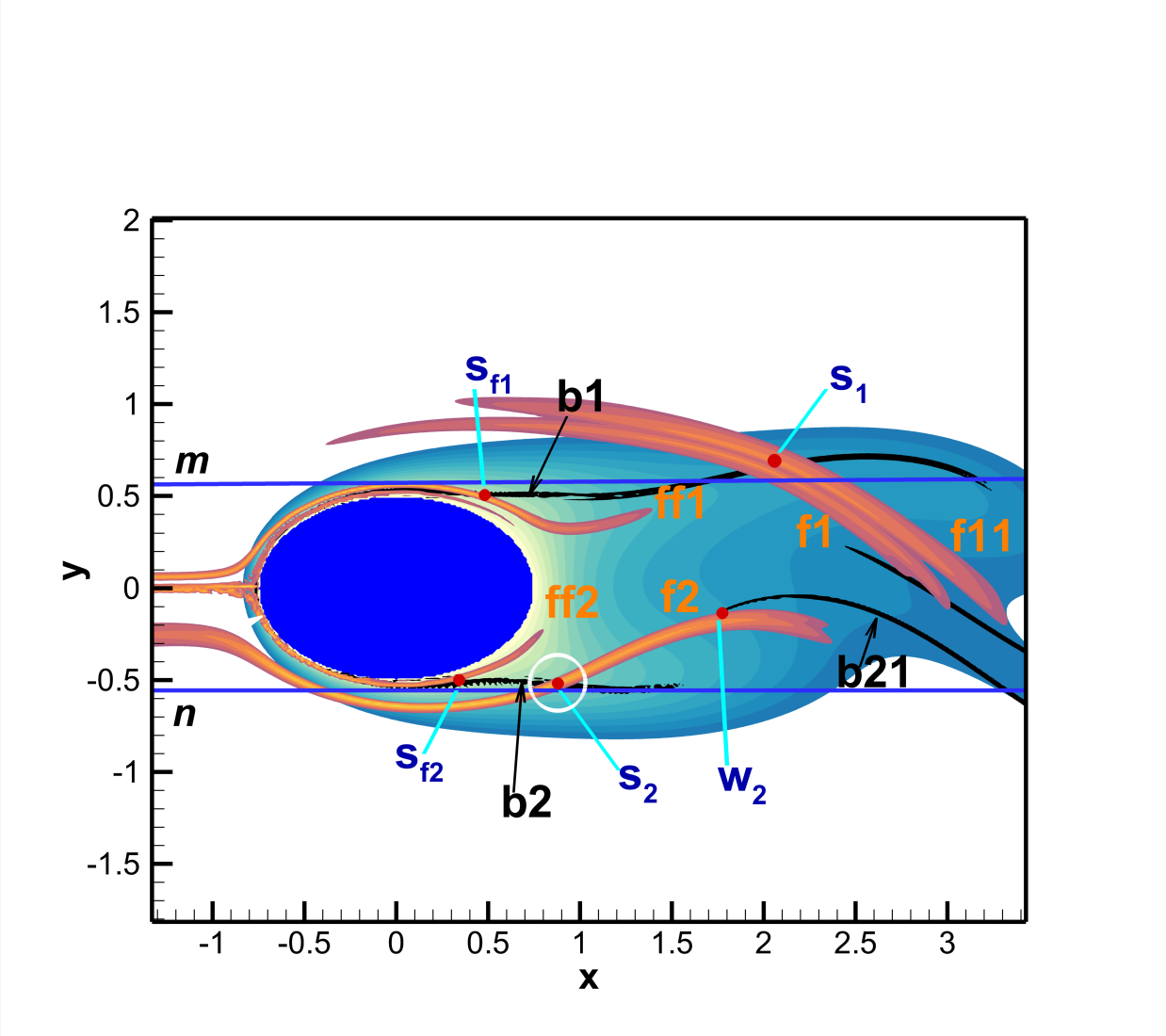}
    }

\end{figure}

\begin{figure}[htbp]\ContinuedFloat
    \centering
    \subfloat[$t^* = 0.688$\label{fig: LCS0IM6}]{%
        \includegraphics[width=0.49\linewidth]{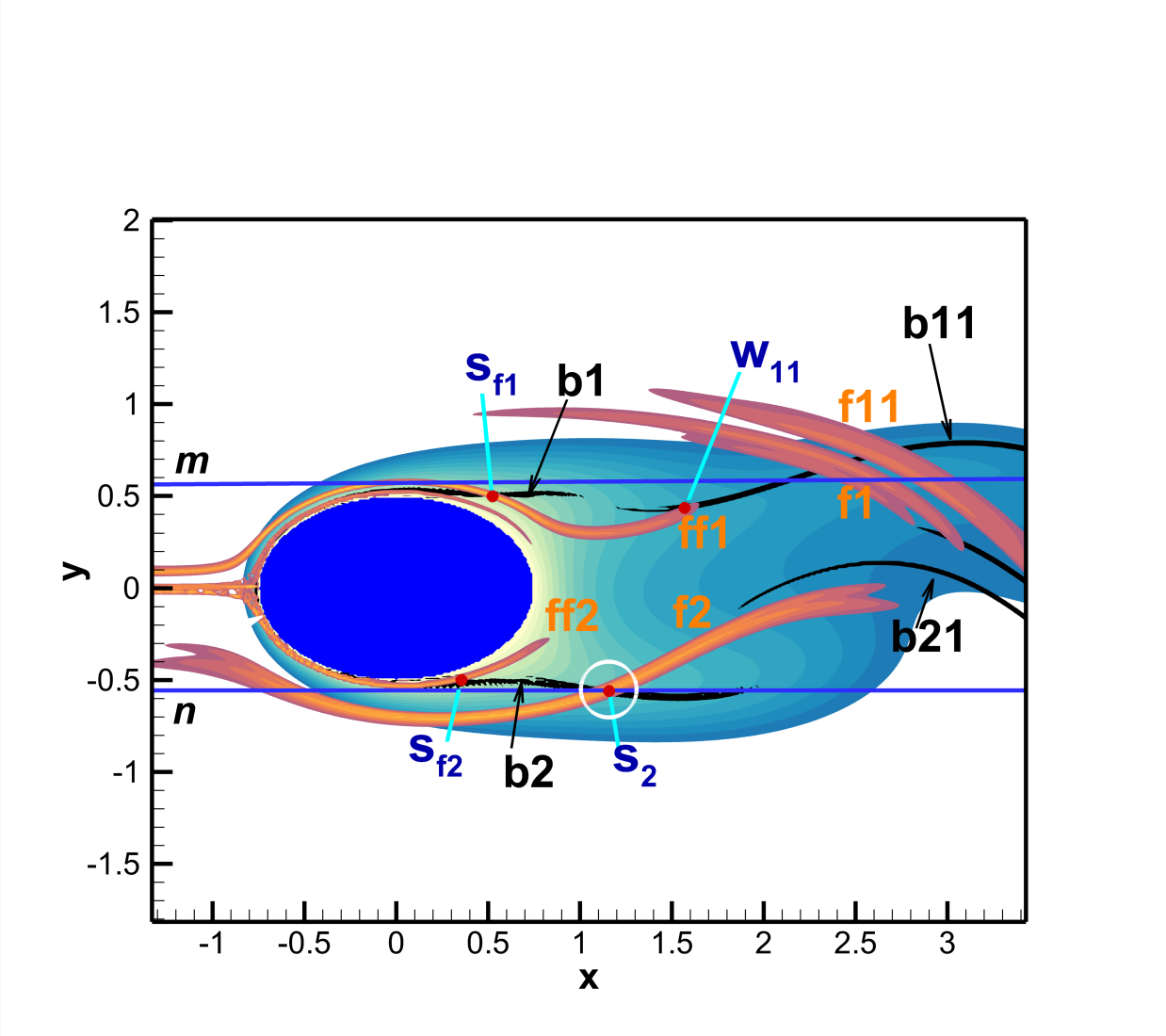}
    }\hfill
    \subfloat[$t^* = 1$\label{fig: LCS0IM7}]{%
        \includegraphics[width=0.49\linewidth]{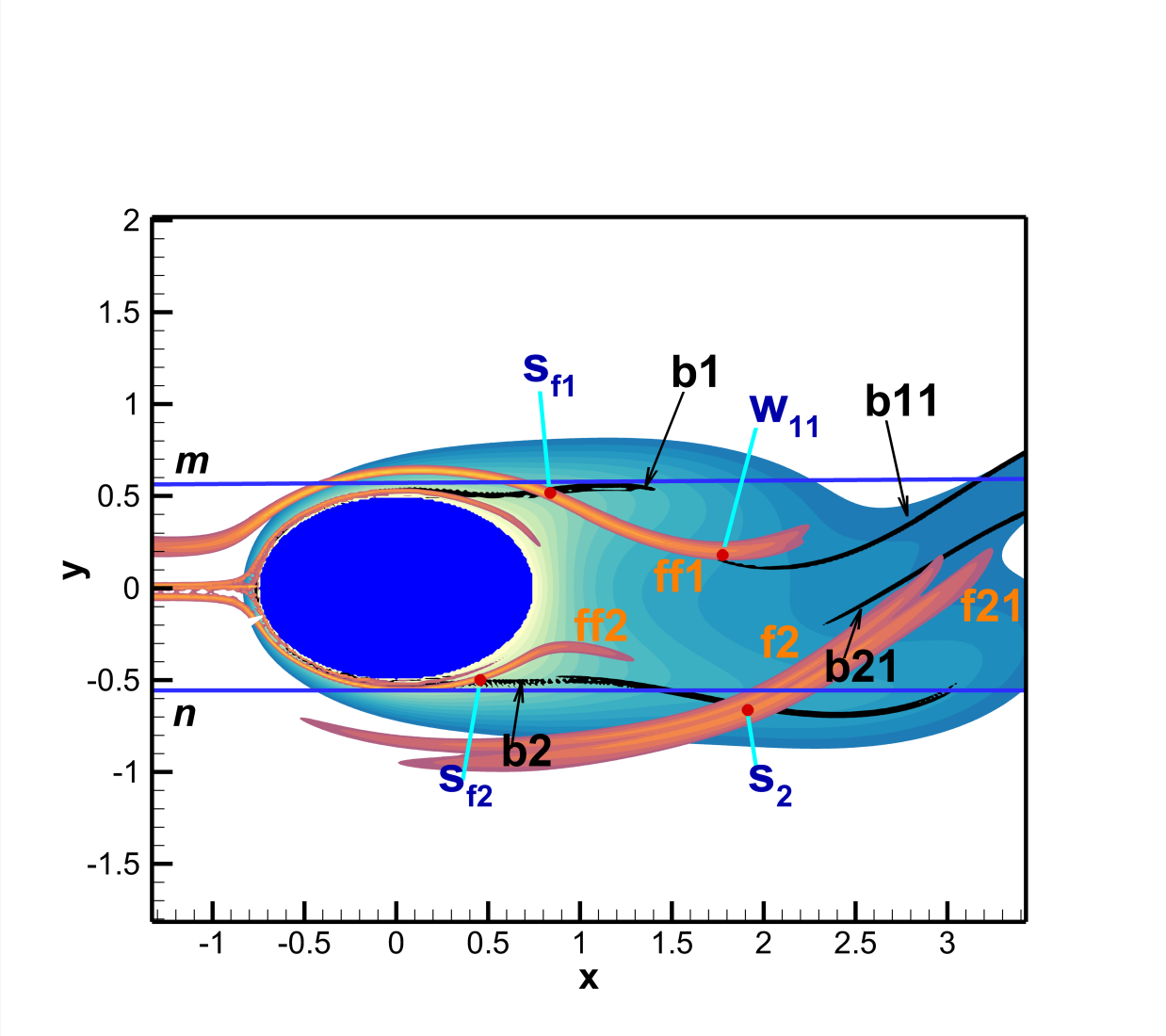}
    }
\caption{Instantaneous temperature contours and LCSs around the elliptical cylinder at $\theta = 0^\circ$ for $Re = 100$ at : (a) $t^* = 0$ , (b) $t^* = 0.114$ , (c) $t^* = 0.246$ , (d) $t^* = 0.409$ , (e) $t^* = 0.54$ , (f) $t^* = 0.688$ , (g) $t^* = 1$.}
\label{fig: LCS0}

\end{figure}

\begin{figure}[htbp]
    \centering
    \subfloat[$t^* = 0$\label{fig: LCS15IM1}]{%
        \includegraphics[width=0.49\linewidth]{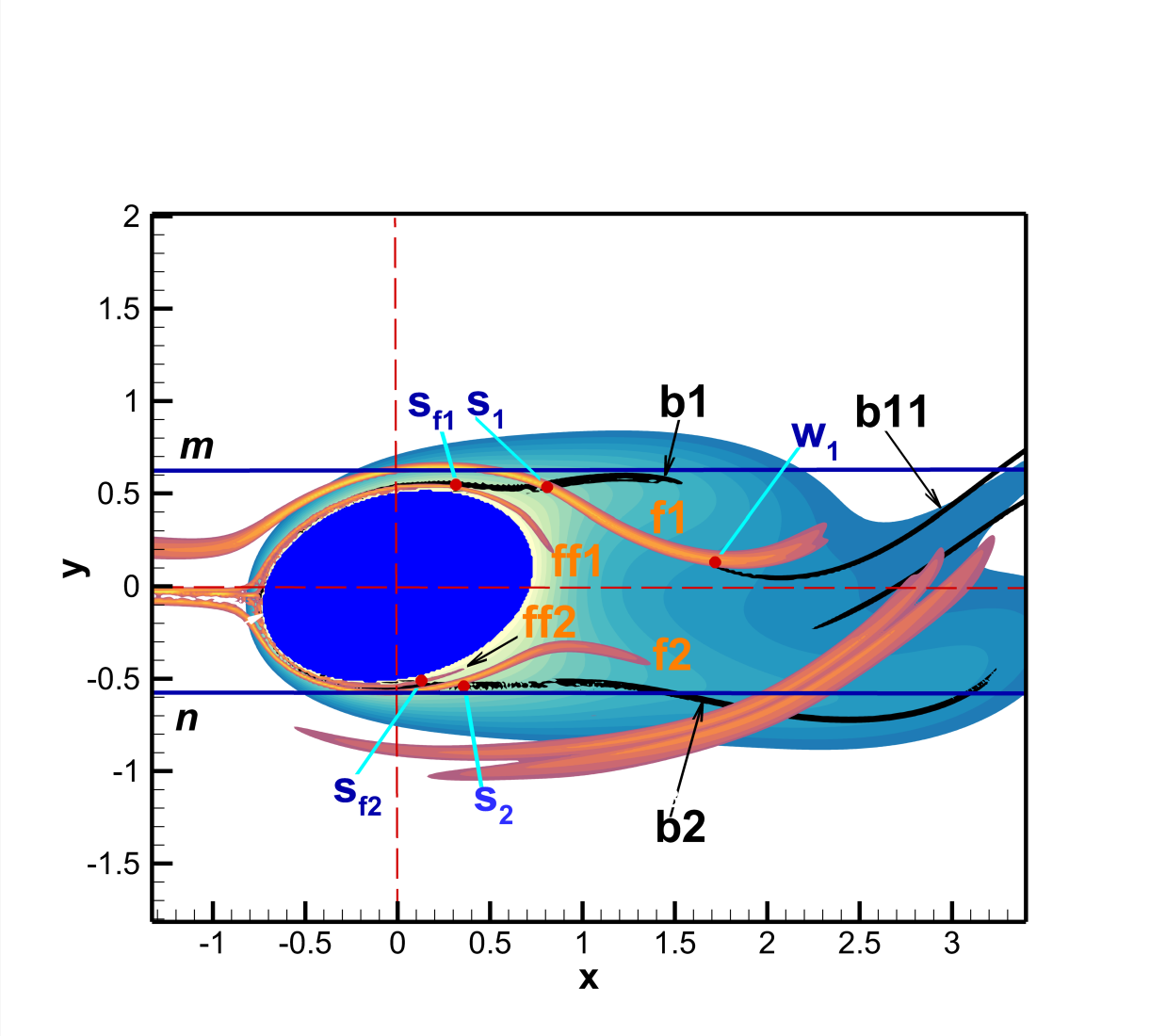}
    }\hfill
    \subfloat[$t^* = 0.129$\label{fig: LCS15IM2}]{%
        \includegraphics[width=0.49\linewidth]{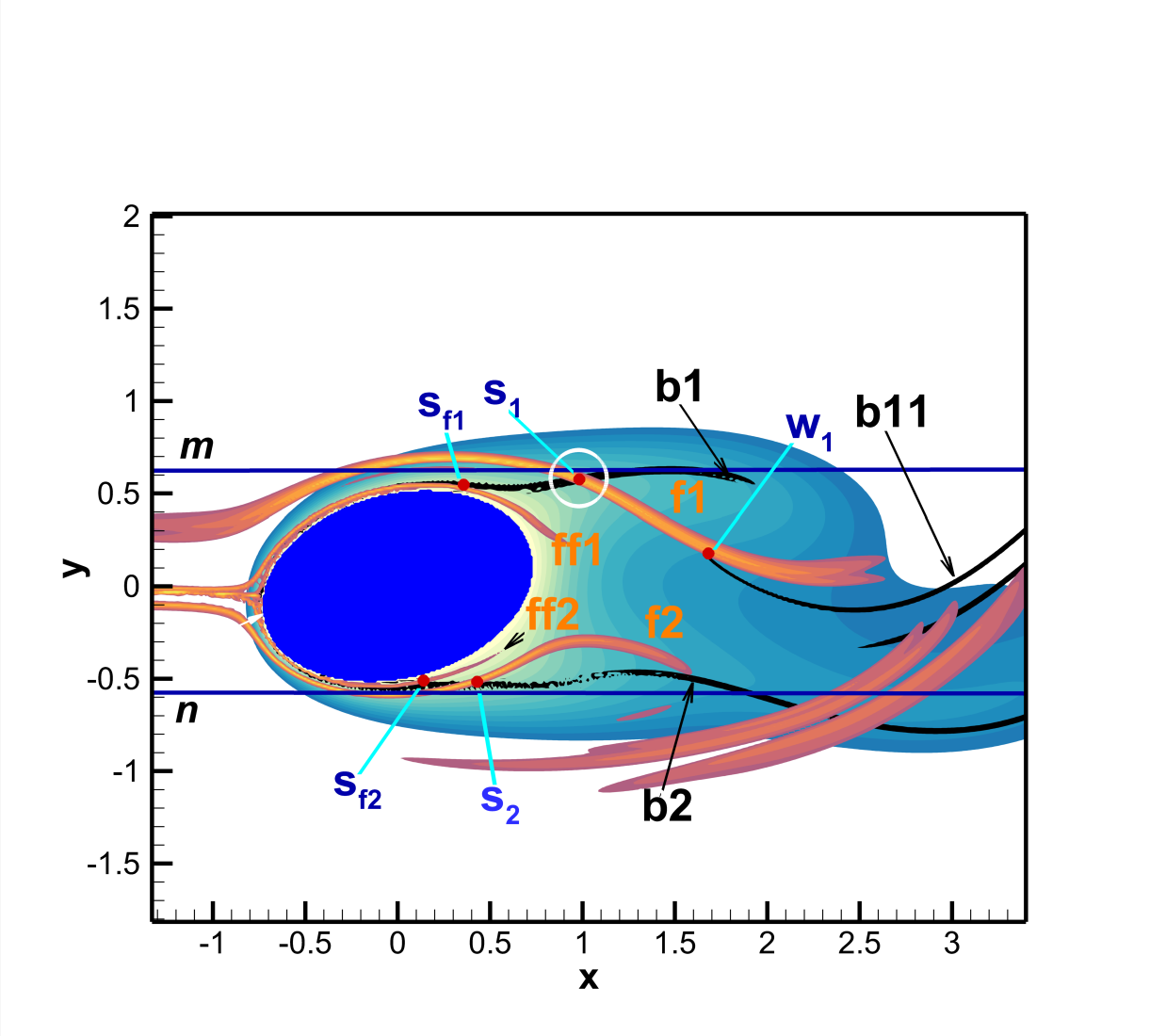}
    }\par\medskip

    \subfloat[$t^* = 0.274$\label{fig: LCS15IM3}]{%
        \includegraphics[width=0.49\linewidth]{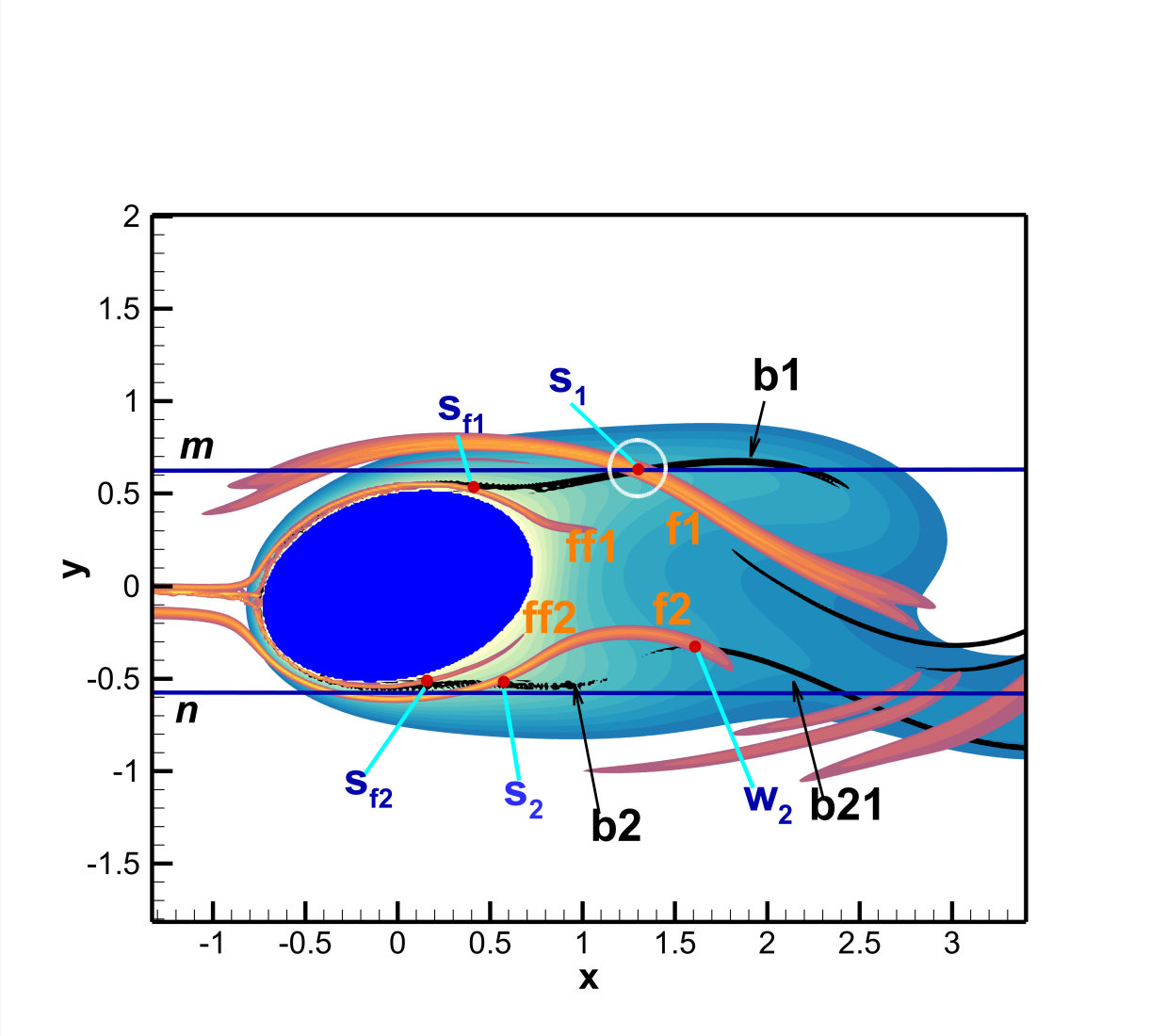}
    }\hfill
    \subfloat[$t^* = 0.403$\label{fig: LCS15IM4}]{%
        \includegraphics[width=0.49\linewidth]{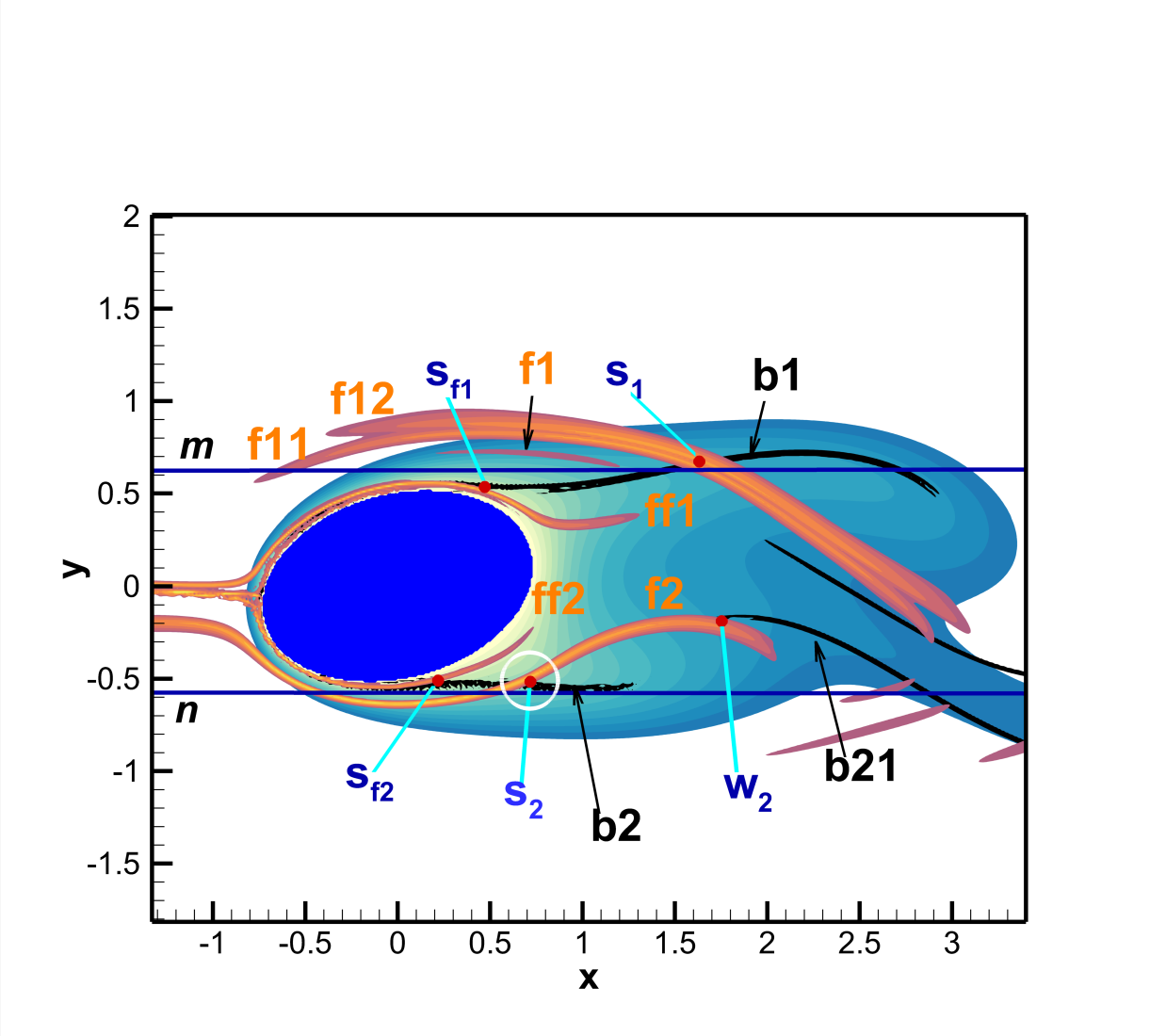}
    }\par\medskip

    \subfloat[$t^* = 0.532$\label{fig: LCS15IM5}]{%
        \includegraphics[width=0.49\linewidth]{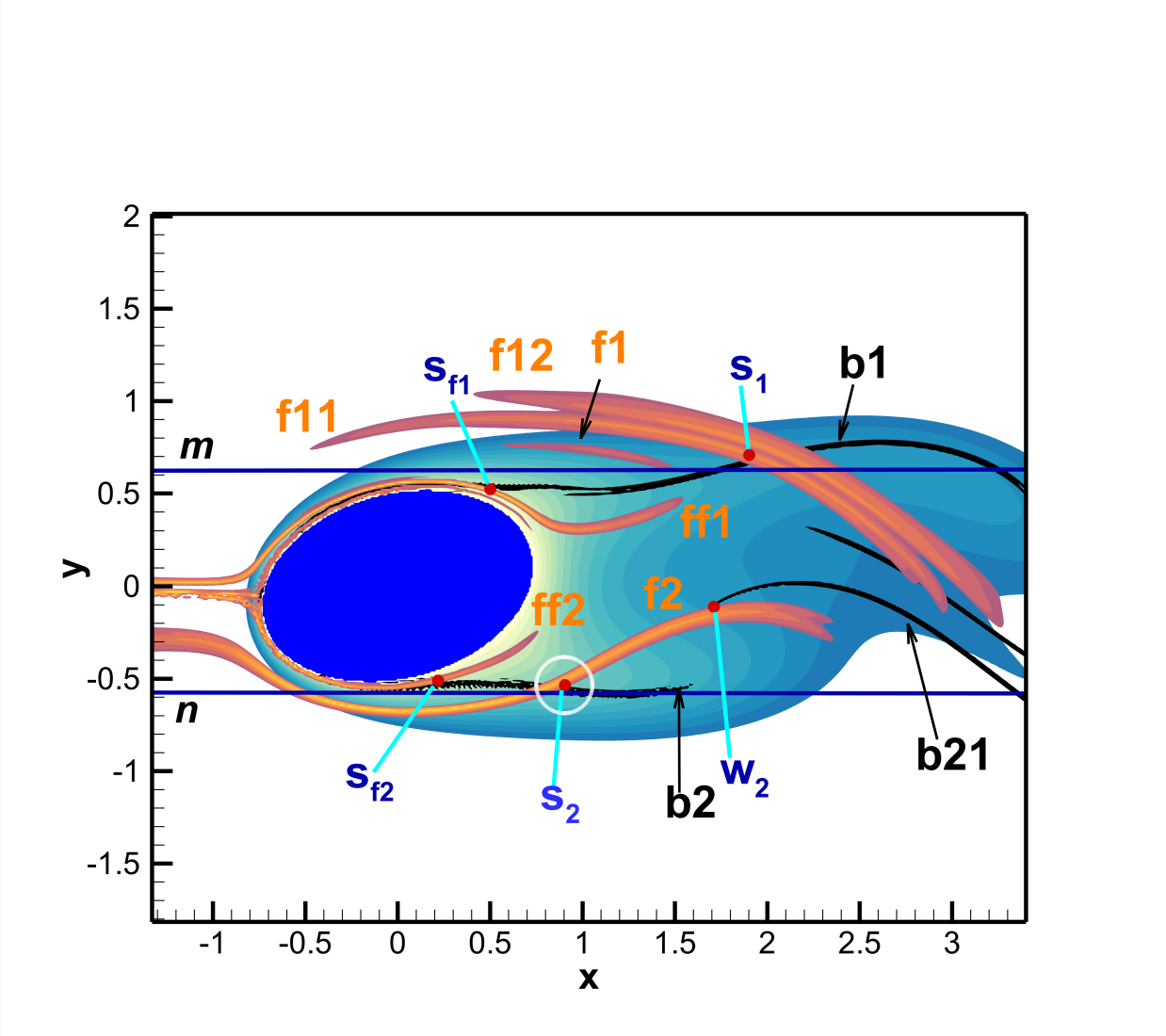}
    }

\end{figure}

\begin{figure}[htbp]\ContinuedFloat
    \centering
    \subfloat[$t^* = 0.645$\label{fig: LCS15IM6}]{%
        \includegraphics[width=0.49\linewidth]{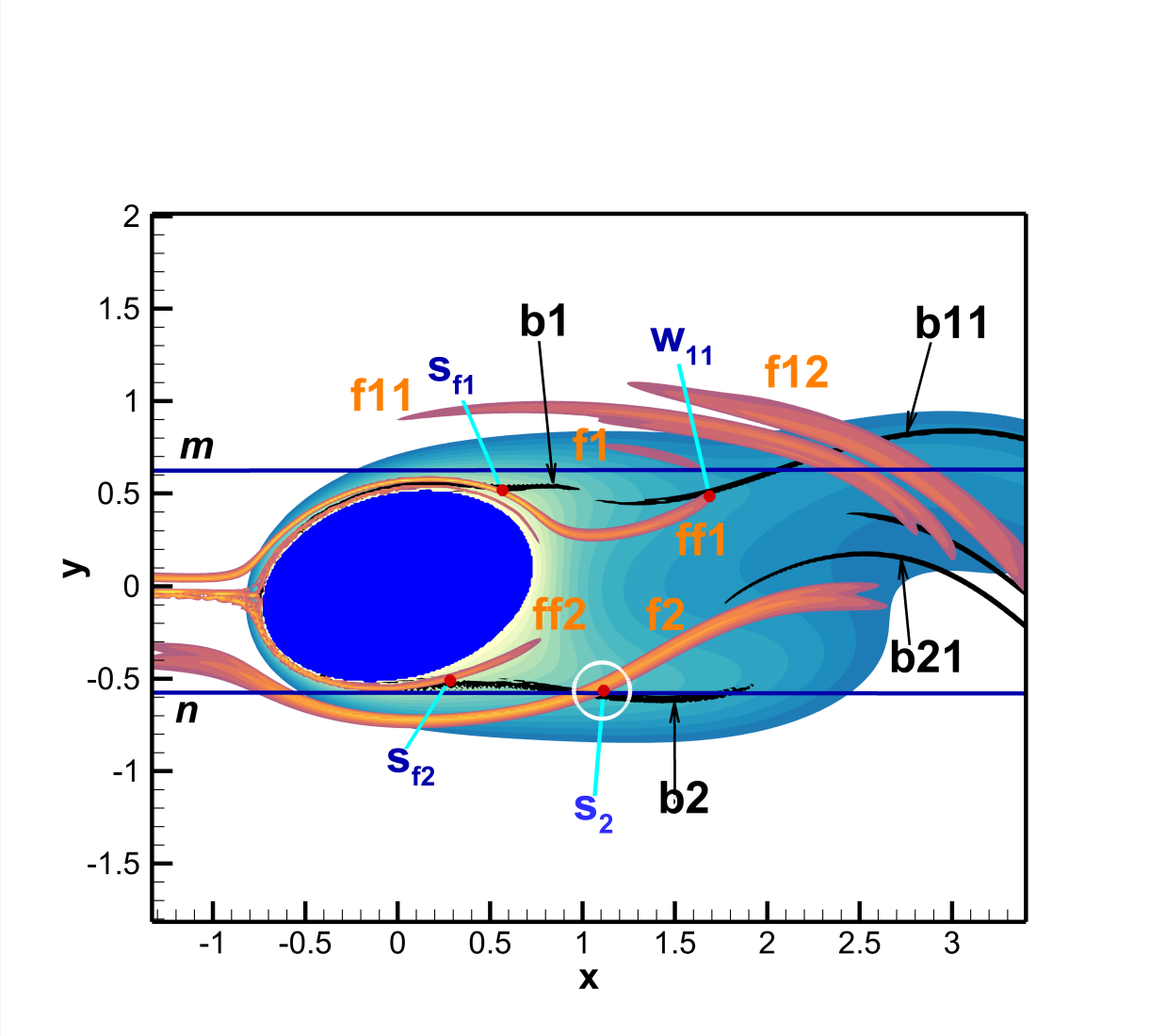}
    }\hfill
    \subfloat[$t^* = 1$\label{fig: LCS15IM7}]{%
        \includegraphics[width=0.49\linewidth]{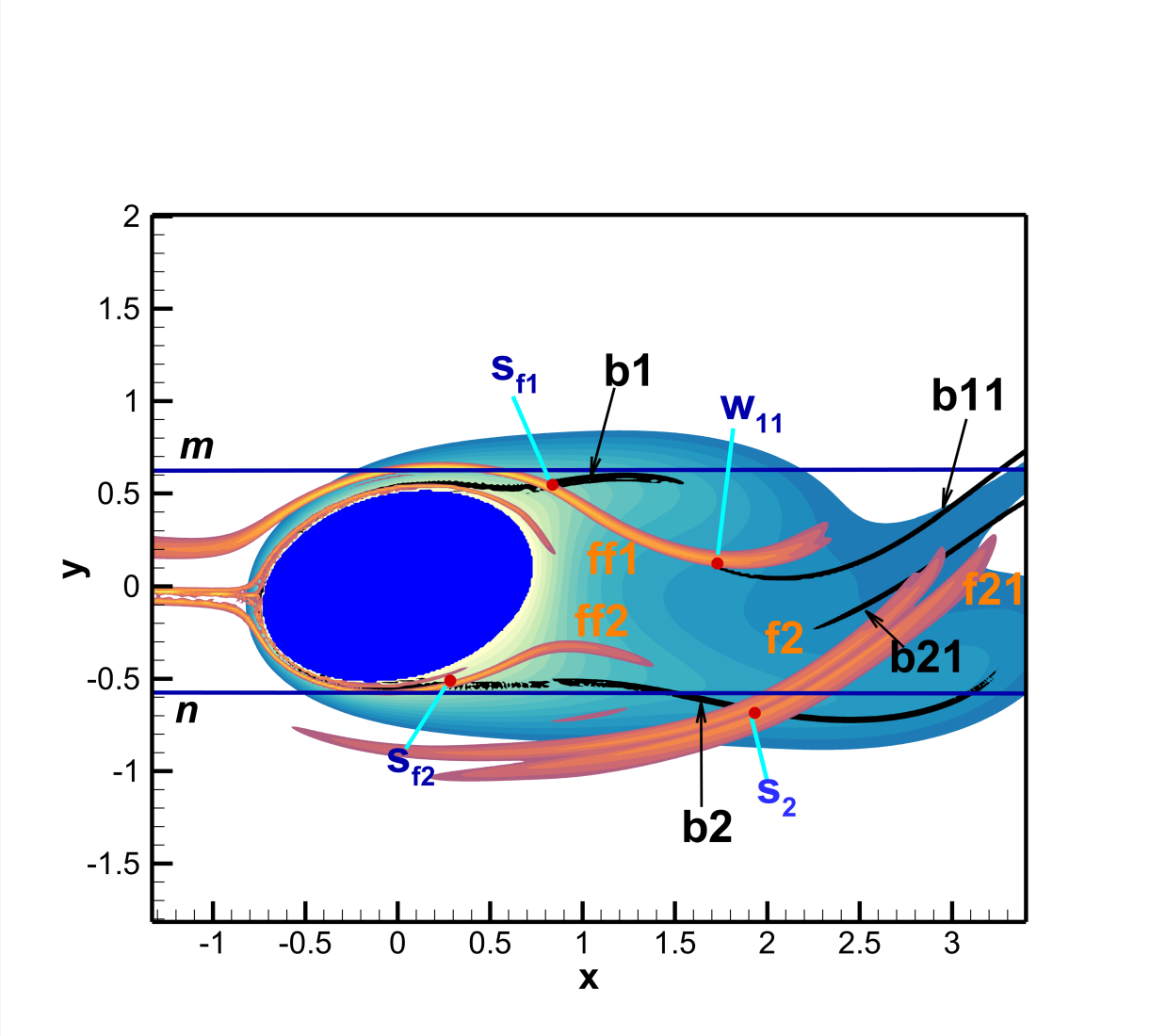}
    }
\caption{Instantaneous temperature contours and LCSs around the elliptical cylinder at $\theta = 15^\circ$ for $Re = 100$ at : (a) $t^* = 0$ , (b) $t^* = 0.129$ , (c) $t^* = 0.274$ , (d) $t^* = 0.403$ , (e) $t^* = 0.532$ , (f) $t^* = 0.645$ , (g) $t^* = 1$.}
\label{fig: LCS15}

\end{figure}

\begin{figure}[htbp]
    \centering
    \subfloat[$t^* = 0$\label{fig: LCS45IM1}]{%
        \includegraphics[width=0.49\linewidth]{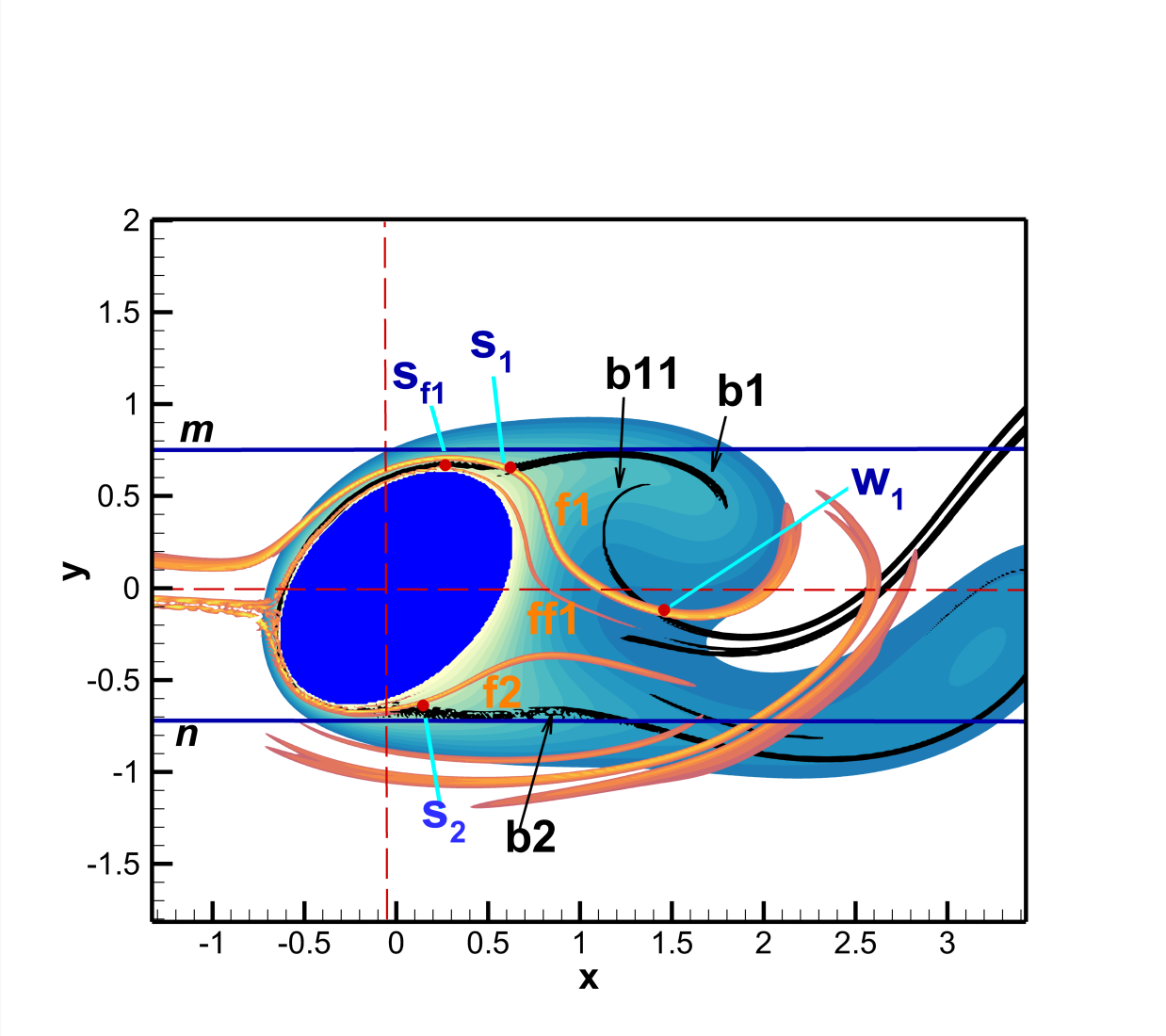}
    }\hfill
    \subfloat[$t^* = 0.176$\label{fig: LCS45IM2}]{%
        \includegraphics[width=0.49\linewidth]{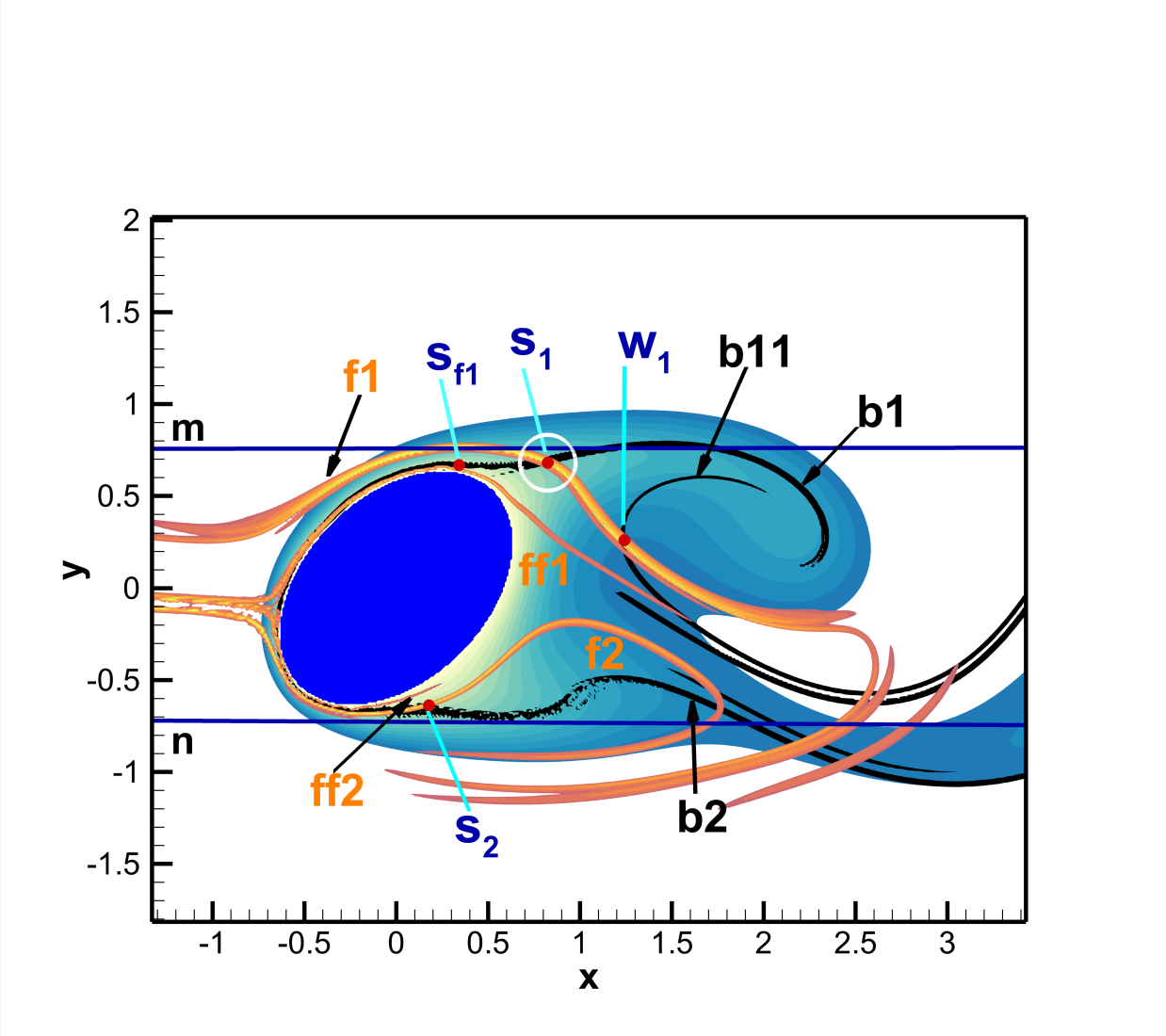}
    }\par\medskip

    \subfloat[$t^* = 0.308$\label{fig: LCS45IM3}]{%
        \includegraphics[width=0.49\linewidth]{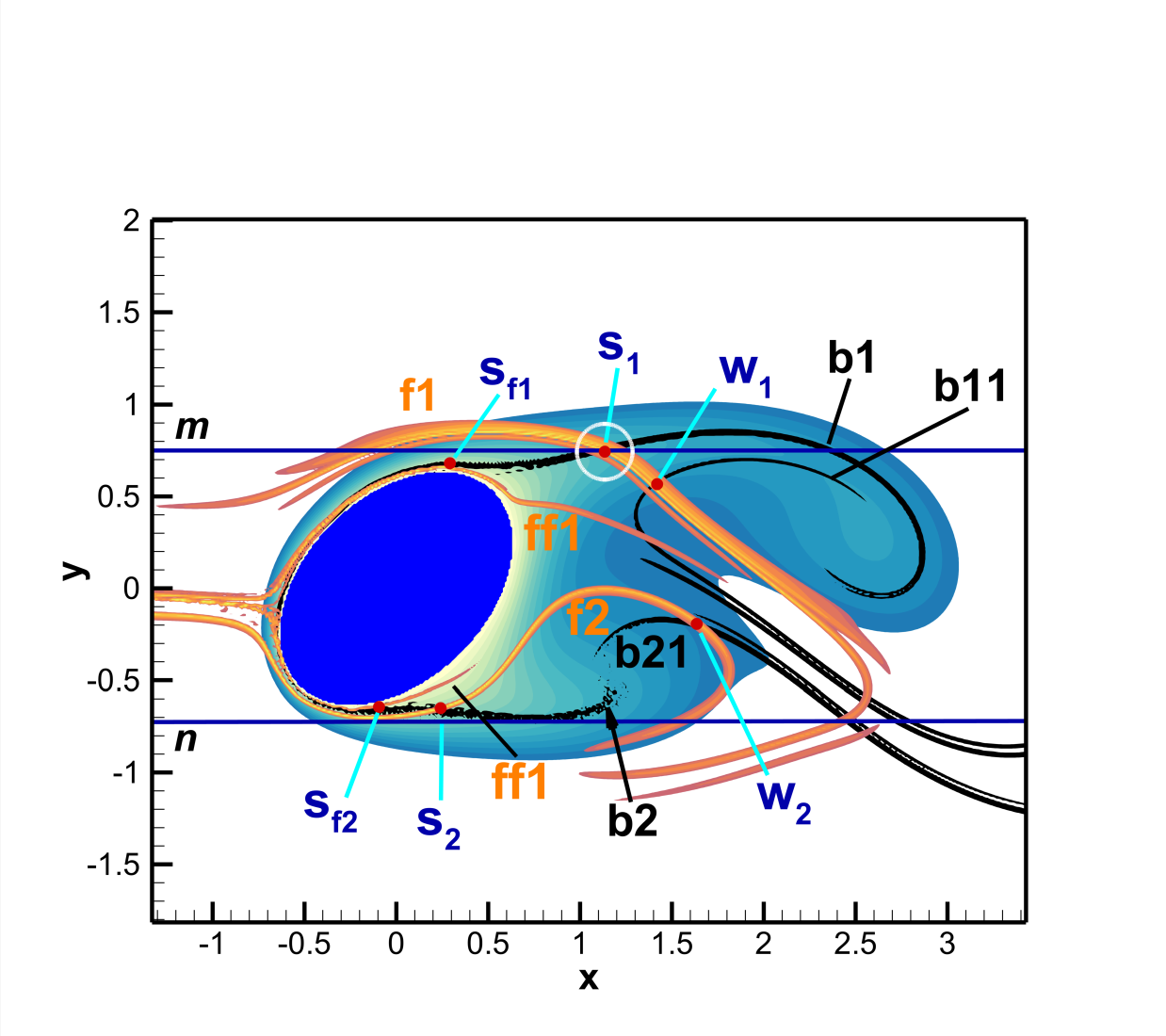}
    }\hfill
    \subfloat[$t^* = 0.485$\label{fig: LCS45IM4}]{%
        \includegraphics[width=0.49\linewidth]{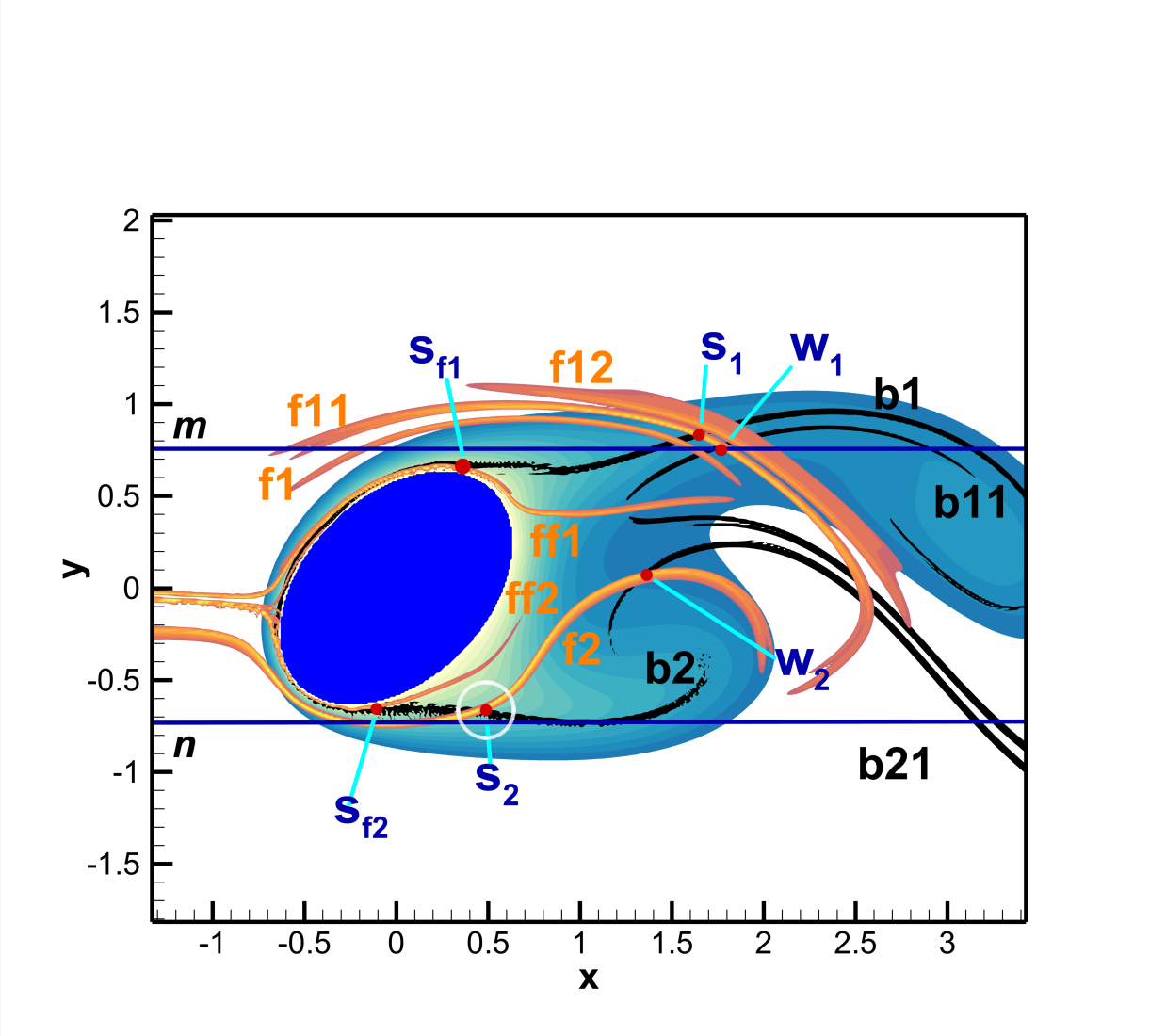}
    }\par\medskip

    \subfloat[$t^* = 0.661$\label{fig: LCS45IM5}]{%
        \includegraphics[width=0.49\linewidth]{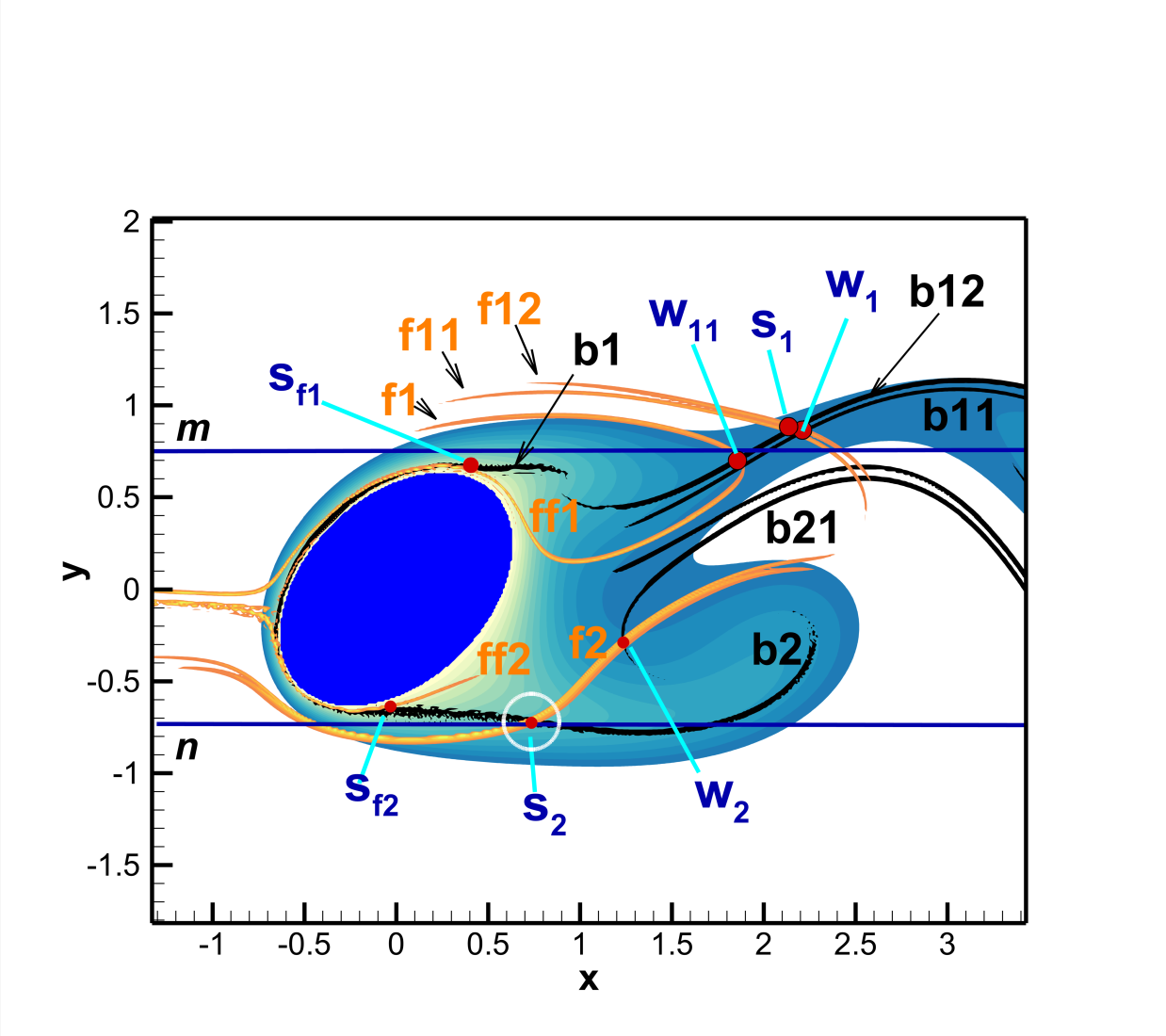}
    }
\end{figure}

\begin{figure}[htbp]\ContinuedFloat
    \centering
    \subfloat[$t^* = 0.823$\label{fig: LCS45IM6}]{%
        \includegraphics[width=0.49\linewidth]{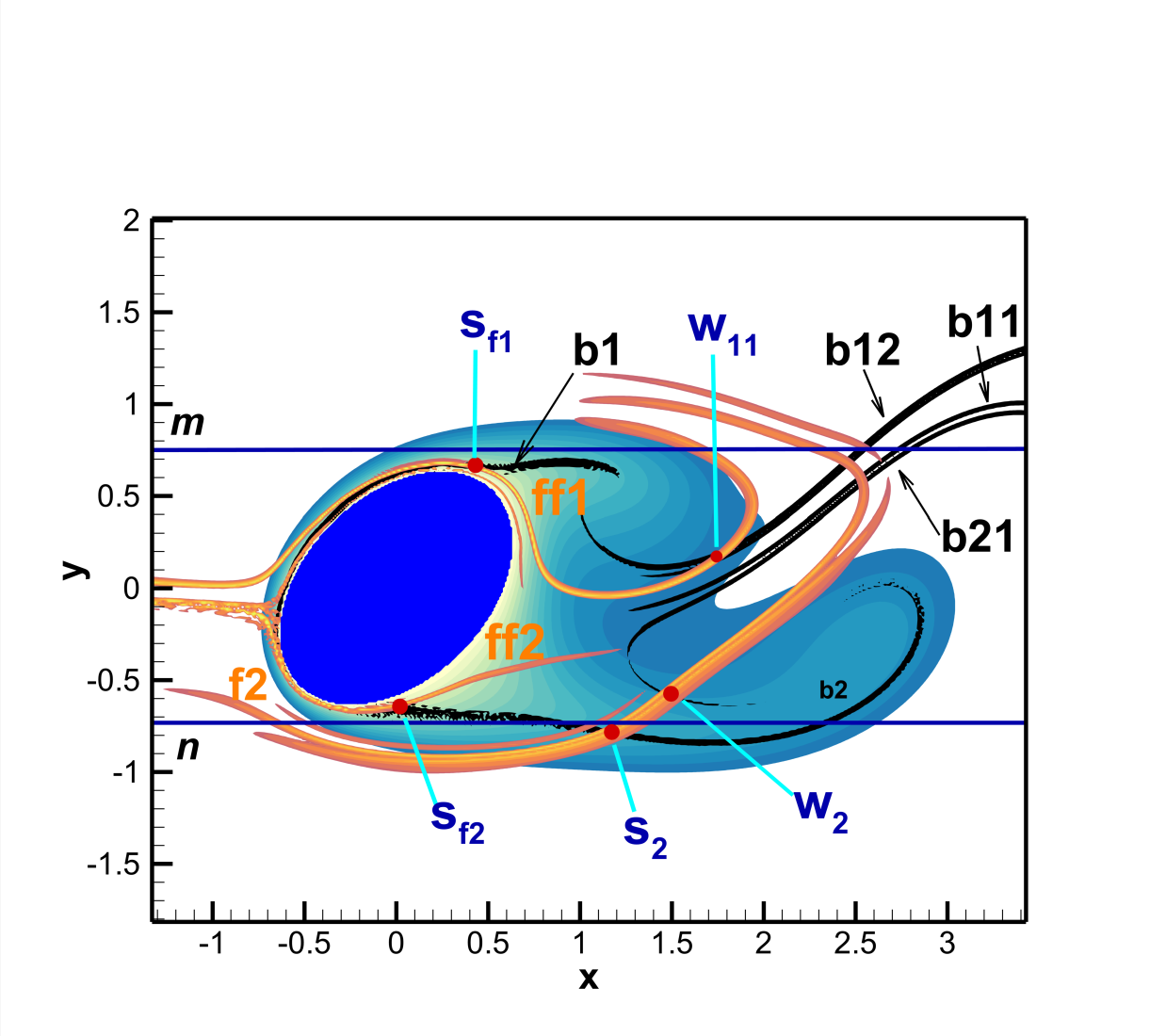}
    }\hfill
    \subfloat[$t^* = 1$\label{fig: LCS45IM7}]{%
        \includegraphics[width=0.49\linewidth]{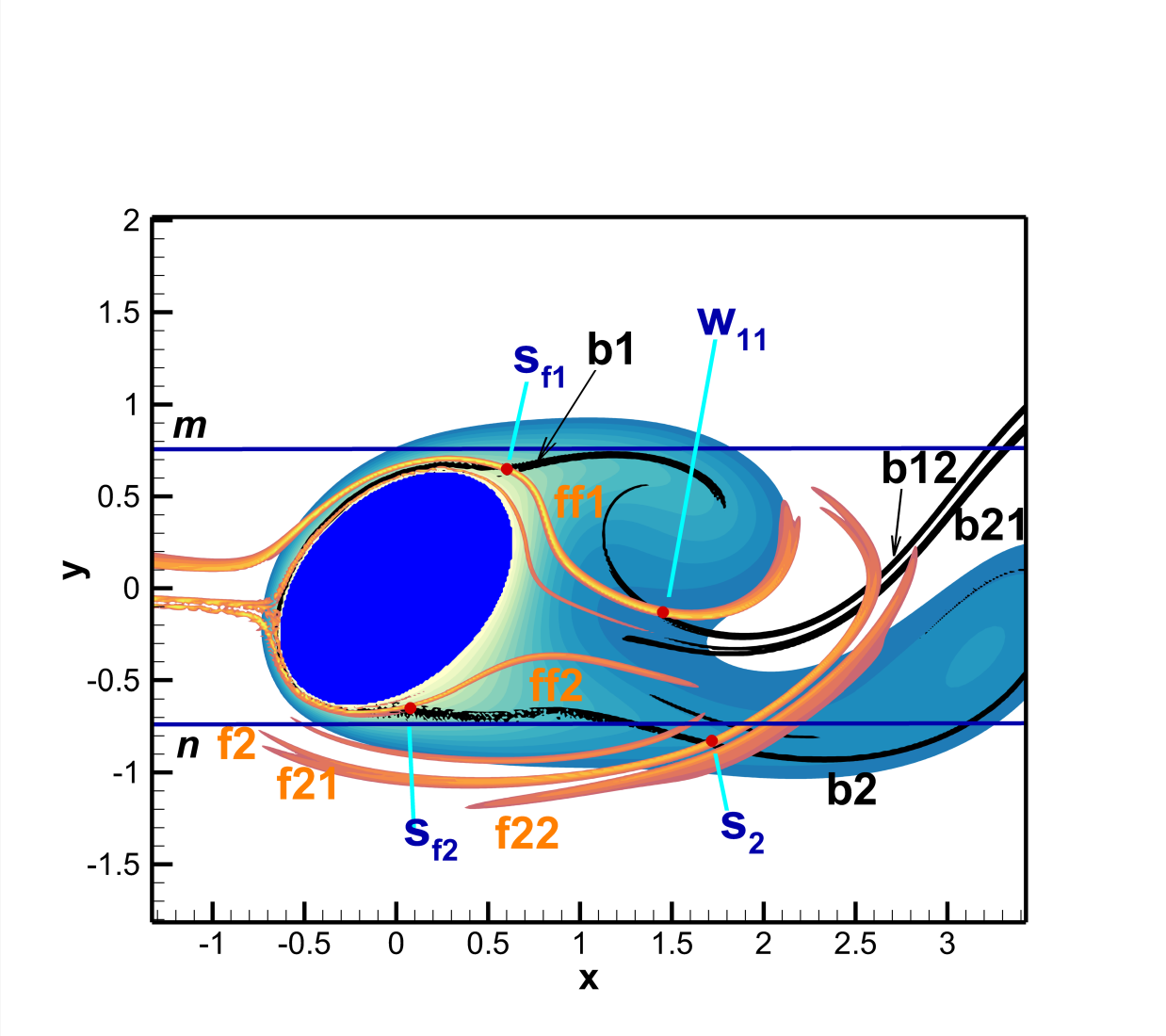}
    }

\caption{Instantaneous temperature contours and LCSs around the elliptical cylinder at $\theta = 45^\circ$ and $Re = 100$ at : (a) $t^* = 0$ , (b) $t^* = 0.176$ , (c) $t^* = 0.308$ , (d) $t^* = 0.485$ , (e) $t^* = 0.661$ , (f) $t^* = 0.823$ , (g) $t^* = 1$.}
\label{fig: LCS45}
\end{figure}

 The active saddle points at $t^* = 0$ for $\theta=0^\circ$, in accordance with the definition, are $s_1, s_{\text{f1}}, s_{\text{f2}}$ and $s_2$, as shown in Fig. \ref{fig: LCS0IM1}. As the non-dimensional time $t^*$ progresses within the time interval $[0, 0.246]$, saddle point $s_1$ undergoes a gradual downstream migration accompanied by a slight upward displacement, as illustrated in Fig. \ref{fig: LCS0IM1} through \ref{fig: LCS0IM3}. To facilitate the tracking of saddle points $s_1$ and $s_2$ in the near-wake region, upper boundary $m$ and lower boundary $n$ of the observational domain are used as reference. At $t^*=0.246$, saddle point $s_1$ intersects with the boundary $m$, highlighted by the white circle marking a significant event in its trajectory evolution in Fig. \ref{fig: LCS0IM3}. This intersection highlights a notable feature in the temporal evolution of the Nusselt number. Specifically, the time at which saddle point $s_1$ intersects the line $m$ corresponds to the first occurrence of a relative maximum in the temporal evolution of Nusselt number, as depicted in Fig. \ref{fig:NU0}.\\ 

Fig. \ref{fig: TrajectoryCombined} illustrates the trajectory of active saddle points in one vortex shedding time period ($T$), for nclination angles $\theta=0^\circ, 15^\circ, 30^\circ, 45^\circ, 60^\circ, 75^\circ$, and $90^\circ$. In the time interval from $t^*=0$ to  $t^*= 0.246$, the trajectories of the active saddle points are systematically tracked for $\theta = 0^\circ$ in the current period. As observed in Fig. \ref{fig:track0}, the trajectory of saddle point $s_1$ exhibits a clear trend of decreasing normal distance from the boundary $m$, whereas the saddle points $s_{\text{f1}}, s_{\text{f2}}$ and $s_2$ display motion that remains approximately parallel to boundaries $m$ and $n$, as illustrated in Fig. \ref{fig:track0}. The parallel displacement of $s_{\text{f1}}$ and the oblique displacement of saddle $s_1$ along a positive slope trajectory corresponds to the slow detachment of LCS ff1 and shedding of LCS f1 from the upper surface of the elliptical cylinder respectively, whereas the parallel displacement of $s_{2}, s_{\text{f2}}$ leads to slow detachment of LCS f2 and ff2 from the lower surface of the cylinder. During this time interval, the LCS b2 breaks down and branches into LCS b21 and the LCS f2 goes under elongatation followed by the intersection with the branch LCS b21 at $t^*=0.246$. The breaking of attracting LCS b2 around the lower surface of the ellipse entrains the mainstream fluid particles from the open cavity in the wake region. The combined effect of breaking of LCS b2 and the elongatation of the LCS f2 lead to the formation of a coherent pocket of hot fluid particles bounded by LCS f2 and LCS b2 highlighted in green box as illustrated in Fig. \ref{fig: LCS0IM3}. Saddle points on the surface of the cylinder advects this concentrated coherent packet of thermally energized fluid particles in the flow.\\

From $t^*=0.246$ onwards, saddle point $s_2$ track is highlighted with the white circle as seen from Fig. \ref{fig: LCS0IM4} to Fig. \ref{fig: LCS0IM5}. The LCS f1 sheds away further downstream and a new LCS branch is observed, denoted as LCS f11. Intersection of LCS f1 and LCS b1, i.e., saddle point $s_1$ also travels downstream and upwards leaving the observational domain. Consequently, saddle point $s_1$ ceases to act as an active saddle point beyond $t^* = 0.246$. Since the framework only incorporates the trajectories of active saddle points, the trajectory pattern of $s_1$ will not be examined further in the Heat transfer analysis. At $t^* = 0.246$, the active saddle points identified within the domain are  $s_{\text{f1}}, s_{\text{f2}}$ and $s_2$. In Fig. \ref{fig:track0}, the trajectories of these active saddle points from $t^* = 0.246$ to $t^* = 0.409$ are observed to align approximately parallel to the boundary lines $m$ and $n$ denoted as $t^*=\bar{t}$. After  $t^* = 0.409$, a noticeable change in  the trajectory of saddle point $s_2$ is observed, as illustrated in Fig.~\ref{fig:track0}. We denote the time instant $t^* = 0.409$ as  $t^*= \bar{t}$. \\

From $t^*=\bar{t}$ onwards, the attracting LCS b1 breaks into a branch denoted as LCS b11. With the elongation of LCS b2 and further downstream convection of LCS f2, saddle point $s_2$, initially convecting parallel to the boundaries of the observational domain now begins to convect downwards exhibiting a decreasing normal distance from the boundary n. The remaining active saddle points  $s_{\text{f1}}, s_{\text{f2}}$ persist in near horizontal motion but the slight downward progression of saddle point $s_2$ leads to its intersection with the boundary $n$ in seen in Fig. \ref{fig: LCS0IM6}. This parallel motion of $s_{\text{f1}}, s_{\text{f2}}$ and downward oblique motion of the saddle point $s_2$ is tracked until $s_2$ intersects with the lower boundary $n$ of the observational domain at $t^*=0.688$ as observed in Fig. \ref{fig:track0}. Notably, during this interval—from the onset of downward motion of $s_2$ to its intersection with lower boundary n, sufficient detachment of LCS f2 took place and the concentrated region of high-temperature fluid bounded by the LCS f2 and the LCS b2 subsequently begins to shed into the wake whereas LCS ff1 and LCS ff2 are still in its slow detachment phase. \\

\begin{figure}[htbp]
    \centering
    \subfloat[$\theta = 0^\circ$\label{fig:track0}]{%
        \includegraphics[width=0.49\linewidth]{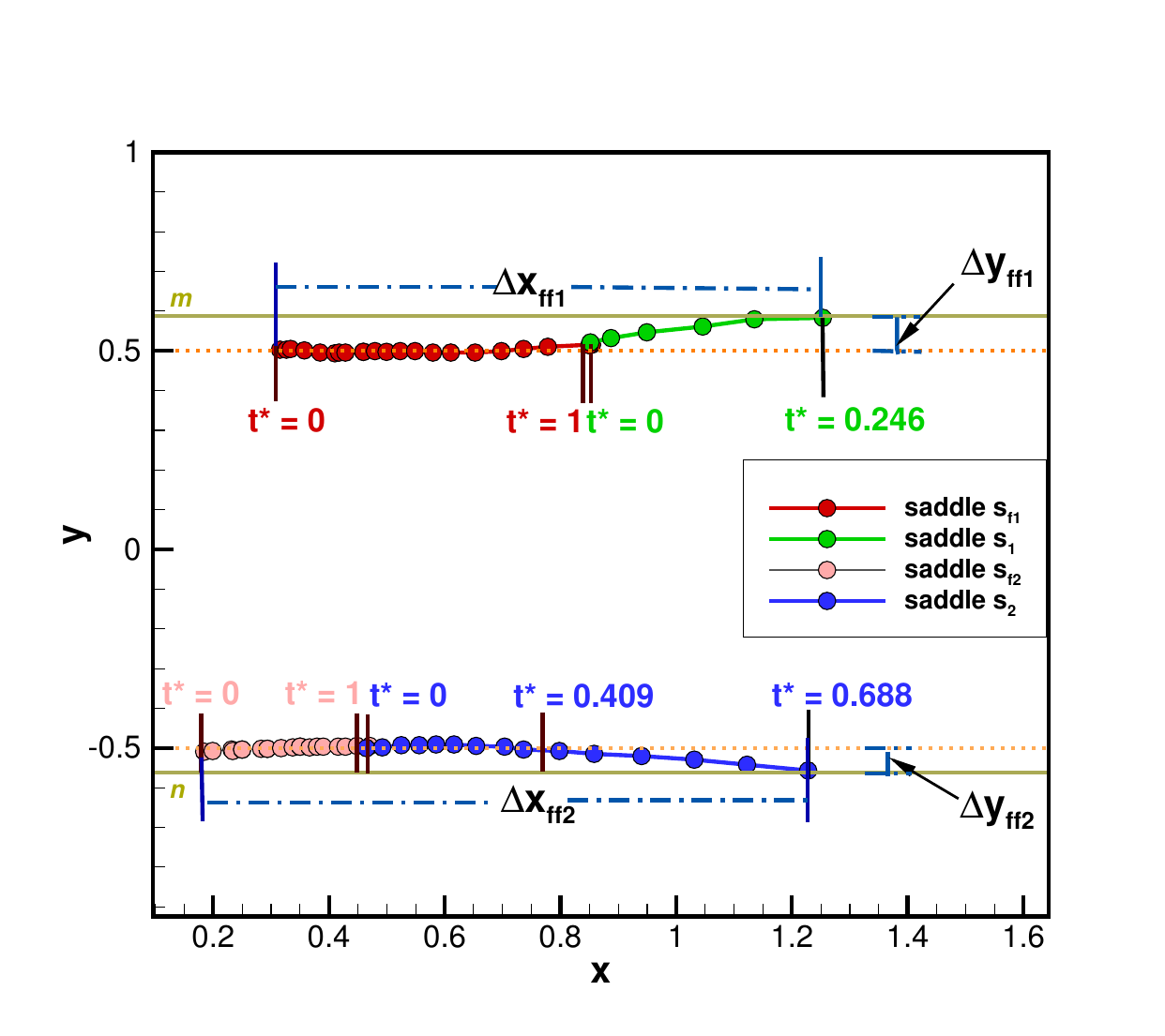}
    }\hfill
    \subfloat[$\theta = 15^\circ$\label{fig:track15}]{%
        \includegraphics[width=0.49\linewidth]{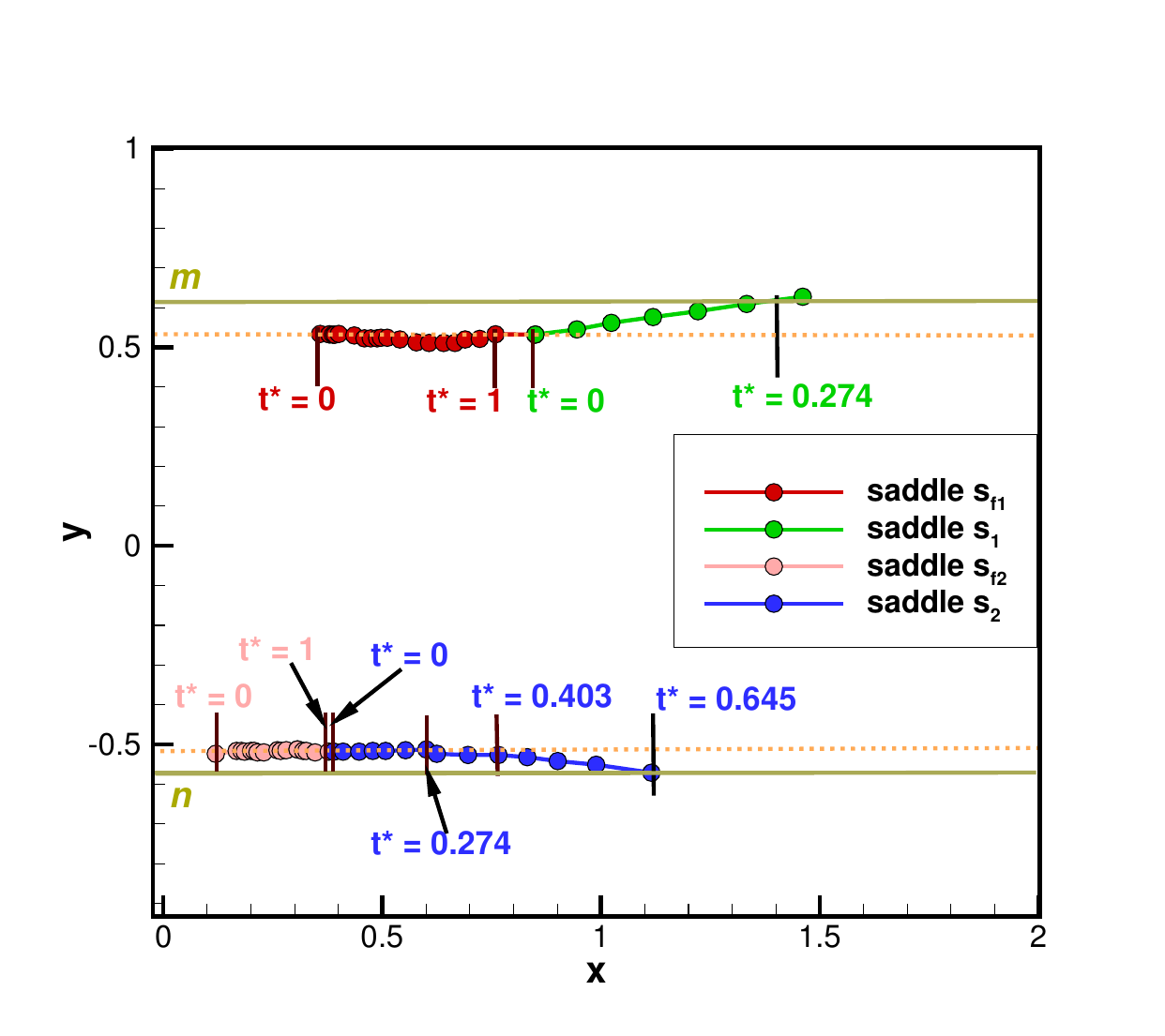}
    }\par\medskip

    \subfloat[$\theta = 30^\circ$\label{fig:track30}]{%
        \includegraphics[width=0.49\linewidth]{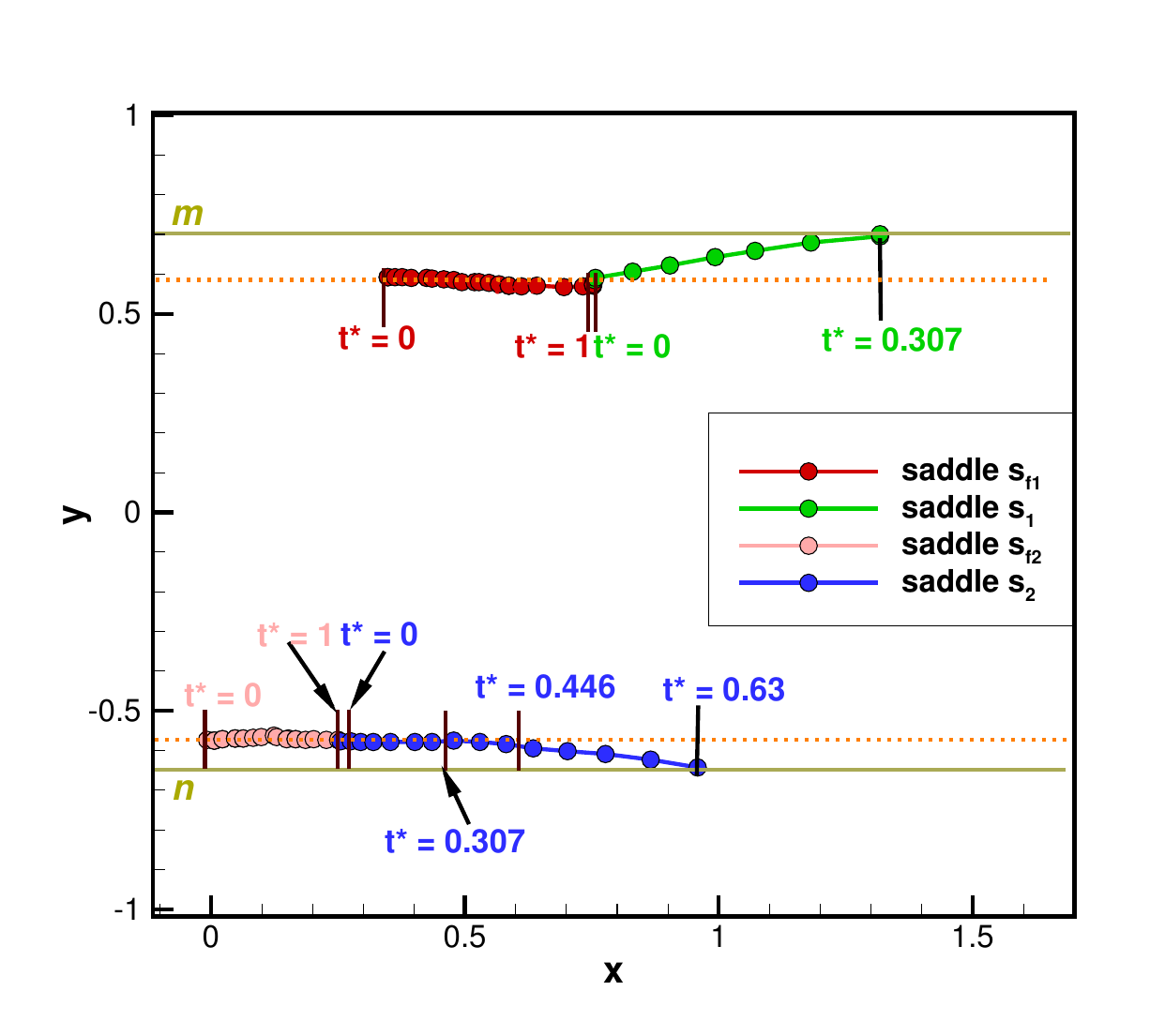}
    }\hfill
    \subfloat[$\theta = 45^\circ$\label{fig:track45}]{%
        \includegraphics[width=0.49\linewidth]{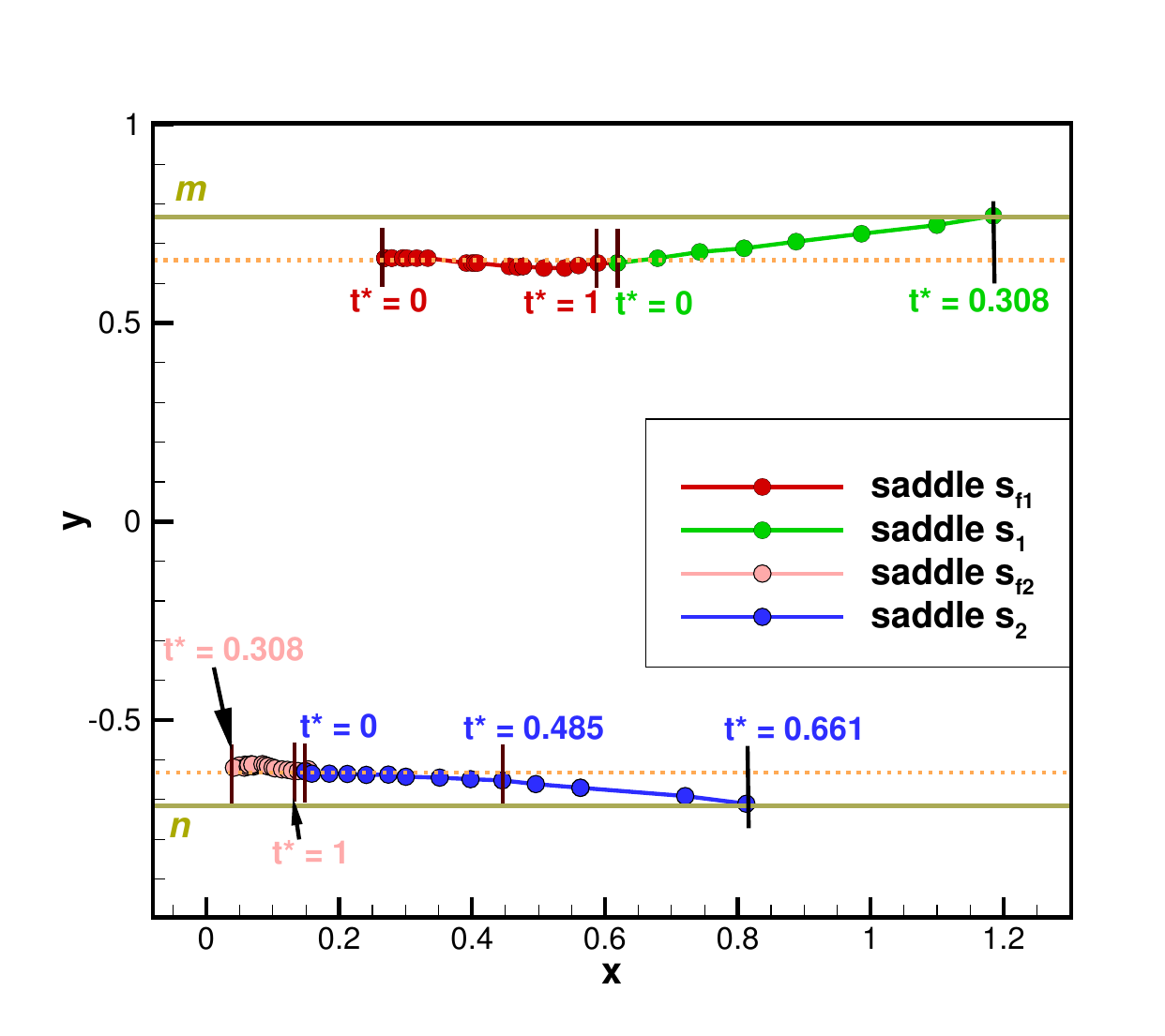}
    }

\end{figure}

\begin{figure}[htbp]\ContinuedFloat
    \centering
    \subfloat[$\theta = 60^\circ$\label{fig:track60}]{%
        \includegraphics[width=0.49\linewidth]{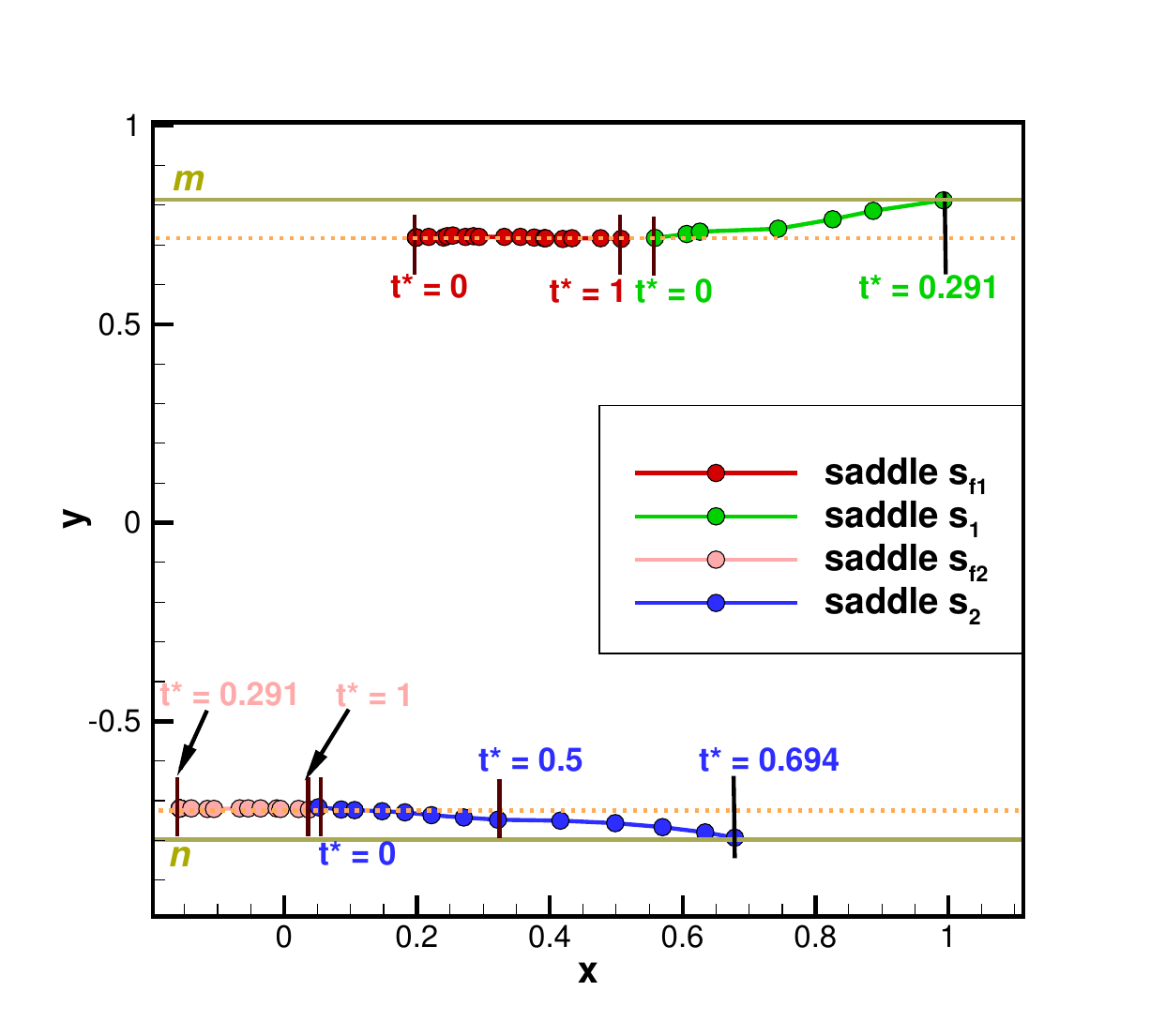}
    }\hfill
    \subfloat[$\theta = 75^\circ$\label{fig:track75}]{%
        \includegraphics[width=0.49\linewidth]{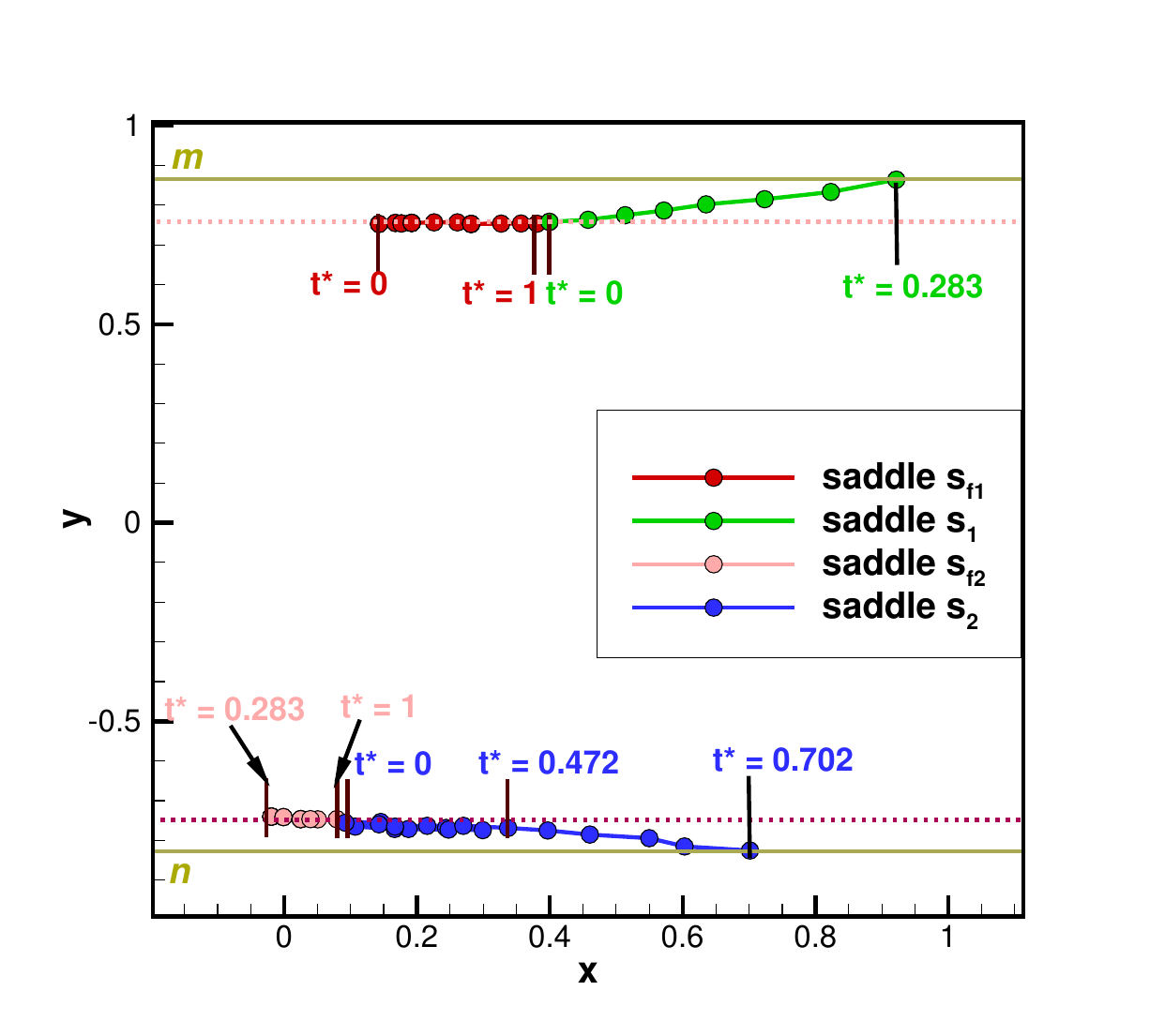}
    }\par\medskip

    \subfloat[$\theta = 90^\circ$\label{fig:track90}]{%
        \includegraphics[width=0.49\linewidth]{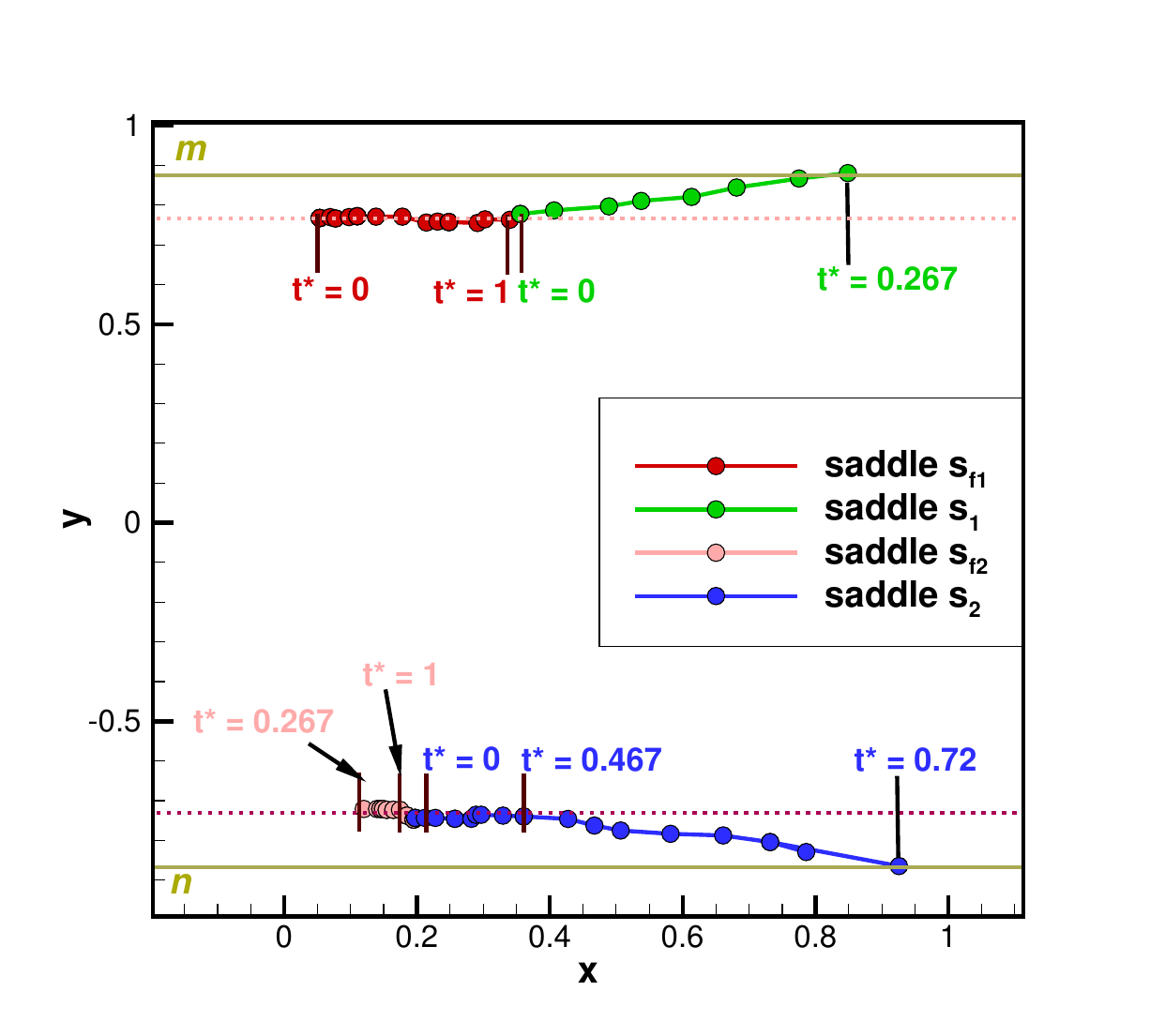}
    }
    \caption{Trajectory of active saddle points in flow past an elliptical cylinder for $Re = 100$ at : (a) $\theta = 0^\circ$, (b) $\theta = 15^\circ$, (c) $\theta = 30^\circ$, (d) $\theta = 45^\circ$, (e) $\theta = 60^\circ$, (f) $\theta = 75^\circ$, (g) $\theta = 90^\circ$}
\label{fig: TrajectoryCombined}
\end{figure}

Slow detachment and shedding of the repelling LCSs governed by the trajectories of the active saddle points $s_{\text{f1}},  s_{\text{f2}}$, $s_1$ and $s_2$ can be linked to the convective heat transfer rate of the elliptical cylinder. Let $(x_k(t^*), y_k(t^*))$ denote the coordinates of saddle point $k$ at time $t^*$ in two-dimensional space. Let $t^*=t_{k,l}$ denote the time instant at which saddle point $k$ intersects with the boundary $l$ of the observational domain. As $t^*$ progresses from $t^*=0$, the saddle point $s_1$ lifts off from the cylinder. The downstream and upwards progression of saddle point $s_1$ initiates the downstream motion and subsequent shedding of the repelling LCS f1 from the upper surface of the elliptical cylinder. The shedding of the LCS f1 from the cylinder's surface contributes to significant convection of heated fluid particles in the wake thus enhancing heat transfer rate. In contrast, the repelling LCSs ff1, ff2 and f2 are in slow detachment phase leading to the negligible contribution to heat convection in this time interval. This is attributed to the near horizontal displacement of active saddle points $s_{\text{f1}},  s_{\text{f2}}$ and $s_2$ respectively. Consequently, a net enhancement in the heat transfer rate on the upper surface of the cylinder is observed, primarily due to the downstream and upward shedding of the repelling LCS f1. This trend continues till $s_1$ reaches the upper boundary of  the observational domain at $t^*=0.246 = t_{{s_1},m}$. For $t^* = t_{{s_1},m}$ onwards, LCS f1 sheds away from the cylinder and advects out of the observational domain. As a result, subsequent shedding of LCS f1 occurs outside the domain of influence and has negligible influence on further enhancing the heat transfer rate. The remaining active saddle points $s_{\text{f1}},  s_{\text{f2}}$ and $s_2$ in the observational domain now governs the heat transfer dynamics. The trajectory of these saddle points remains closely parallel to flow direction resulting in the the slow and prolonged detachment of LCSs ff1, ff2 anf f2 thereby leading to a net weakened heat transfer rate. For $t^* \in [0, t_{{s_1},m}]$, the flow exhibits enhanced convective heat transfer whereas a decline in convective heat transfer is observed in the interval $[t_{{s_1},m}, \bar{t}]$. This transition highlights $ t^*=t_{{s_1},m} $ as a critical time instant, marking the shift from positively sloped profile to a negatively sloped profile in the temporal evolution of Nusselt number. Notably, \( t^* = 0.246 = t_{{s_1},m} \) coincides with the time at which the Nusselt number attains its first local maximum, as shown in Fig.~\ref{fig:NU0}.\\

From $ t^*=\bar{t}$ onwards, the onset of the shedding phase for LCS f2 is marked by saddle point $s_2$ beginning its downward oblique motion and enhancing heat transfer. The trajectories of the saddle points $s_{\text{f1}},  s_{\text{f2}}$ and $s_2$ are tracked until $t^*=t_{{s_2},n}$. The transition from weakened convective transfer in interval $[t_{{s_1},m}, \bar{t}]$ to a regime of stronger convective transfer in the interval $[\bar{t}, t_{{s_2},n} ]$ identifies the instant $t^*=\bar{t}$ as the second critical time point in the thermal transport cycle. Consequently, the transition at $t^*=\bar{t}$ underscores a shift in the slope of the Nusselt number profile. Significantly,  $t^*=\bar{t}$ aligns with the time at which the Nusselt number reaches its intermediate local minimum as shown in Fig.~\ref{fig:NU0}. The increasing trend in the $Nu$ profile persists until  $t^*=t_{{s_2},n}$.\\

During the time interval $[t_{{s_2},n}, 1]$, following the departure of saddle point $s_2$ from the observational domain, the active saddle points that remain within the domain are $s_{\text{f1}}$ and $s_{\text{f2}}$. In Fig. \ref{fig: LCS0IM6} to Fig. \ref{fig: LCS0IM7}, the LCS disappears completely after shedding in the far wake and the LCS ff1 and LCS ff2 has reaches the initial location of LCS f1 and f2 respectively.  In Fig. \ref{fig:track0}, both of the remaining active saddle points exhibit trajectories approximately parallel to that of the boundaries $m$ and $n$.  As a result of this parallel convection, a continued decrease in the Nusselt number is observed in Fig. \ref{fig:NU0}. These observations unfolds the correlation between the trajectories of active saddle points and the temporal trend of the Nusselt number. When the active saddle points convect predominantly parallel to the flow direction, a decline in the Nusselt number is observed and when any of the active saddle points migrates obliquely with respect to the flow direction, the Nusselt number exhibits an increasing trend, reflecting enhanced thermal convection. This relationship persists and can be consistently identified in subsequent time intervals as well. \\

 The active saddle points identified across all angular configurations at $t^* = 0$ are $s_{\text{f1}},  s_{\text{f2}}$, $s_1$ and $s_2$. For $\theta = 15^\circ$ and $\theta = 30^\circ$, the $Nu$ profiles exhibit notable differences from the $\theta = 0^\circ$ case, particularly in the absence of an intermediate local minimum, as illustrated in Figs.~\ref{fig:NU15} and~\ref{fig:NU30}. The Lagrangian framework developed for $\theta = 0^\circ$ to analyze convective heat transfer is adapted to account for this distinction, offering a consistent approach to interpreting the heat transfer dynamics under varying inclination angles. The framework defined for $\theta = 0^\circ$, including the LCS configurations, the construction of the observational domain, and the labeling conventions for LCSs and saddle points, is consistently applied to study convective heat transfer for $\theta = 15^\circ$ and $\theta = 30^\circ$.  Note that the configurations of LCSs and the isotherms for $\theta =15^\circ$ and $\theta = 30^\circ$ are qualitatively same and hence illustrations provided in this paper are only for $\theta =15^\circ$, shown in Fig. \ref{fig: LCS15}.\\

The trajectories of the active saddle points observed in Fig. \ref{fig:track15} and Fig. \ref{fig:track30} are qualitatively same as that in the case $\theta = 0^\circ$, with LCS f1 beginning its shedding phase due to the onset of oblique advection of saddle point $s_1$ and LCSs ff1, ff2 and f2 undergoing horizontal detachment due to parallel advection of remaining active saddle points $s_{\text{f1}}, s_{\text{f2}}$ and $s_2$ as shown from Fig.~\ref{fig: LCS15IM1} to Fig.~\ref{fig: LCS15IM3}. As discussed before, the observations of the trajectories of the active saddle point are recorded visually till $t^*=t_{{s_1},m} = 0.274$ for $\theta = 15^\circ$ and $t^*=t_{{s_1},m} = 0.307$ for $\theta = 30^\circ$. For $t^* \in [0, t_{{s_1},m}]$, shedding of LCS f1 dominates the heat convection leading to a rise in Nu profile in this interval. At $t^*= t_{{s_1},m}$, the active saddle points are $s_{\text{f1}}$, $s_{\text{f2}}$ and $s_2$. From $t^* = t_{{s_1},m}$ onwards, the trajectories of saddle points $s_{\text{f1}}$ and $s_{\text{f2}}$ exhibit nearly parallel displacement, while saddle point $s_2$ begins to deviate marginally in a downward, negatively oblique direction as seen in Fig. \ref{fig:track15} and Fig. \ref{fig:track30}. This deviation corresponds to the local onset of the shedding phase of LCS f2 at $t^* = t_{{s_1},m}$. However, due to the minimal downward migration, the shedding of LCS f2 is comparatively less vigorous than that of LCS f1 during the $t^* \in [0, t_{{s_1},m}]$. As a result, a slight decline in the slope of the convective heat transfer rate is anticipated at $t^* = t_{{s_1},m}$. This behavior is reflected in Figs.~\ref{fig:NU15} and~\ref{fig:NU30}, where a noticeable change in slope is observed at $t^* = t_{{s_1},m}$, indicated by solid and dotted black lines. As time progresses further, the shedding of LCS f2 intensifies, resulting in enhanced convective heat transfer, which is marked by an increase in the slope of the Nusselt number profile at $t^* = 0.403$ for $\theta = 15^\circ$ and $t^* = 0.446$ for $\theta = 30^\circ$. This time instant is denoted as \( t^* = \bar{t} \) in the evolution of the Nusselt number. In the interval $[\bar{t},  t_{{s_2},n}]$, from Fig.~\ref{fig: LCS15IM4} to Fig.~\ref{fig: LCS15IM6}, the shedding of LCS f2 continues leading to an increase in convective heat transfer and is tracked until the intersection of saddle point $s_2$ and the lower boundary $n$ is observed at \( t^* = t_{{s_2},n} \). In the time interval \( [ t_{{s_2},n}, 1] \), the active saddle points within the domain are $s_{\text{f1}}$ and $s_{\text{f2}}$, as shown in Fig.~\ref{fig: LCS15IM6}. In Figs.~\ref{fig:track15} and~\ref{fig:track30}, the trajectories of these saddle points display nearly parallel advection, which leads to a reduction in convective heat transfer in this interval. As a result, the time instant \( t^* = t_{{s_2},n} \) marks a change in the slope of the Nusselt number profile. Notably, \( t^* = t_{{s_2},n}\) coincides with the second local maximum of the Nusselt number for both $\theta = 15^\circ$ and $\theta = 30^\circ$, as shown in Figures~\ref{fig:NU15} and~\ref{fig:NU30}. \\

We will extend the established framework for inclination angles $\theta = 45^\circ, 60^\circ, 75^\circ$ and $90^\circ $. Since the underlying dynamics across these inclination angles exhibit consistent behavior, the figures presented will focus only on the illustrations of \( \theta = 45^\circ \). In Fig.~\ref{fig: LCS45}, it can be noted that contours of LCSs delineate the hot fluid regions from the mainstream fluid more distinctly than for  $\theta = 0^\circ$ and  $15^\circ$. A similar sequence of events as that of $\theta = 0^\circ$ is observed for these higher inclination angles in the evolution of active saddle points and the associated repelling LCSs over the interval $t^* = 0$ to \( t^* = 1 \), as shown in Fig.~\ref{fig: LCS45}. At \( t^* = 0 \), the forward-time FTLE field, post-filtering at 50\% of the maximum FTLE value, does not reveal a prominent FTLE ridge between LCS f2 and the lower surface of the cylinder. Consequently, the intersection responsible for the formation of saddle point $s_{\text{f2}}$ is not markedly visible at the initial time, so the active saddle points comprise of \( s_1 \), $s_{\text{f1}}$ and $s_{\text{f2}}$. As shown in Fig.~\ref{fig: TrajectoryCombined}, across all inclination angles, the trajectory of \( s_1 \) exhibits a consistent downstream and outward progression away from the cylinder surface. This motion corresponds to the shedding of LCS f1 from the upper surface, which, in turn, promotes enhanced convective transport of thermal energy in the wake. The saddle points $s_{\text{f1}}$ and $s_{\text{f2}}$ predominantly follow trajectories nearly parallel to boundaries $m$ and $n$ respectively, indicating a prolonged detachment phase of the associated LCSs ff1 and f2 which contributes minimally to convection as illustrated in Fig.~\ref{fig: LCS45IM1} to Fig.~\ref{fig: LCS45IM3}. These events leads to the net enhancement in convective heat transfer on the upper surface of the cylinder in the interval $[0, t_{{s_1},m}]$ which aligns with the increasing trend in the Nusselt number, as observed in Fig.~\ref{fig:NU45}. Once saddle point \( s_1 \) reaches the upper boundary $m$ and exits the observational domain at \( t^* = t_{s_1,m} \) in Fig.~\ref{fig: LCS45IM4}, the dynamics shift. The appearance of a visually distinct LCS \( \text{ff}_2 \) on the cylinder surface and the emergence of saddle point  $s_{\text{f2}}$ in Fig.~\ref{fig: LCS45IM3} mark the shift to a regime governed by the active saddle points $s_{\text{f1}}$ and $s_{\text{f2}}$, and \( s_2 \). Their near-horizontal progression sustains the slow detachment of LCSs ff1, ff2 and f2, resulting in a weakened convective heat transfer rate until saddle point \( s_2 \) begins to migrate with a progressively decreasing normal distance from boundary \( n \). The time instant of the onset of the downward progression of the saddle point $s_2$ is denoted as \( t^* = \bar{t} \).  Hence, in the interval $[t_{{s_1},m}, \bar{t}]$, a decreasing trend in $Nu$ profile is observed. This transition at \( t^* = t_{s_1,m} \) consistently aligns with the first local maximum in the Nusselt number profiles across all angular configurations as shown from Fig.~\ref{fig:NU45} to Fig.~\ref{fig:NU90}, further reinforcing \( t^* = t_{s_1,m} \) as a critical temporal marker in the convective heat transfer process.\\

The transition of saddle point $s_2$  from parallel displacement to an oblique trajectory at \( t^* = \bar{t} \) marks the transition of LCS f2 from its slow detachment phase to an active shedding phase shown in Fig.~\ref{fig: LCS45IM4} and Fig.~\ref{fig: LCS45IM5}. This corresponds to a shift from weak to strong convective heat transfer on the lower surface of the cylinder, and is reflected as a change in the slope of the Nusselt number profile in Fig.~\ref{fig:NU45}. Notably, this time instant also coincides with the intermediate local minimum of the Nusselt number curve, indicating the onset of a new convective regime as shown in Fig.~\ref{fig:NU45} to Fig.~\ref{fig:NU90}.\\

As the flow evolves beyond $t^* = \bar{t}$, saddle point \( s_2 \) continues its trajectory toward boundary \( n \), ultimately intersecting it at \( t^* = t_{s_2,n} \), which signals the vigorous shedding of LCS f2 and enhancement in heat convection rate in time interval $[\bar{t}, t_{s_2,n}]$  as shown in Fig.~\ref{fig: LCS45IM5} and Fig.~\ref{fig: LCS45IM6}. The active saddle points identified in the interval $[t_{s_2,n}, 1]$ are $s_{\text{f1}}$ and $s_{\text{f2}}$ and their corresponding LCSs ff1 and ff2 dominates the convective heat transfer process. Both the saddle points continues to advect nearly parallel to boundaries $m$ and $n$ respectively shown in  Fig.~\ref{fig:track45} and the LCSs ff1 and ff2 undergo slow detachment. This gradual detachment results in a decline in the overall heat transfer rate in the interval $[t_{s_2,n}, 1]$. As observed in Fig.~\ref{fig:NU45}, the time instant \( t^* = t_{s_2,n} \) corresponds to the second local maximum in the Nusselt number profile, marking the culmination of the shedding of LCS f2. \\

This sequence of saddle point dynamics and corresponding transitions in LCS behavior underscores the robustness of the correlation between Lagrangian saddle points and unsteady convective heat transfer. Defining an observational domain and the concept of active saddle points identified an influential region in the flow field bounded by lines $m$ and $n$. This region is critical in governing convective behavior, as it is within this region that the trajectories of active saddle points consistently align with the heat transfer patterns of the heated elliptical cylinder across all angular configurations. This spatial confinement enables predictive insight into when and where significant changes in heat transfer will occur, offering a diagnostic tool for interpreting and anticipating thermal transport dynamics in unsteady bluff body wakes.
The proposed saddle point–driven LCS framework consistently captures heat transfer dynamics across all inclination angles considered in this study, confirming its generality in application.\\

\subsection{Geometric Investigation of Saddle Point Dynamics in the Near-Wake Flow Field and Nusselt Number Trends}

Across the inclination angles $\theta = 0^\circ, 15^\circ, 30^\circ, 45^\circ, 60^\circ, 75^\circ$ and, $90^\circ $, additional insightful observations arise from examining specific saddle points located in the wake region, denoted as $w_1$ (Intersection of LCSs f1 and b11), $ w_2 $ (Intersection of LCSs f2 and b21), and $w_{11}$ (Intersection of LCSs ff1 and b11) as shown in Fig.~\ref{fig: LCS0} to Fig.~\ref{fig: LCS45}. Since the qualitative behaviour of saddle point trajectories remains consistent, the illustrations provided will only be of \( \theta = 45^\circ \). To synthesize the observed correlation between trajectories of saddle points in the wake and thermal behavior, we present a trend-based formulation that qualitatively captures the temporal evolution of the Nusselt number. Rather than predicting absolute values of $Nu$, this representation focuses on the pattern of changes—such as monotonic increases or decreases and the occurrence of slope shifts in response to the spatial trajectories of saddle points. As pointed out earlier, $(x_k(t^*), y_k(t^*))$ denote the coordinates of saddle point $k$ at time $t^*$. The corresponding formulation is given by-

\[
N(t^*) = 
\begin{cases} 
y_{w_1}(t^*) & ,  0 \leq t^* \leq t_{s_1,m} \\
-y_{w_2}(t^*) & ,  t_{s_1,m} \leq t^* \leq t_{s_2,n} \\
y_{w_{11}}(t^*) & , t_{s_2,n} \leq t^* \leq 1
\end{cases}
\]

\begin{figure}[htbp]
    \centering
    \subfloat[$\theta = 0^\circ$\label{fig: emp0}]{%
        \includegraphics[width=0.49\linewidth]{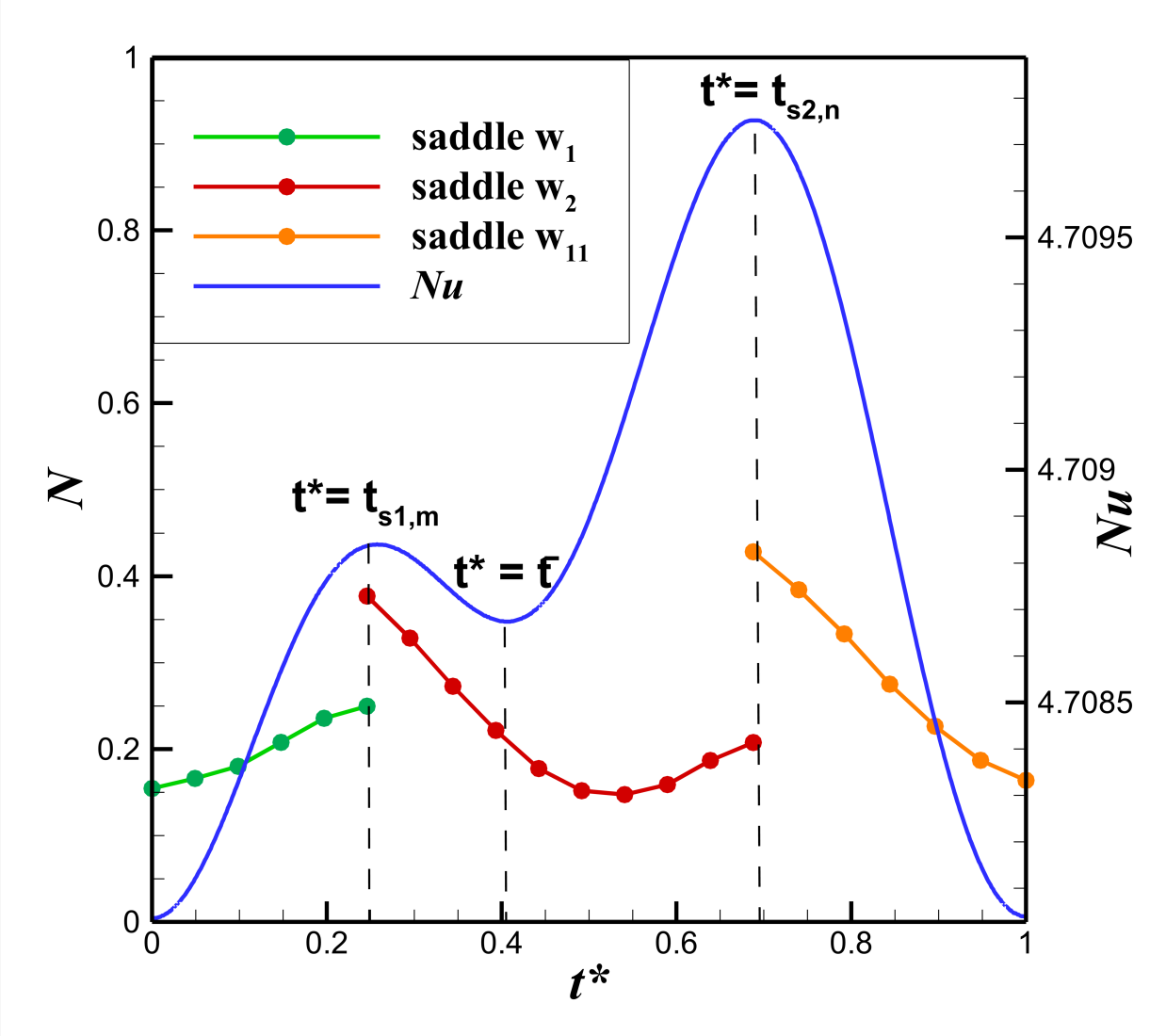}
    }\hfill
    \subfloat[$\theta = 15^\circ$\label{fig: emp15}]{%
        \includegraphics[width=0.49\linewidth]{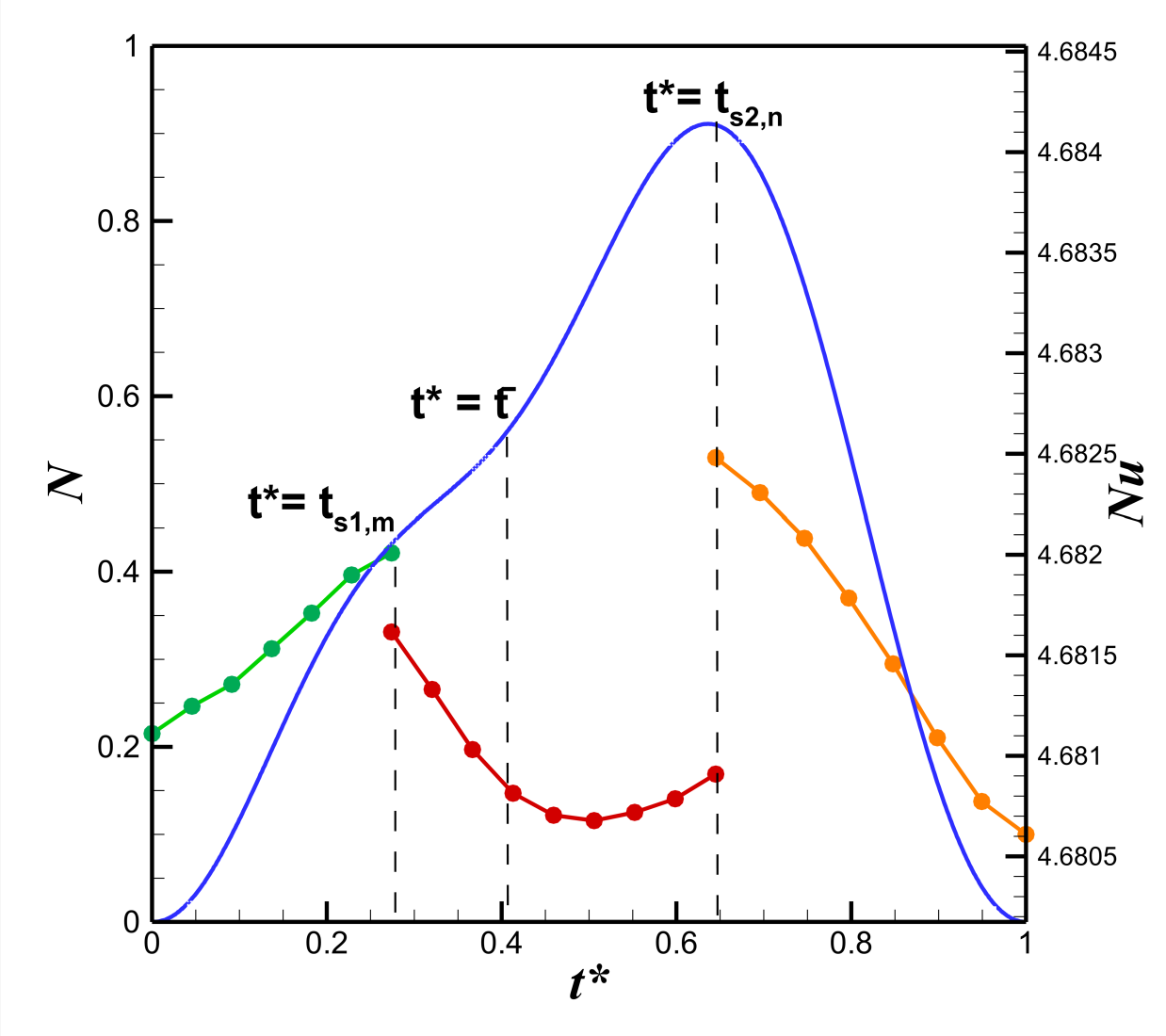}
    }\par\medskip

    \subfloat[$\theta = 30^\circ$\label{fig: emp30}]{%
        \includegraphics[width=0.49\linewidth]{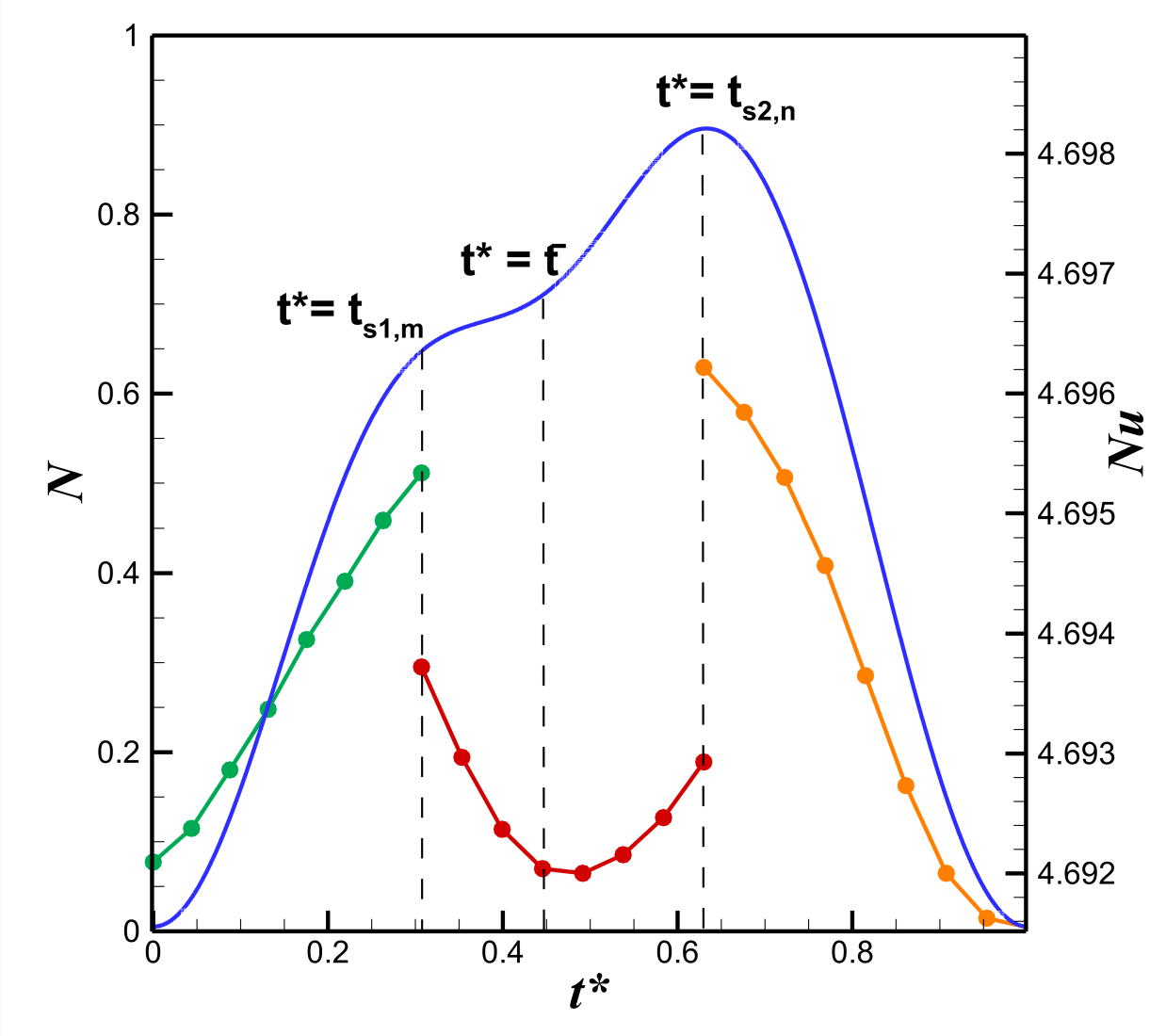}
    }\hfill
    \subfloat[$\theta = 45^\circ$\label{fig: emp45}]{%
        \includegraphics[width=0.49\linewidth]{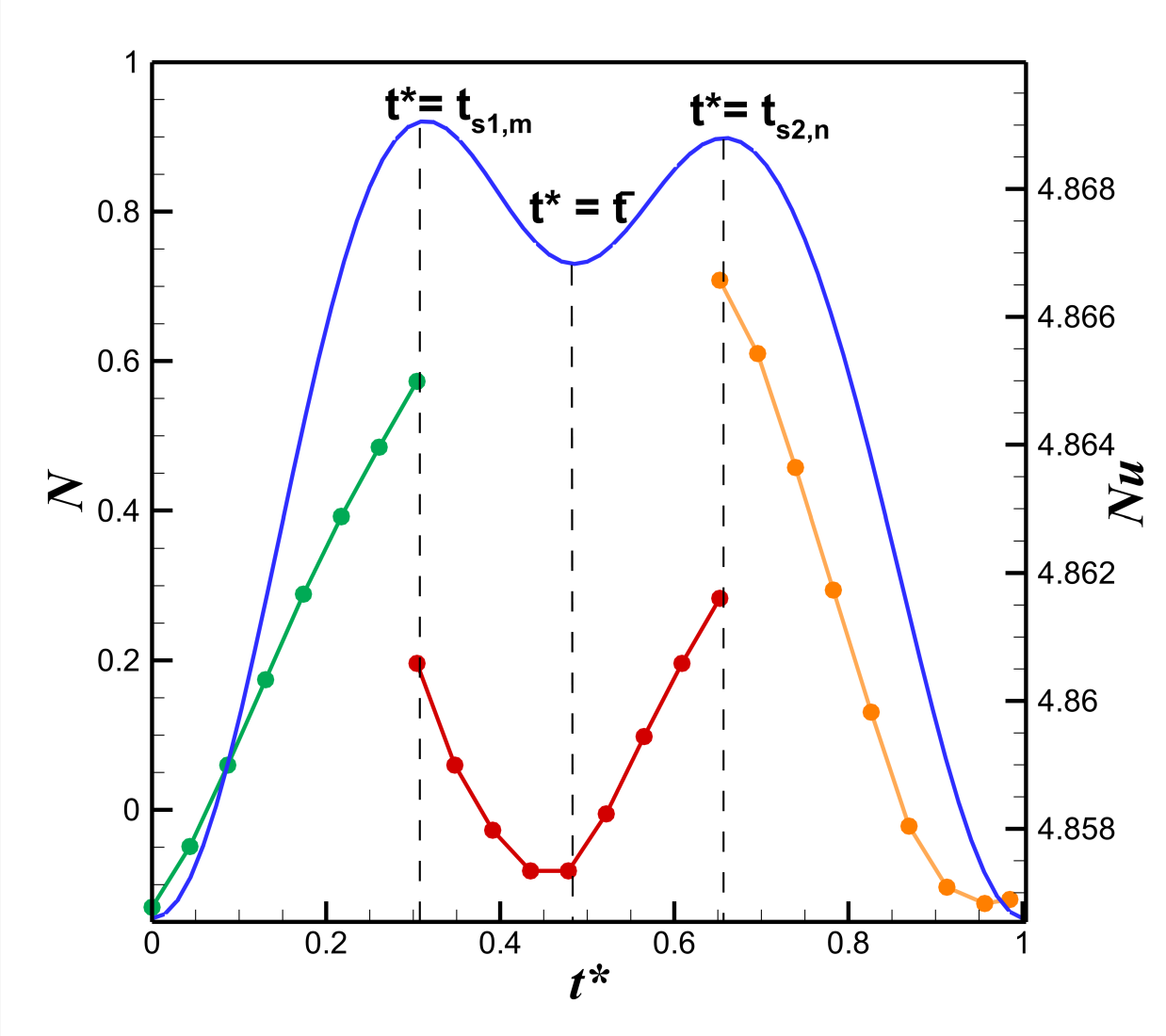}
    }\par\medskip

    \subfloat[$\theta = 60^\circ$\label{fig: emp60}]{%
        \includegraphics[width=0.49\linewidth]{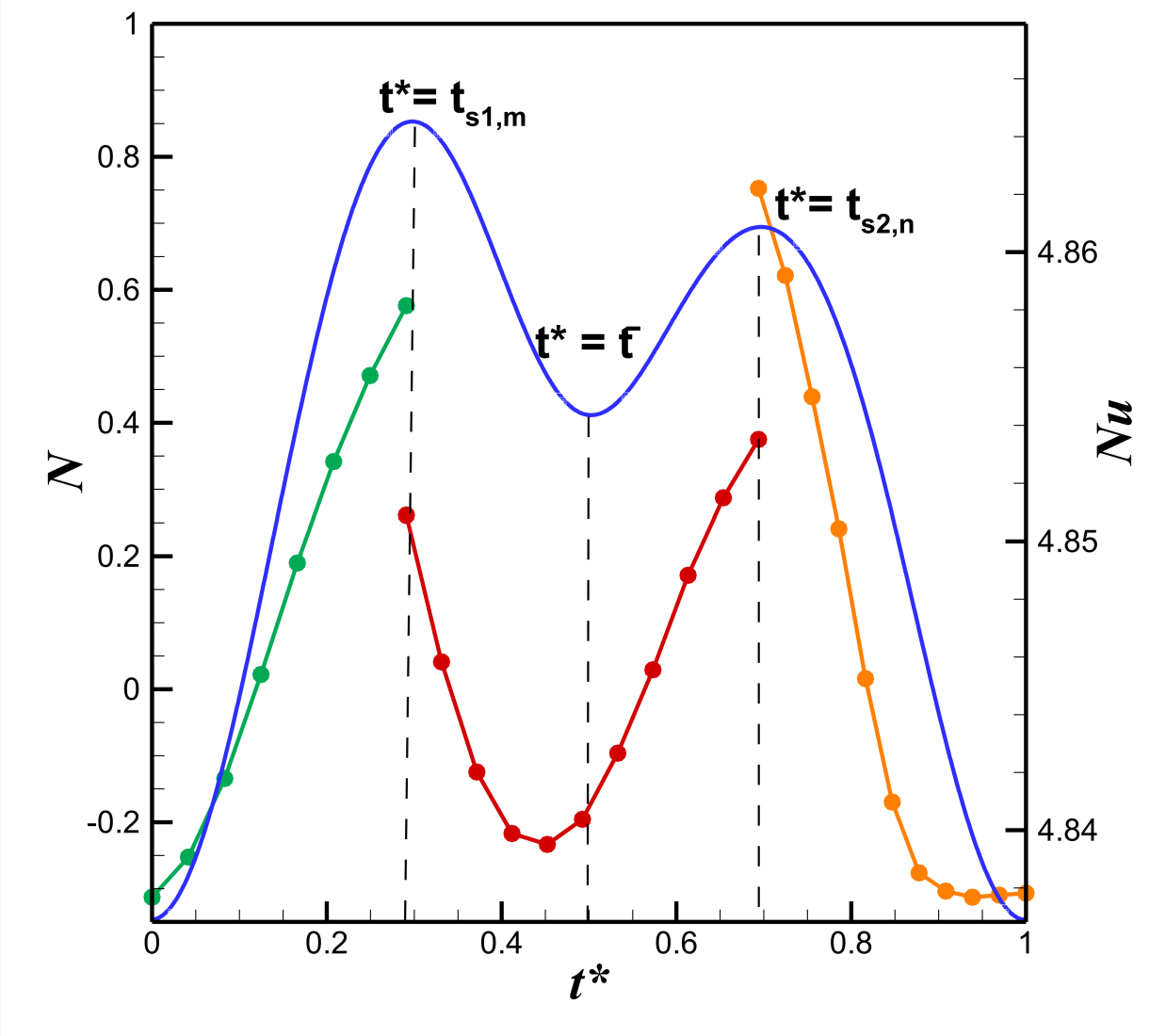}
    }

\end{figure}

\begin{figure}[htbp]\ContinuedFloat
    \centering
    \subfloat[$\theta = 75^\circ$\label{fig: emp75}]{%
        \includegraphics[width=0.49\linewidth]{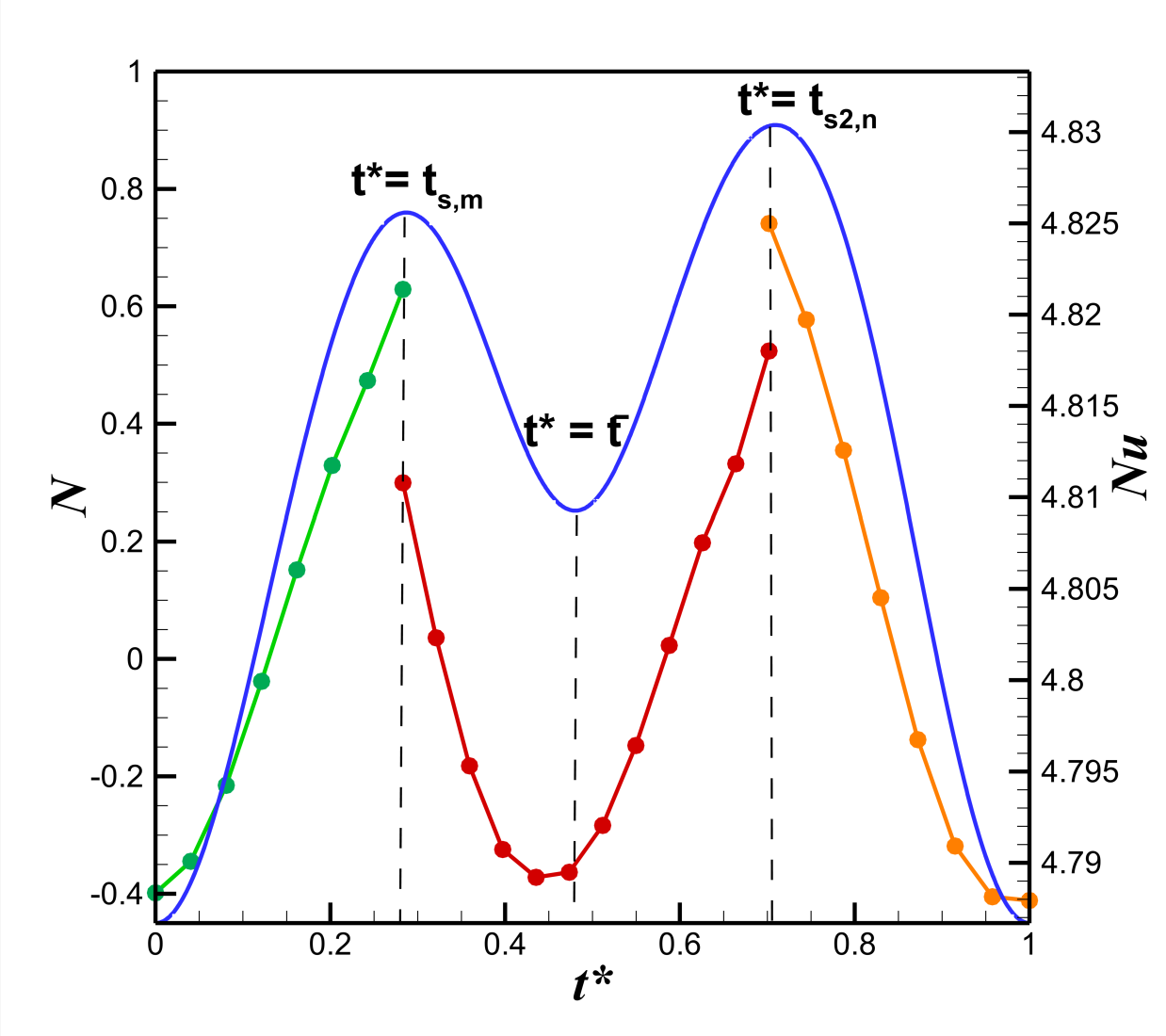}
    }\hfill
    \subfloat[$\theta = 90^\circ$\label{fig: emp90}]{%
        \includegraphics[width=0.49\linewidth]{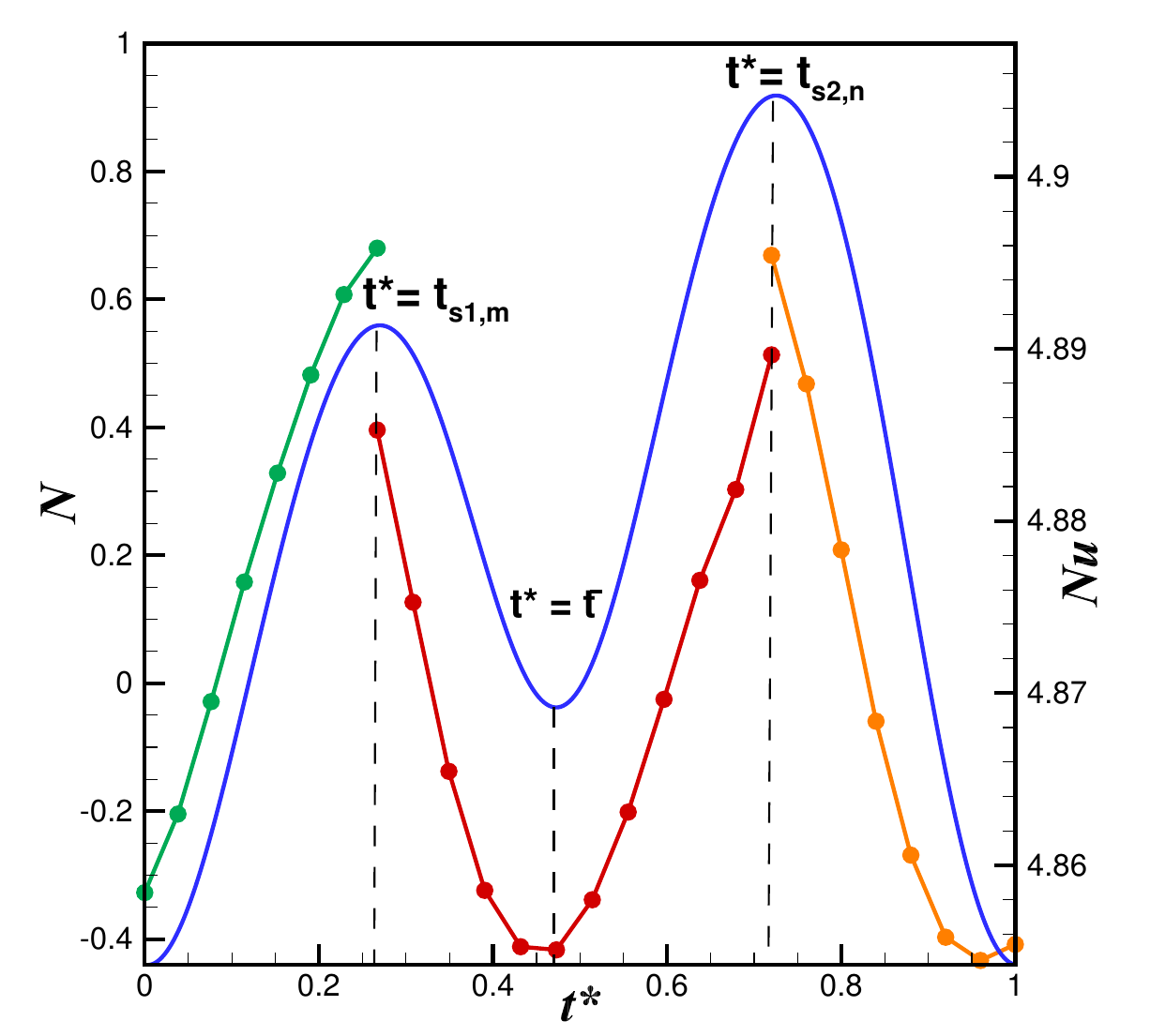}
    }
  \caption{Graphical representation of Temporal function $N(t^*)$ with $Nu$ profile (blue) for $Re = 100$ at : (a) $\theta = 0^\circ$, (b) $\theta = 15^\circ$, (c) $\theta = 30^\circ$, (d) $\theta = 45^\circ$, (e) $\theta = 60^\circ$, (f) $\theta = 75^\circ$, (g) $\theta = 90^\circ$. Colour map of Fig. 12(a) is applied in Figs. 12(b) to 12(g).}
    \label{fig: empirical}
 \end{figure} 

Fig. \ref{fig: empirical} illustrates the graphical represenataion of the function $N(t^*)$ for all inclination angles considered. For lower inclination angles $\theta = 0^\circ$, $15^\circ$ and $30^\circ$, the formation of saddle points in the wake are not very prominent. As the location of saddle points in the flow is independent of the FTLE threshold \cite{rockwood2017detecting}. To locate $w_1, w_2$ and $w_{11}$ in the flow field, the FTLE Threshold is reduced from $50\%$ to $40\%$, and the extraction of these saddle points is carried out. For all angular configurations between $t^*=0$ and $t^*= t_{s_1,m}$ , saddle point $w_1$ initially exhibits upstream motion, progressing against the flow direction as illustrated in Fig. \ref{fig: LCS45IM1} to Fig. \ref{fig: LCS45IM2}. and then begins to convect downstream, simultaneously increasing its distance from the centerline, thus increasing $y_{w_1}$. Thus, $N(t^*)$ increases in the interval $[0,  t_{s_1,m}]$, which aligns with the increasing trend in $Nu$ plot as shown in Fig. \ref{fig: empirical}. The vertical upliftment of saddle point $w_1$ catalyses the shedding of LCS f1, which leads to the increment in heat transfer rate in this interval. \\

In the interval $[t_{s_1,m}, t_{{s_2},n}]$, as seen from Fig. \ref{fig: LCS45IM3} to Fig. \ref{fig: LCS45IM5}, saddle  $w_1$ continues its downstream motion. The repelling LCS f1 fragments into multiple branches and sheds into the far wake region. The saddle point $w_2$ ascends upstream and reaches its maximum vertical position before beginning to drop. After reaching a local maximum vertical position, saddle point $w_2$ begins to descend in the upstream direction. Also, $N(t^*)$ equates to the reflection (w.r.t the x-axis) of the vertical motion of saddle point $w_2$, denoted as $-y_{w_2}$. For $\theta = 15^\circ$ and $30^\circ$, $Nu$ exhibits an increasing trend in the interval $[t_{s_1,m}, t_{{s_2},n}]$ with a slight reduction in the slope at $t^*= \bar{t}$ and finally achieving a relative maximum at $t^*=t_{{s_2},n}$, indicating a more gradual transition in convective heat transfer behavior, shown in Fig.~\ref{fig:NU15} and Fig.~\ref{fig:NU30}. For $t^* \in [t_{s_1,m}, t_{{s_2},n}]$, the negative slope region of $N(t^*)$ corresponds to the reduced slope in $Nu$ profile and the positive slope region of $N(t^*)$ reflects the increased slope of $Nu$. For $\theta = 30^\circ$, the zero slope time instant for 
$N(t^*)$ aligns with the shift to an increased slope region in $Nu$ profile as highlighted in Fig.~\ref{fig: emp30}. For $\theta = 0^\circ$, $45^\circ$, $60^\circ$, $75^\circ$, and $90^\circ$,  the $Nu$ exhibits a decreasing trend from $t^* = t_{s_1,m}$, attains a relative minimum at $t^*= \bar{t}$ and then increases till $t^*=t_{{s_2},n}$ as shown in Fig.~\ref{fig:SurfaceNu}. $N(t^*)$ closely resembles $Nu$ plot for these angles and follows the same montonicity trend, first decreasing, then reaching a local minimum and then increasing within the interval, as seen in Fig. \ref{fig: empirical}.  A striking observation for these inclination angles is the alignment of the local minima of $Nu$ and of the function $N(t^*)$ at the time instant when saddle $w_2$ reaches its peak elevation prior to beginning to drop, i.e., at $t^*= \bar{t}$.  Note that in $[t_{s_1,m},  \bar{t}]$, near the upper surface of the elliptical cylinder, sufficient separation of LCS f2  in horizontal direction does not occur due to the slow convection of saddle $s_2$ parallel to boundary $n$ and and the hinderance in vertical shedding is caused by the upward movement of saddle $w_2$. Additionally, the elongation of the repelling LCS ff1 near the trailing edge acts as a barrier, preventing the mainstream flow from effectively interacting with the cylinder surface. As a result, the convective heat exchange rate ultimately decreases, leading to a decline in $Nu$. Following the time instant  $t^*= \bar{t}$ as observed from Fig. \ref{fig: LCS45IM4} to Fig. \ref{fig: LCS45IM5}, LCS f1 lose the top two branches downstream while branching and the closest branch connects with LCS ff1 and starts falling downwards. The downward vertical motion of saddle $w_2$ facilitated the shedding of LCS f2 resulting in the release of this thermally energized fluid region into the far wake, thus, enhancing the rate of convective heat exchange near the lower surface of the elliptical cylinder. As a result, $Nu$ reaches its second local maximum. Additionally, the formation of a new saddle point, $w_{11}$ is observed during this interval.\\

During the period $t^*=t_{{s_2},n}$ to $t^*=1.0$, saddle point $w_2$ continues its downward motion and $w_{11}$ falls downwards in an arc-like fashion and obtains the initial position of saddle point $s_1$ at $t^*=1.0$ as seen from Fig. \ref{fig: LCS45IM5} to Fig. \ref{fig: LCS45IM7} . The vertical downward displacement of saddle $w_{11}$ and the elongation of the repelling LCS ff1 near the upper surface of the  cylinder corresponds to very slow detachment of LCS ff1. This results in insufficient separation and delayed shedding of LCS ff1 hindering the release of thermally energized fluid into the wake. As a consequence, the rate of convective heat transfer diminishes, reflected by a decreasing trend in the Nusselt number during this time interval. This decrement in $Nu$ profile is captured accurately by the graph of $N(t^*)$. This temporal alignment of $N(t^*)$ highlights the dynamic interplay between saddle point kinematics and local convective heat transfer.\\

\subsection{Quantifying Threshold Displacement of Dominant LCSs for Enhanced Heat Transfer efficiency}

To further substantiate the correlation between LCSs and convective heat transfer behaviour, we present a detailed analysis of the spatial displacements of key LCSs to associate with the prominent peaks and slope transitions observed in the Nusselt number profiles for each angle of inclination. Specifically, we track both the horizontal and vertical displacements of dominant repelling LCSs by following the trajectories of their associated active saddle points. This approach enables the identification of threshold separation distances between repelling LCSs and the elliptical cylinder surface that are indicative of enhanced convective heat transfer.  The horizontal distance traversed by saddle point $k$ from its initial location at $t^*=0$ to its location at the point of intersection with the boundary $l$ is denoted as  $\Delta x_{k, l} = x_{k}(t_{k,l})-x_{k}(0)$. Similarly, the vertical distance traversed by saddle point $k$ from its initial location at $t^*=0$ to its location at the point of intersection with the boundary $l$ is denoted as $\Delta y_{k, l} = y_{k}(t_{{k},l})-y_{k}(0)$.\\

Since LCSs f1 and f2 evolve into LCSs $\text{ff}_1$ and $\text{ff}_2$, respectively, over time period $T $. This analysis utilizes only the repelling  LCSs $\text{ff}_1$ and $\text{ff}_2$ as dominant LCSs to quantify the the spatial extent these structures must traverse downstream to correspond with maximized heat transfer rates in two vortex shedding time period $2T$. The boundaries of the observational domain, denoted by \( m \) and \( n \), serve as reference lines for this quantification.

The vertical displacement of these LCSs is characterized by the intersection of the active saddle points \( s_1 \) and \( s_2 \) with boundaries \( m \) and \( n \), respectively. Accordingly, we define:
\[
\Delta y_{\text{ff}_1} = \Delta y_{s_1, m} \quad \Delta y_{\text{ff}_2} = \Delta y_{s_2, n} 
\]

The horizontal displacement reflects both the net shedding distance and the gradual downstream migration of the slowly detaching LCSs \( \text{ff}_1 \) and \( \text{ff}_2 \), represented by the cumulative motion of their associated saddle points \( s_{\text{f1}} \) and \( s_{\text{f2}} \). For $\theta = 0^\circ, 15^\circ, 30^\circ$, these are given by:
\[
\Delta x_{\text{ff}_1} = \Delta x_{s_1, m} + \left[ x_{s_{\text{f1}}}(1) - x_{s_{\text{f1}}}(0) \right], \quad
\Delta x_{\text{ff}_2} = \Delta x_{s_2, n} + \left[ x_{s_{\text{f2}}}(1) - x_{s_{\text{f2}}}(0) \right].
\]

As for $\theta = 45^\circ, 60^\circ, 75^\circ, 90^\circ$, saddle point \( s_{\text{f2}} \) emerges prominently at $t^* = t_{s_1,m}$. So, $\Delta x_{\text{ff}_2}$ for these angles is given by:
\[
\Delta x_{\text{ff}_2} = \Delta x_{s_2, n} + \left[ x_{s_{\text{f2}}}(1) - x_{s_{\text{f2}}}(t_{s_1,m}) \right].
\]

Together, these displacement metrics capture the spatial extent of LCS evolution that leads to significant entrainment of thermal structures and optimizes heat transfer rates. By quantifying these distances, we establish spatial thresholds for LCS–surface separation that correlate with convective efficiency. eveloping. By leveraging these displacement metrics, it is possible to forecast the spatial configuration of LCSs that maximize heat transfer in a reduced-order predictive models of unsteady convection. This enables the prediction of isotherms configuration corresponding to enhanced convective heat transfer in the unsteady wake of bluff bodies without resorting to full-scale numerical simulations. 

The comparative data for all inclination angles investigated is summarized below in Table~\ref{tab:geometry-data}. 

\begin{table}[h!]
\centering
\caption{Displacement metrics for Varying Inclination Angles}
\label{tab:geometry-data}
\begin{tabular}{cccccccc}
\toprule
$\theta$ & 0$^\circ$ & 15$^\circ$ & 30$^\circ$ & 45$^\circ$ & 60$^\circ$ & 75$^\circ$ & 90$^\circ$ \\
\midrule
$\Delta x_{\text{ff}_1}$   & 0.9496   & 1.052  & 0.9795  & 0.9232  & 0.8082   & 0.7818 & 0.7976\\
$\Delta y_{\text{ff}_1}$   & 0.0855   & 0.0905  & 0.1196  & 0.108  & 0.0963  & 0.108 & 0.1051 \\
$\Delta x_{\text{ff}_2}$ & 1.0477   & 0.9937  & 0.9466  & 0.7784  & 0.516   & 0.6783  & 0.8119\\
$\Delta y_{\text{ff}_2}$  & 0.059   & 0.0642  & 0.0671  & 0.0817  & 0.0992   & 0.0788  & 0.1372\\
\bottomrule

\end{tabular}
\end{table}

\subsection{Sensitivity analysis of the Observational Domain}

\begin{figure}[htbp]
    \centering
    \subfloat[$40\%$\label{fig:40m}]{\includegraphics[width=0.45\linewidth]{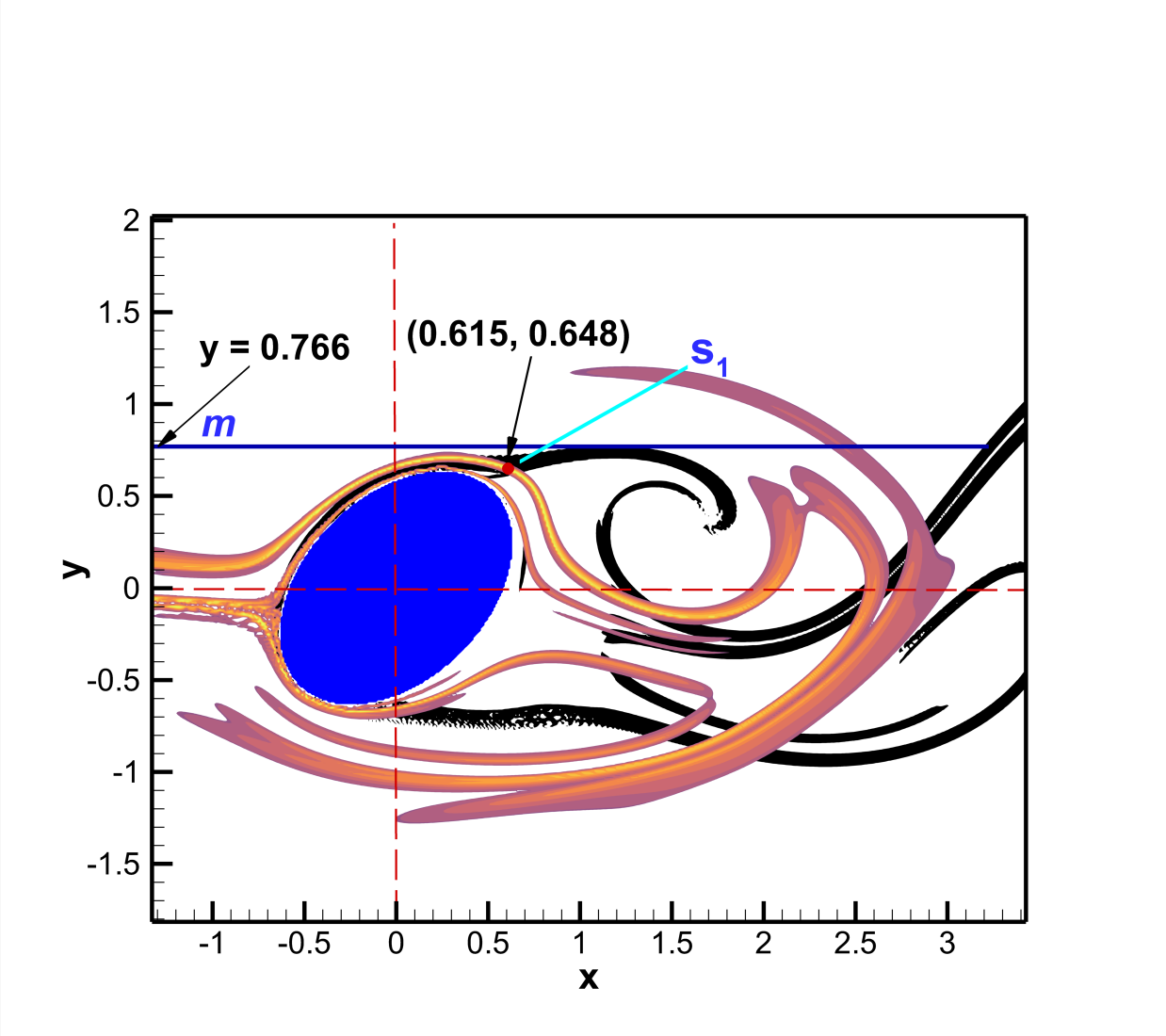}}\hfill
    \subfloat[$45\%$\label{fig:45m}]{\includegraphics[width=0.45\linewidth]{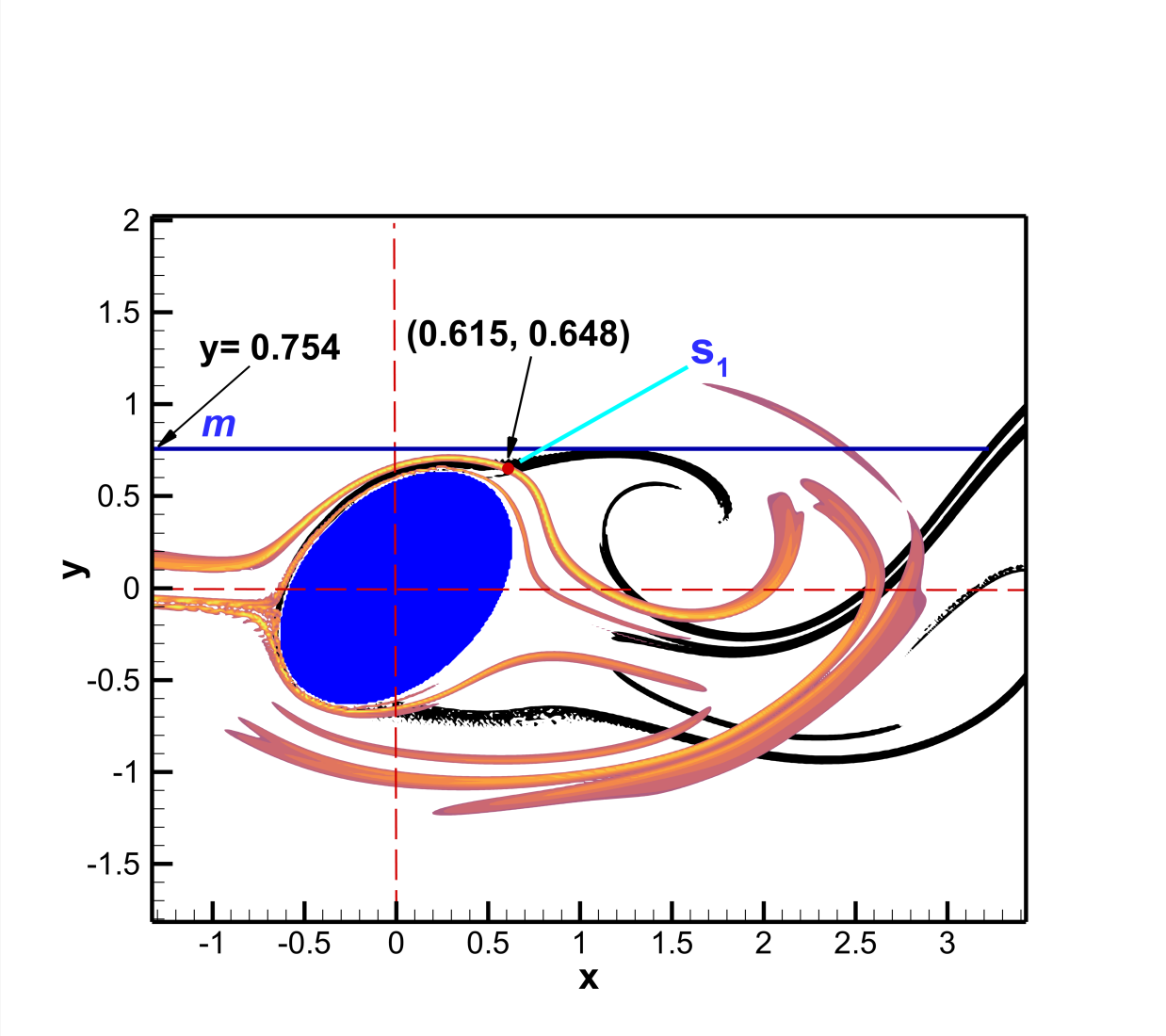}}\par\medskip
    \subfloat[$50\%$\label{fig:50m}]{\includegraphics[width=0.45\linewidth]{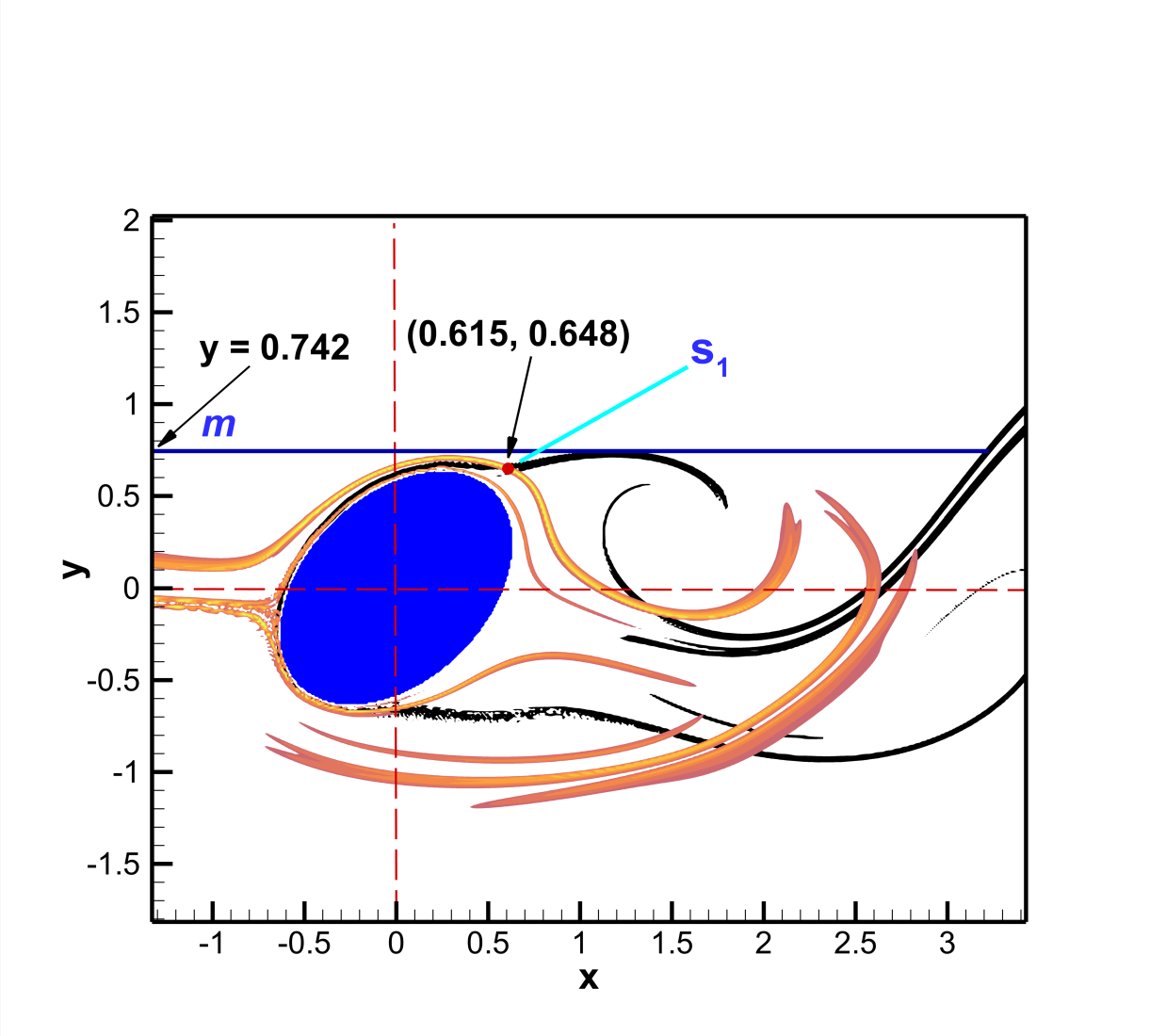}}\hfill
    \subfloat[$55\%$\label{fig:55m}]{\includegraphics[width=0.45\linewidth]{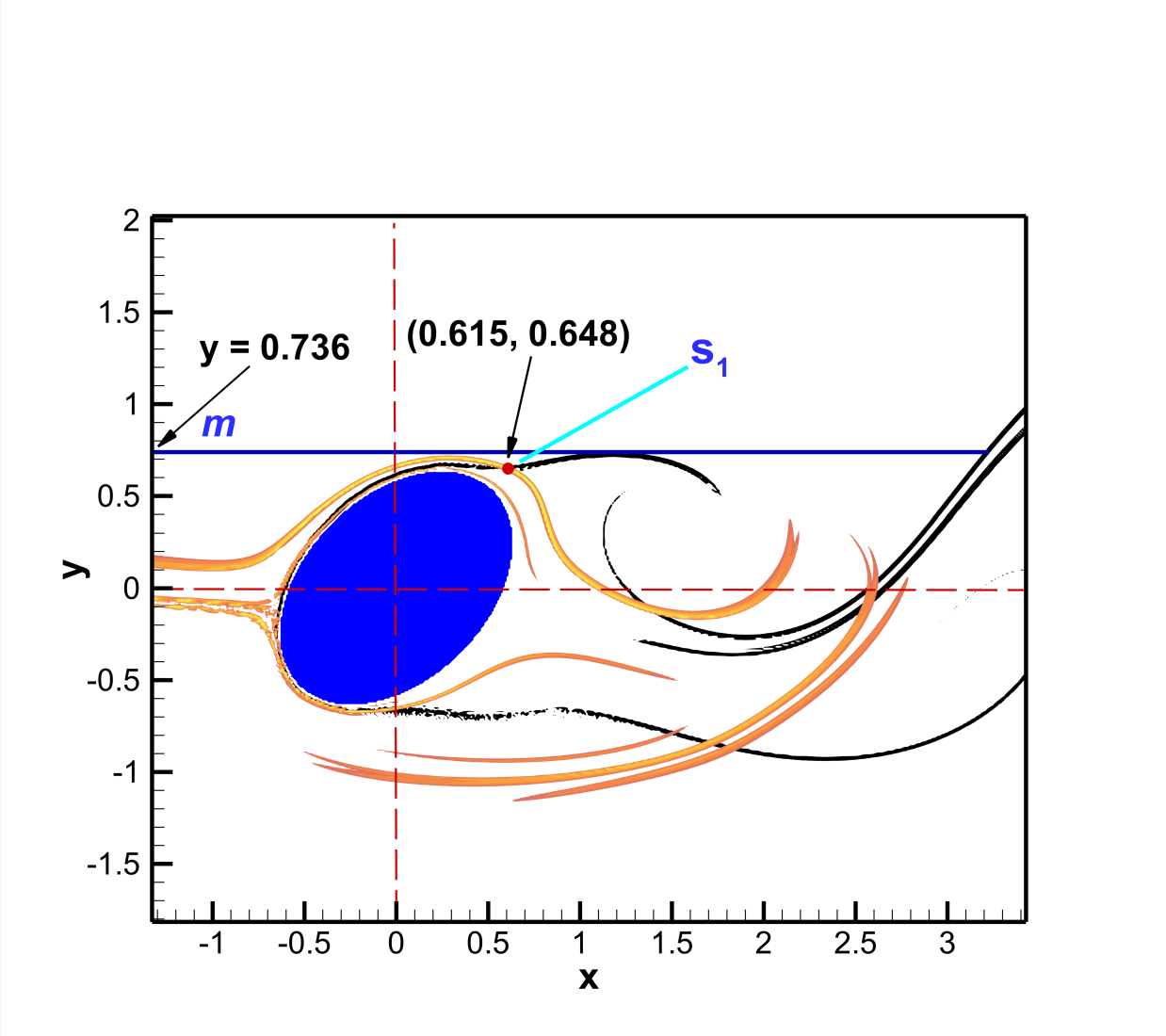}}\par\medskip
    \subfloat[$60\%$\label{fig:60m}]{\includegraphics[width=0.45\linewidth]{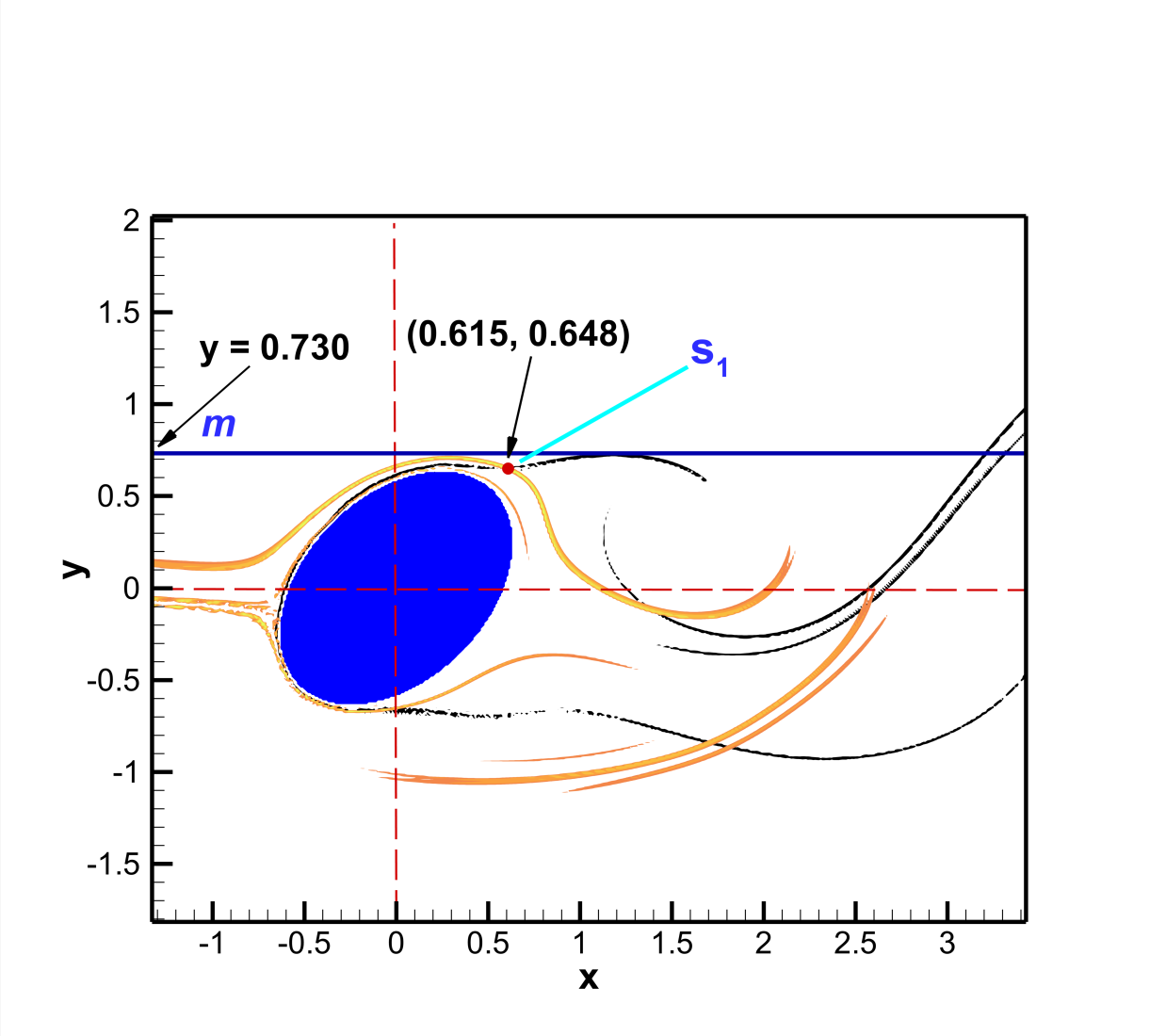}}\hfill
    \subfloat[$65\%$\label{fig:65m}]{\includegraphics[width=0.45\linewidth]{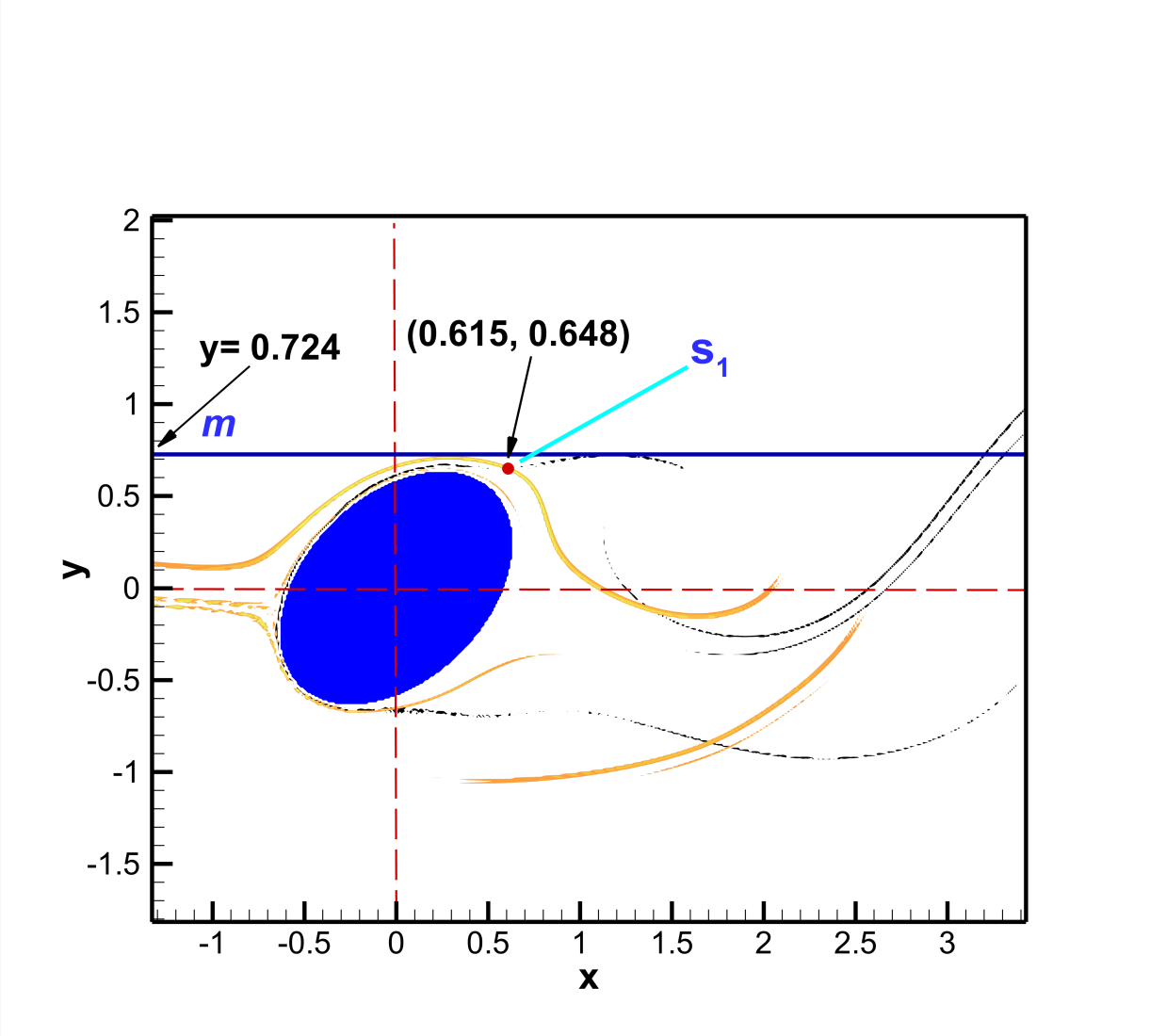}} 
    \caption{Location of line $m$ and saddle point $s_1$ with FTLE plots at $t^* = 0$ for $\theta=45^\circ$ at $Re=100$ with FTLE threshold (in $\%$) at : (a) $40\%$ , (b) $45\%$ , (c) $50\%$ , (d) $55\%$ , (e) $60\%$ , (f) $65\%$.}
    \label{fig:app_btle}
\end{figure}

\begin{figure}[htbp]
    \centering
    \subfloat[$40\%$\label{fig:40n}]{\includegraphics[width=0.45\linewidth]{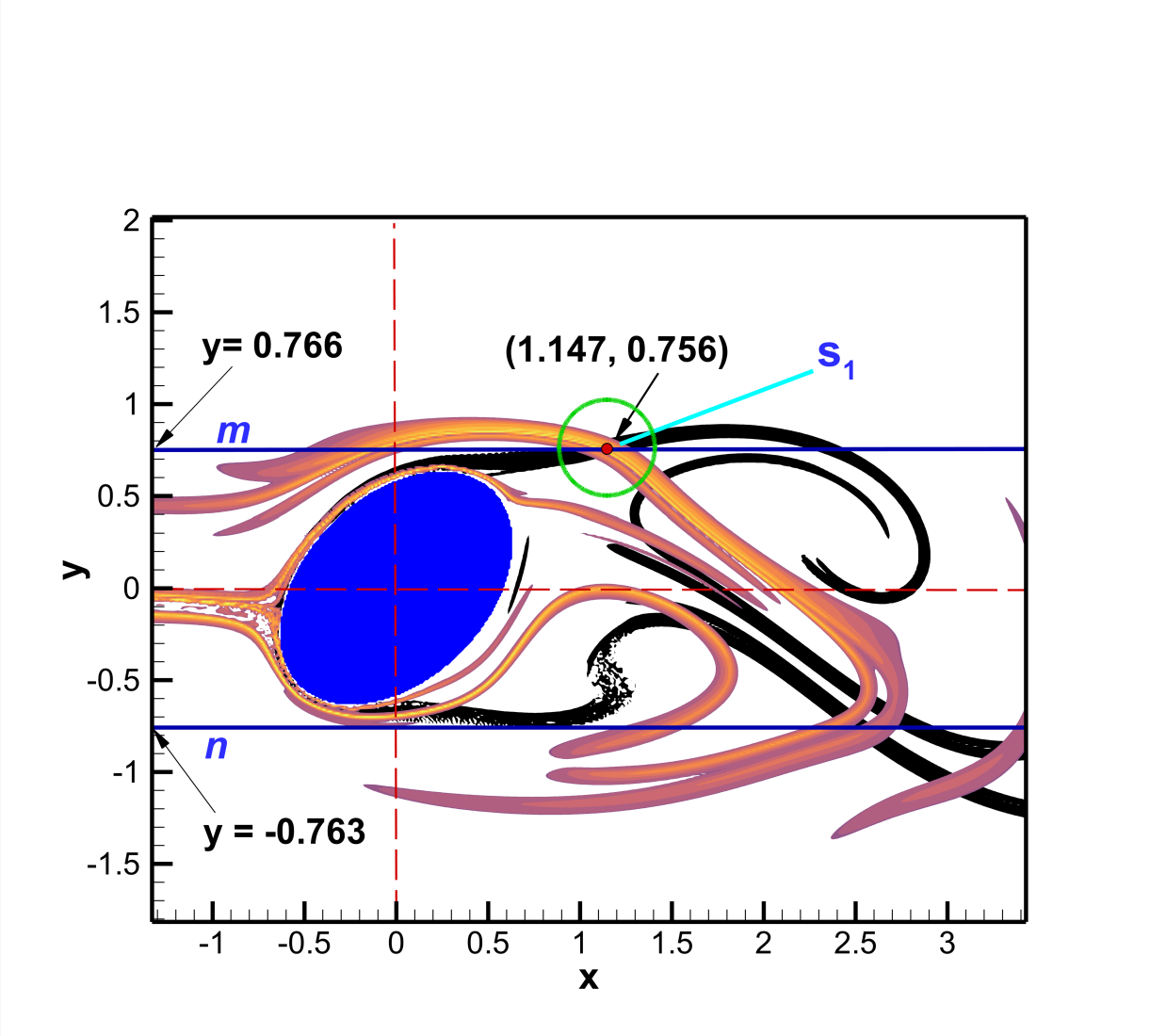}}\hfill
    \subfloat[$45\%$\label{fig:45n}]{\includegraphics[width=0.45\linewidth]{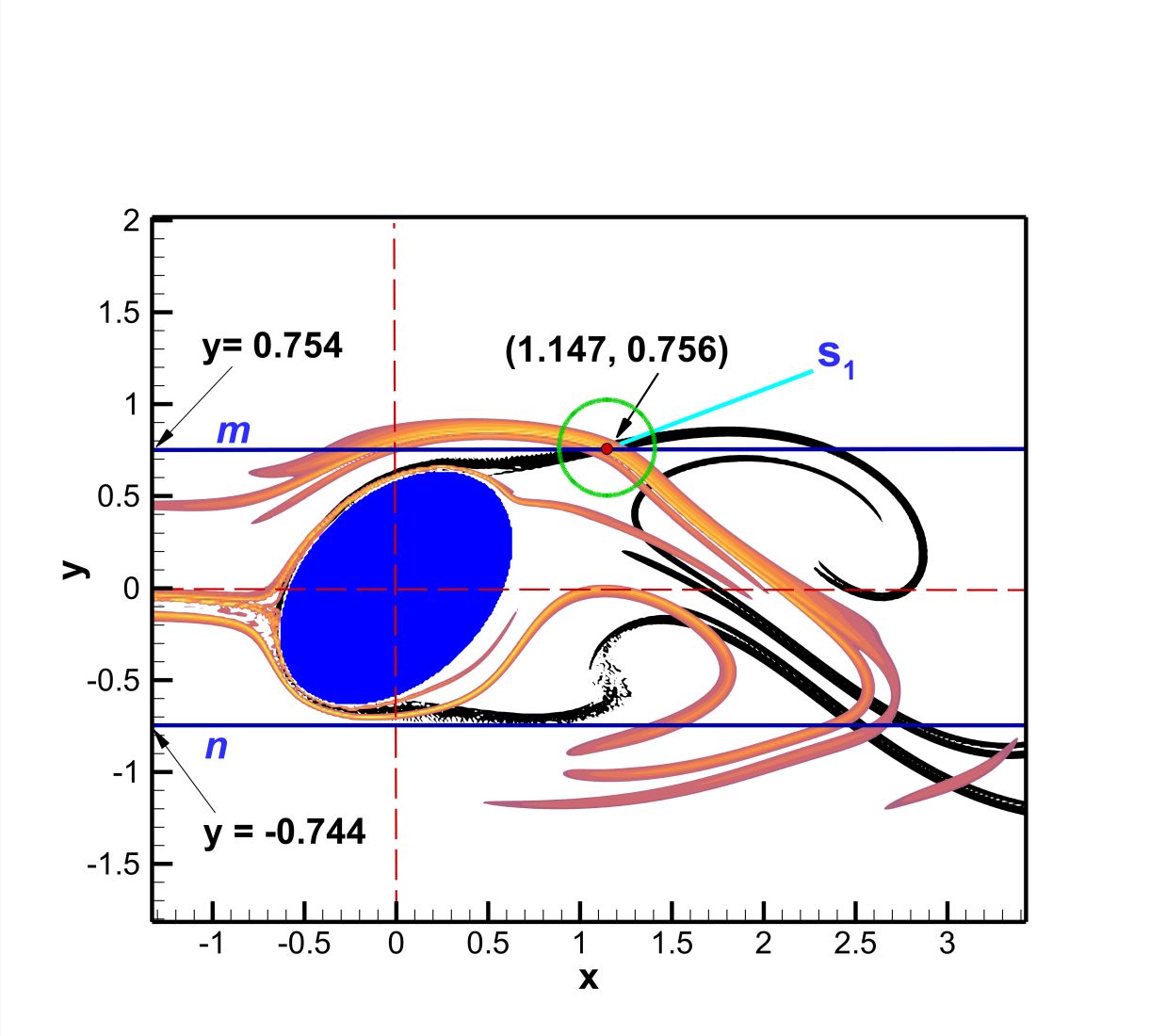}}\par\medskip
    \subfloat[$50\%$\label{fig:50n}]{\includegraphics[width=0.45\linewidth]{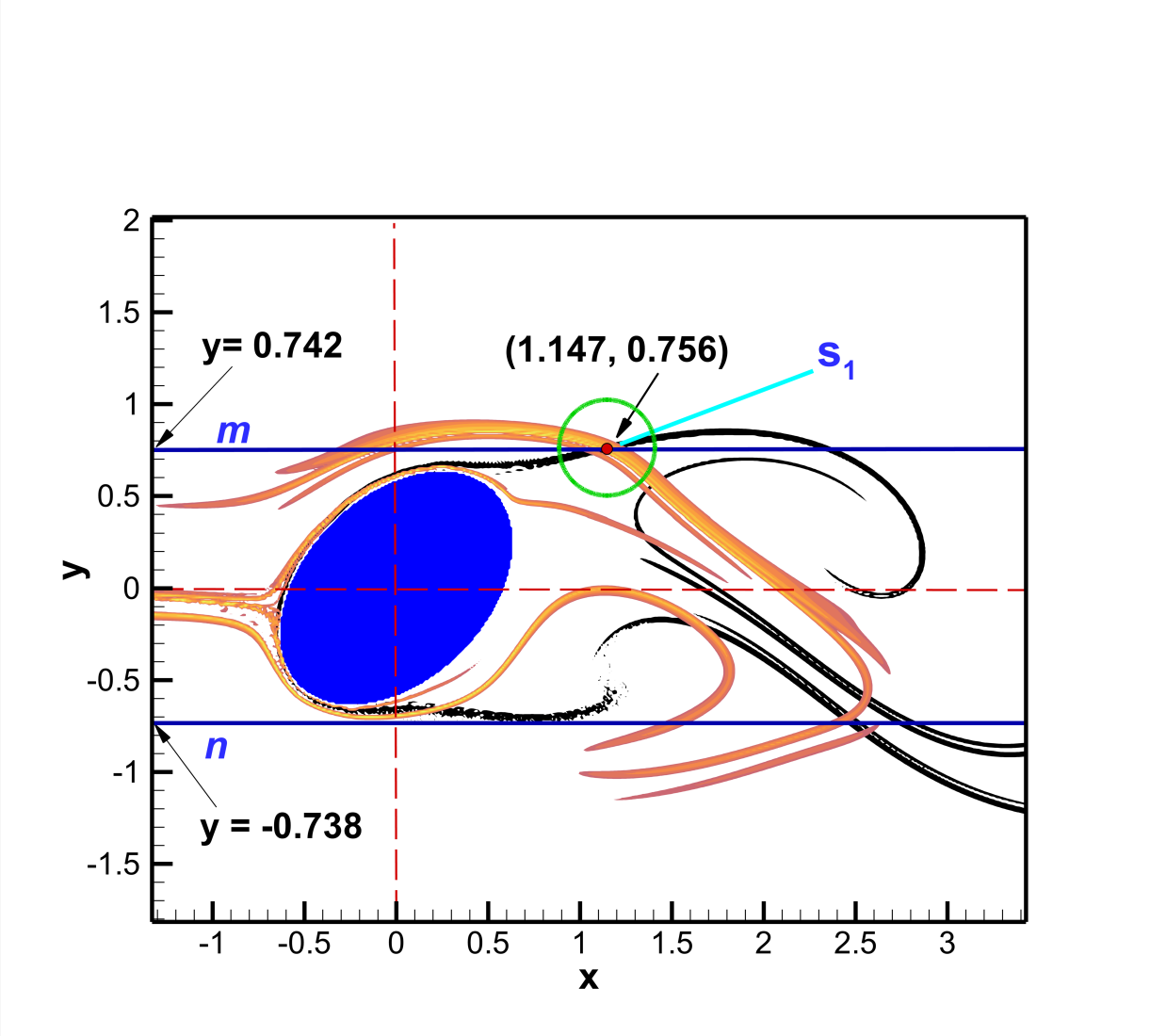}}\hfill
    \subfloat[$55\%$\label{fig:55n}]{\includegraphics[width=0.45\linewidth]{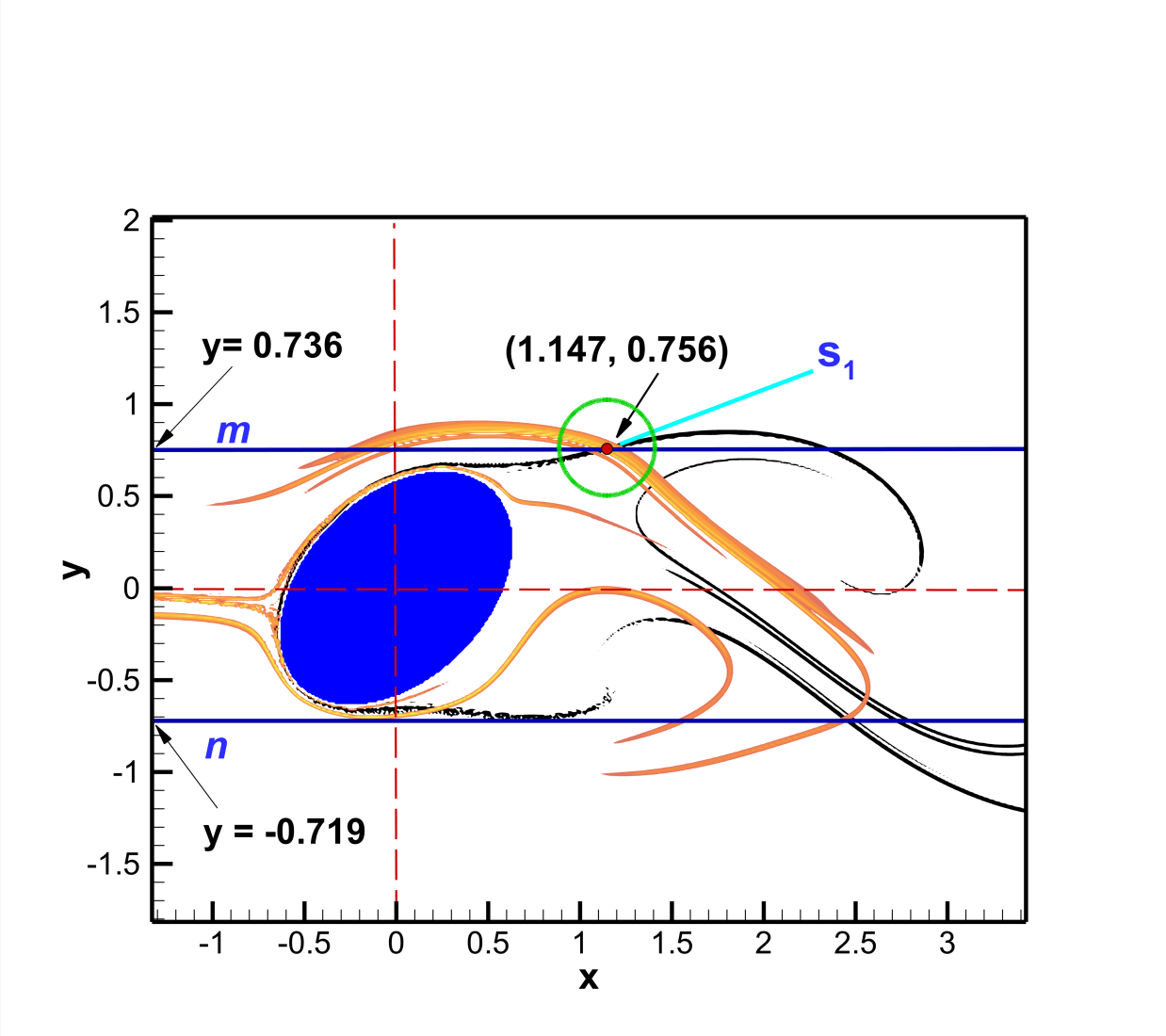}}\par\medskip
    \subfloat[$60\%$\label{fig:60n}]{\includegraphics[width=0.45\linewidth]{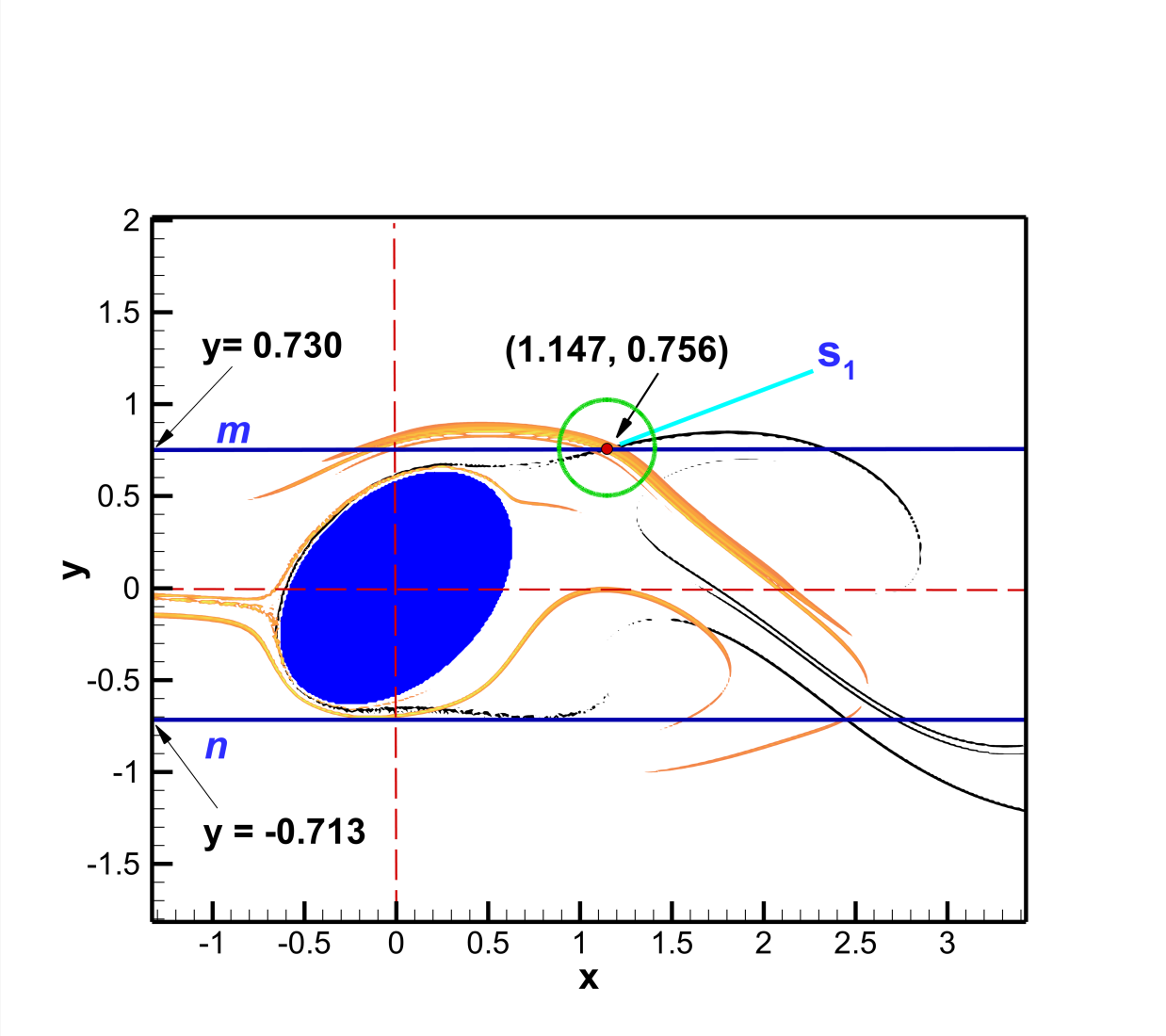}}\hfill
    \subfloat[$65\%$\label{fig:65n}]{\includegraphics[width=0.45\linewidth]{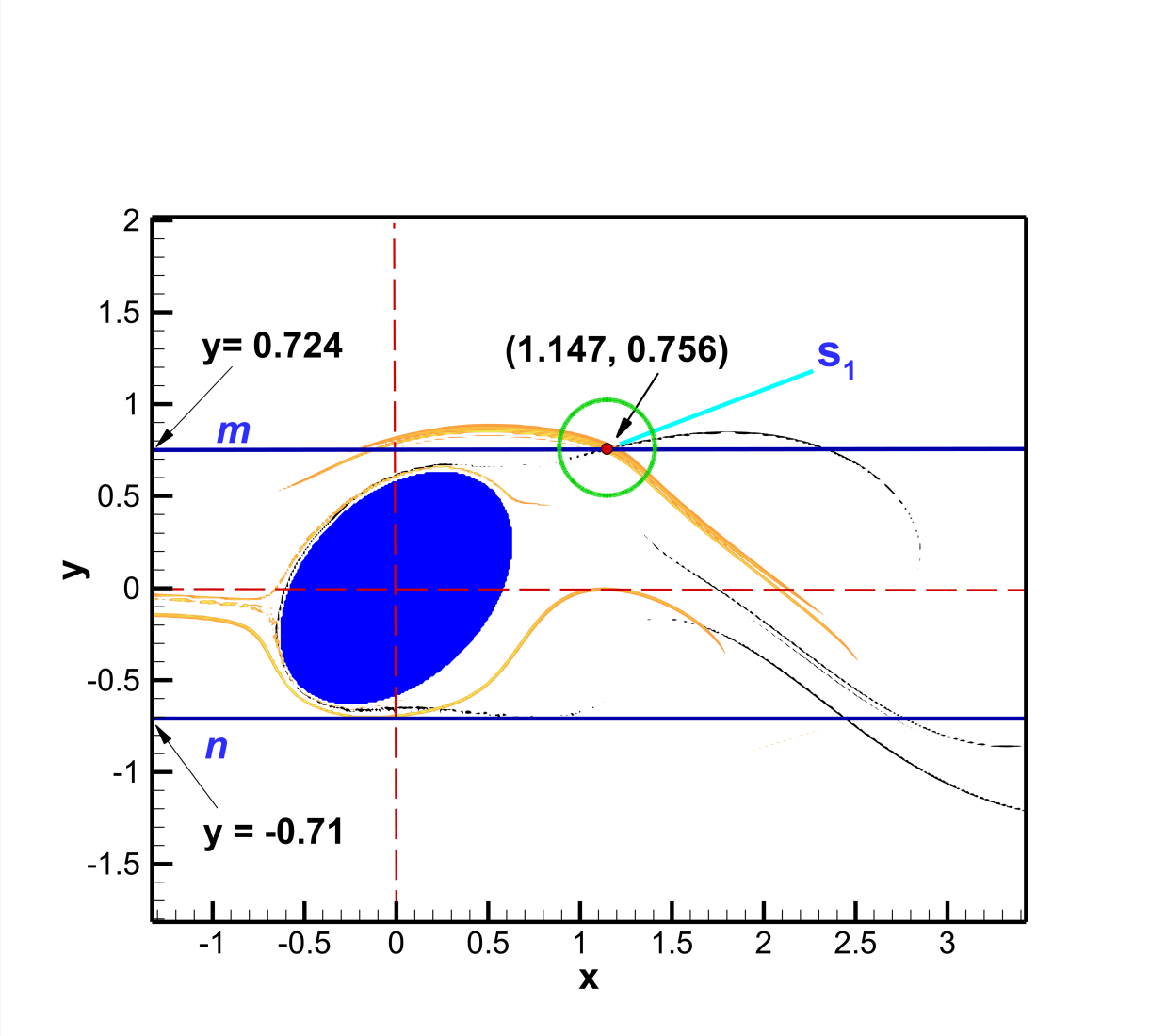}}
    \caption{Location of line $m$, $n$ and saddle point $s_1$ with FTLE plots at $t^* = t_{s_1,m}$ for $\theta=45^\circ$ at $Re=100$ with FTLE threshold (in $\%$) at : (a) $40\%$ , (b) $45\%$ , (c) $50\%$ , (d) $55\%$ , (e) $60\%$ , (f) $65\%$.}
    \label{fig:app_ftle}
\end{figure}

In order to extract LCSs, both forward-time FTLE (fFTLE) and backward-time FTLE (bFTLE) scalar fields are filtered by applying a user defined threshold. The placement of bounadry lines $m$ and $n$ of the observational domain is sensitive to the chosen threshold for the bFTLE scalar field, as the FTLE threshold percentage affects the apparent thickness of the attracting LCSs (b1 and b2), as demonstrated in Fig.~\ref{fig:app_btle}. Since attracting and repelling LCSs correspond to ridges in the FTLE field, varying the FTLE threshold alters their visibility and clarity. For illustrative purposes, we explore FTLE threshold values ranging from $40\%$ to $65\%$. At thresholds above $65\%$, the FTLE ridges become dimnished for effective interpretation. Accordingly, Figs.~\ref{fig:app_btle} and \ref{fig:app_ftle} present filtered bFTLE and fFTLE fields for flow past an elliptical cylinder inclined at $\theta = 45^\circ$ at varying thresholds to demonstrate the impact of this filtering on LCS identification and the positioning of boundaries of the observational domain $m$ and $n$.\\

To accurately identify the intersection of fFTLE and bFTLE ridges, it is essential to minimize the visual thickness of the FTLE ridges. Due to this, a filtering threshold of $60\%$ of the maximum FTLE value is applied to reduce ridge thickness and identifying saddle points while preserving key structural features. The intersections of these saddle points with boundary lines $m$ and $n$ are critical observations within the defined observational domain. As shown in Figs. \ref{fig:app_btle} and \ref{fig:app_ftle}, the locations of saddle points—determined by the intersections of fFTLE and bFTLE ridges—remain robust against variations in the FTLE threshold. It is noteworthy that varying the FTLE ridge extraction threshold influences only the visual thickness of the identified structures, without affecting their spatial location or topology \cite{rockwood2017detecting}. The position of the boundaries $m$ and $n$ shift slightly with increasing threshold percentage and the location of the intersection between line $m$ and saddle point $s_1$ at $t^* = t_{s_1,m}$ remains consistent across Figs. \ref{fig:40n} to \ref{fig:65n}. The table below summarizes the sensitivity of boundary lines $m$ and $n$ with respect to different FTLE filtering thresholds. Note that since lines $m$ and $n$ are parallel to $x$-axis, their location will be given by $y$ coordinate as shown in Fig.~\ref{fig:app_btle} and \ref{fig:app_ftle}.

\begin{table}[h!]
\centering
\caption{Location of the boundary lines $m$ and $n$ for varying FTLE thresholds}
\begin{tabular}{ccc}
\hline
\makecell{FTLE Threshold\\(in \%)} & Line $m$ & Line $n$ \\
\hline
40\% & 0.766 & -0.763 \\
\hline
 45\% & 0.754 & -0.744 \\
\hline
 50\% & 0.742 & -0.738 \\
\hline
 55\% & 0.736 & -0.719  \\
\hline
 60\% & 0.730 & -0.713\\
\hline
 65\% & 0.724 & -0.71 \\
\hline
\end{tabular}
\label{tab:line_locations_transposed}
\end{table}

\section{Concluding Remarks}\label{sec:conclusion}

This work is concerned with the exploration of the dynamics of forced convective heat transfer in the unsteady wake of an inclined elliptical cylinder using a Lagrangian framework rooted in Finite-Time Lyapunov Exponent (FTLE) analysis. By focusing on the evolution of Lagrangian saddle points—formed at the intersection of repelling and attracting Lagrangian Coherent Structures (LCSs)—we have established a novel correlation between flow kinematics and surface thermal response. A key outcome of this study is the definition of an observational domain, bounded by prominent LCS ridges, within which the motion of carefully identified \textit{active saddle points} provides a robust interpretation of convective heat transfer trends. To the best of our knowledge, this is the first demonstration of a direct correlation between Lagrangian saddle point trajectories and the temporal evolution of the surface-averaged Nusselt number in bluff body wakes. To this end, we compute both forward-time and backward-time Finite-Time Lyapunov Exponent (FTLE) fields, filtering them with adaptive thresholds to isolate the most dynamically relevant structures. The framework built around this concept successfully captures essential features of the Nusselt number profiles across all examined inclination angles ($\theta = 0^\circ$ to $90^\circ$). For intermediate angles ($\theta = 15^\circ, 30^\circ$), it reveals slope transitions, while for higher inclinations, it accurately locates local peaks and troughs in heat transfer performance. The nature of saddle point displacement was shown to govern heat transport efficiency. Oblique migration of active saddle points, associated with the vigorous shedding of repelling LCSs, leads to enhanced convective transfer, whereas parallel displacement—linked to slower detachment dynamics—results in weakened heat removal. This geometric insight introduces a useful physical distinction between effective and subdued thermal episodes in the flow. A time-dependent function based on the motion of specific saddle points was constructed, reflecting key qualitative features of the Nusselt number evolution—such as slope changes and trend reversals—thereby providing a reduced-order representation of the thermal behavior. Such models could provide rapid estimates of Nusselt number evolution in varying angular configurations, enabling efficient design-space exploration and real-time control in engineering applications where computational resources or time are constrained. To deepen the predictive power of the framework, the study quantified the threshold distances repelling LCSs must travel from the surface to trigger enhanced heat transfer. These displacement metrics, derived from saddle trajectories, serve as an indicator of thermal activity and offer practical relevance in engineering flow control or surface cooling strategies.

In summary, the proposed Lagrangian framework not only provides a new tool to interpret unsteady convective processes but also bridges coherent structure analysis and thermal performance prediction and establishes a reproducible methodology for associating transport structures in the flow with surface-level heat transfer phenomena. This contributes significantly to the broader understanding of flow–thermal interaction in geometrically complex and time-dependent fluid systems. The motion of these active saddles exhibits a strong correlation with slope transitions in the temporal profile of the surface-averaged Nusselt number, thereby establishing a physically interpretable connection between Lagrangian saddle dynamics and unsteady surface heat transfer behavior in engineering and geophysical flows.

\end{document}